\documentclass[10pt]{article}
\usepackage[a4paper,textwidth=6.5in,top=1in,bottom=1in]{geometry}
\usepackage[T1]{fontenc}
\usepackage{lmodern}
\usepackage[round]{natbib}
\usepackage{amsmath,amssymb,amsthm}
\usepackage{thmtools,thm-restate}
\newif\ifrecalling
\DeclareRobustCommand{\restatetitle}[1]{\ifrecalling restated\else #1\fi}
\newcommand{\recall}[1]{{\recallingtrue #1*}}
\usepackage{mathrsfs}
\usepackage{booktabs}
\usepackage{graphicx}
\usepackage{xcolor}
\usepackage[colorlinks=true,linkcolor=blue!60!black,citecolor=blue!60!black,urlcolor=blue!60!black]{hyperref}
\usepackage{url}

\newtheorem*{thmAinf}{Theorem A (informal)}
\newtheorem*{thmBinf}{Theorem B (informal)}
\newtheorem{theorem}{Theorem}[section]
\newtheorem{lemma}[theorem]{Lemma}
\newtheorem{proposition}[theorem]{Proposition}

\theoremstyle{definition}
\newtheorem{definition}[theorem]{Definition}
\newtheorem{remark}[theorem]{Remark}
\numberwithin{equation}{section}

\newcommand{\R}{\mathbb{R}}
\newcommand{\C}{\mathbb{C}}
\newcommand{\N}{\mathbb{N}}
\newcommand{\Hs}{\mathcal{H}}
\newcommand{\Qs}{\mathcal{Q}}
\newcommand{\Ac}{\mathcal{A}}
\newcommand{\Kc}{\mathcal{K}}
\newcommand{\Lc}{\mathcal{L}}
\newcommand{\Sc}{\mathcal{S}}
\newcommand{\Ls}{\mathscr{L}}
\newcommand{\Pm}{\mathbb{P}}
\newcommand{\ip}[2]{\langle #1,#2\rangle}
\newcommand{\En}{\mathsf{E}}
\newcommand{\Mo}{\mathsf{M}}
\newcommand{\Sph}{S^{2n-1}}
\DeclareMathOperator{\conv}{conv}
\DeclareMathOperator{\sgn}{sgn}
\DeclareMathOperator{\dist}{dist}
\DeclareMathOperator{\ran}{ran}
\DeclareMathOperator{\spn}{span}
\DeclareMathOperator{\Real}{Re}

\renewcommand{\Re}{\Real}

\title{Stationary points of sliced $\mathcal W_1$ tomography with neural quantum states: finite and infinite width for networks of Gaussian units with trainable centers}

\author{Takahiro Kajisa\\[2pt]
{\normalsize Department of Informatics, The University of Electro-Communications, Tokyo, Japan}\\
{\normalsize\texttt{t.kajisa@uec.ac.jp}}}
\date{}

\begin{document}
\maketitle

\begin{abstract}
We consider continuous-variable quantum state tomography in which the wave function of a pure state is represented by a neural network and one minimizes a loss that aggregates over directions the 1-Wasserstein distances between quadrature measurement distributions. The prior work (Kajisa, NeurIPS 2026) showed, for the loss aggregated by the maximum over directions, that every critical point at which the loss is differentiable is a global minimizer, but its proof uses an overparameterization assumption that does not hold at finite width. The aim of this paper is to remove this assumption, and we prove two theorems for one-hidden-layer networks of Gaussian units in which the centers are initialized at random and both the amplitudes and the centers are trained. The aggregation is the $L^q$ norm over directions ($1\le q\le\infty$), which includes sliced $\mathcal W_1$ ($q=1$) and max-sliced $\mathcal W_1$ ($q=\infty$). Theorem~A concerns all $q$. As the width tends to infinity, every limit of a sequence of stationary points whose exponential moments of the photon number and displacements of the centers are bounded either coincides with the target up to a global phase or lies outside a neighborhood of the target that does not depend on the width. Theorem~B corresponds to the theorem of the prior work. For $q=\infty$, if the loss is differentiable at the limit, then the limit coincides with the target even if it lies outside this neighborhood. At finite width as well, the infidelity of a stationary point inside this neighborhood is at most the square of the error of approximating the target by the range of the derivative with respect to the parameters, multiplied by a factor that is a power of a logarithm. In a numerical example with a cat state as the target, the infidelity at the point where the minimization stopped is between $1.05$ and $1.63$ times the square of the error of approximating the target by the linearization of the network along the directions used in the minimization.
\end{abstract}

\section{Introduction}\label{sec:intro}

Tomography of a continuous-variable quantum state is a nonlinear inverse problem in which a unit vector of $L^2(\R^n)$ is recovered from the distributions of the measured values of quadratures. If a pure state of $n$ optical modes is represented by a unit vector $\phi$, then a measurement by an interferometer and homodyne detection yields, for each direction $\theta\in\Sph$, samples from the distribution $\Pm^\theta_\phi$ (the slice) of the measured values of the quadrature $X_\theta$ \citep{vogelrisken,lvovsky}. For $n=1$ and $\theta=(\cos t,\sin t)$, the density of the slice is the squared modulus $|\mathcal F_t\phi|^2$ of the fractional Fourier transform of angle $t$. Recovering $\phi$ up to a global phase from the family of slices is therefore a phase retrieval problem. In what follows, when two states coincide up to a global phase, we simply say that they coincide.

One method of reconstruction is to represent the state by a family $\psi_\omega$ with parameter $\omega$ and to minimize the discrepancy between the slices of $\psi_\omega$ and the slices of the target $\phi$. We take $\psi_\omega$ to be a wave function represented by a network, that is, a neural network, and we measure the discrepancy between the slices by the quantity obtained by aggregating over the directions the 1-Wasserstein distances in the individual directions. This paper treats the losses in which the aggregation is the $L^q$ norm over directions, namely
\[
\Lc_q(\psi):=\Big(\int_{\Sph}\mathcal W_1\big(\Pm^\theta_\psi,\Pm^\theta_\phi\big)^q\,d\varsigma(\theta)\Big)^{1/q}\qquad(1\le q\le\infty),
\]
where $\varsigma$ is the uniform probability measure and $q=\infty$ is read as the supremum. The loss with $q=1$ is the average over directions and is of the type of the sliced Wasserstein distance \citep{bonneel}, which is widely used in machine learning. The loss with $q=\infty$ is the supremum and is of the type of the max-sliced $\mathcal W_1$ \citep{deshpande}. The slices are quadratic in $\psi$ and $\omega\mapsto\psi_\omega$ is nonlinear, so $\omega\mapsto\Lc_q(\psi_\omega)$ need not be convex. Moreover, $\mathcal W_1$ is a supremum over Kantorovich potentials, so the loss need not be differentiable. At the points where the loss is positive, however, it is regular in the sense of Clarke. The question is therefore how accurate the estimate is at the points where a gradient method can stop, that is, at stationary points in the sense of Clarke \citep{clarke}.

The prior work \citep{kajisa2026sliced} proposed a method that uses the loss with $q=\infty$ and showed, for a single mode, that every critical point at which the loss is differentiable is a global minimizer, that is, a state that coincides with the target up to a global phase (Theorem 2.12 of the prior work). Its proof, however, uses the assumption that the range of the derivative with respect to the parameters contains every Schwartz function in the tangent space of the state (Assumption 2.9 of the prior work). This assumption is an idealization of overparameterization and does not hold for networks of finite width, for which this range is finite-dimensional (Remark 2.10 of the prior work). For one-hidden-layer networks of finite width with Gaussian units, this paper estimates the error of the stationary points for all $q$ without using this assumption.

\subsection{Main results}

A network of width $m$ has the amplitudes $a_j\in\C$ and the centers $c_j\in\R^n$ as its parameter $\omega=(a,c)$ and is given (Definition~\ref{def:rbf}) by
\[
\psi_\omega=\frac{\Psi_\omega}{\|\Psi_\omega\|},\qquad\Psi_\omega=\frac{\varepsilon}{\sqrt m}\sum_{j=1}^ma_j\varphi_{c_j},\qquad\varphi_c(x)=\exp\Big(-\frac{|x-c|^2}{2\sigma^2}\Big).
\]
In this paper, a state represented by a network in this way is called a neural quantum state. The amplitudes, which are the weights of the output layer, and the centers of the units, which are the weights of the hidden layer, are both trained. The error of the estimate is measured by the infidelity $u(\psi):=1-|\ip\phi\psi|^2$. The quantity $\arccos|\ip\phi\psi|$ is the distance between states up to a global phase (the Fubini--Study distance), and $u(\psi)$ is the square of its sine. Writing $\dist$ for the distance with respect to the norm of $L^2(\R^n)$, we have $u(\psi)=\dist(\phi,\C\psi)^2$. Let $\mathcal V(c)$ be the span of the units $\varphi_{c_1},\dots,\varphi_{c_m}$. Since $\C\psi_\omega\subset\mathcal V(c)$, we have $u(\psi_\omega)\ge\dist(\phi,\mathcal V(c))^2$, and the right-hand side is the smallest infidelity that the network can attain when its centers are $c$.

The main results use two quantities. The energy of a state $f$ is $\En(f):=\sum_{j=1}^n(\|x_jf\|^2+\|\partial_jf\|^2)$, which is twice the expectation of the total photon number plus the number of modes. The norm that includes the energy is $\|f\|_{\Qs}:=(\|f\|^2+\En(f))^{1/2}$. The exponential moment of the photon number is, for $\beta>0$, $\Mo_\beta(f):=\sum_ke^{\beta|k|}|\ip{h_k}f|^2$, where the $h_k$ are products of Hermite functions and $|k|$ is the total photon number. In the notation of quantum optics it is $\ip f{e^{\beta\hat N}f}$ (the precise definitions are in Section~\ref{sec:setup} and \eqref{eq:Mbeta}). The main results assume that the exponential moment of the network state is bounded, and Theorem~A and the propositions at finite width make the same assumption on the target as well. The exponential moment is determined by the state alone, and this assumption on the network state can be checked without knowing the target.

\begin{thmAinf}
Draw the initial centers independently from a distribution whose support is the whole space. Then, with probability $1$, the following holds. For every $q\in[1,\infty]$, every sequence of stationary points for $\Lc_q$ of networks whose width tends to infinity, for which the exponential moments are bounded and the sum of the squared displacements of the centers from their initial values is bounded, has a subsequence that converges in the norm that includes the energy. Every limit of such a subsequence either coincides with the target up to a global phase or lies outside a neighborhood of the target. The size of this neighborhood is determined only by $q$, the upper bound on the exponential moment, the energy of the target, and the number of modes.
\end{thmAinf}

The precise statement of Theorem~A (informal) is Theorem~\ref{thm:limit} in Section~\ref{sec:results}. The neighborhood determined by the explicit constants is small (Section~\ref{sec:discussion}).

\paragraph{Sketch of the proof of Theorem~A.}
Theorem~A is obtained from a single inequality for stationary points of finite width by taking the limit in which the width tends to infinity. We write $\delta_\omega$ for the error of approximating $\ip\phi{\psi_\omega}\phi$ by the range of the derivative with respect to the parameters at the point $\omega$, where the range also includes the derivatives in the directions that move the centers (equation~\eqref{eq:deltaomega}). We have $\delta_\omega\le\dist(\phi,\mathcal V(c))$. At a stationary point, the generalized gradient of the loss with respect to the state contains an element that vanishes on the range of the derivative. We apply this element to the element of the range of the derivative that is closest to $\ip\phi{\psi_\omega}\phi$, and we use that, for each direction and each potential, the integral of the potential against the difference of the slices is a difference of quadratic forms of the states. It follows (Lemma~\ref{prop:onepoint}) that at a stationary point
\[
\Lc_q(\psi_\omega)\ \le\ 2\,\Gamma(\psi_\omega)\big(\delta_\omega+u(\psi_\omega)\big).
\]
Here $\Gamma$ is determined by the energies of the target and of the error direction. The overparameterization assumption of the prior work corresponds to $\delta_\omega=0$. At finite width we have $\delta_\omega>0$ in general, but $\delta_\omega$ converges to zero as the width tends to infinity. Indeed, with probability $1$, every nonempty open set contains infinitely many initial centers. Since the sum of the squared displacements is bounded, every nonempty open set contains at least one center of the stationary point if the width is sufficiently large. The set of all states whose exponential moment is at most a given bound is compact in the norm that includes the energy. The loss is continuous in this norm, so taking the limit in the inequality above shows that a limit $\psi_*$ either coincides with the target or satisfies $\Lc_q(\psi_*)\le2\,\Gamma(\psi_*)\,u(\psi_*)$. On the other hand, for a state $\psi$ whose exponential moment is at most the same bound and for which $u$ is sufficiently small, we have (Lemma~\ref{prop:sharp})
\[
\Lc_q(\psi)\ \ge\ \kappa_*(u)\sqrt u.
\]
Here $\kappa_*$ is an explicit function determined by $q$, the upper bound on the exponential moment, the energy of the target, and the number of modes, and as $u\to0$ it is at least a positive constant multiple of a negative power of $\log(1/u)$. This lower bound is a stability estimate for the inverse problem of recovering a state from its family of slices and does not depend on the model. In the range where $u$ is sufficiently small, an upper bound linear in $u$ and a lower bound in terms of $\sqrt u$ are incompatible, so the limit $\psi_*$ either satisfies $u(\psi_*)=0$ or lies outside this range. This argument is the same for all $q$.

\begin{thmBinf}
In the same setting, let $q=\infty$. If the loss is differentiable at a limit of the sequence of stationary points, then that limit coincides with the target even if it lies outside the neighborhood of Theorem~A.
\end{thmBinf}

Theorem~B gives the conclusion of the landscape theorem of the prior work, without the overparameterization assumption, for limits of stationary points of finite width. For $q<\infty$, even where the loss is differentiable, the method of this paper cannot exclude the possibility that the gradients in the individual directions are nonzero while their integral over the directions is zero, so a statement of this form is not obtained. The precise statement of Theorem~B (informal) is Theorem~\ref{thm:landscape} in Section~\ref{sec:results}.

\paragraph{Sketch of the proof of Theorem~B.}
We take the limit in which the width tends to infinity not in the upper bound on the loss but in the stationarity condition itself. Write $\psi_k$ for the states of the sequence of stationary points and $c^{(k)}$ for their centers, and suppose that $\psi_k$ converges to $\psi_*$ in the norm that includes the energy. For $q=\infty$, an element of the generalized gradient of the loss with respect to the state is represented as $h\mapsto2\Re\ip vh$ by an element $v$ of the closed convex hull of the gradients at the pairs of a direction and a potential that attain the maximum of the loss (the active gradients). At a stationary point there is such a $v_k$ for which $h\mapsto\Re\ip{v_k}h$ vanishes on the range of the derivative. The range of the derivative contains the complex subspace $\mathcal V(c^{(k)})$, so $\ip h{v_k}=0$ for all $h\in\mathcal V(c^{(k)})$. Hence, for every square-integrable function $f$,
\[
|\ip f{v_k}|\ \le\ \|v_k\|\,\dist\big(f,\mathcal V(c^{(k)})\big).
\]
The norms $\|v_k\|$ are bounded, and $\dist(f,\mathcal V(c^{(k)}))\to0$ for the same reason as in the sketch of the proof of Theorem~A, so $v_k$ converges weakly to $0$. The active gradients are weakly continuous in the state, the direction, and the potential, so $0$ belongs to the closed convex hull of the active gradients at $\psi_*$. If the loss is differentiable at $\psi_*$, there is only one active gradient, so it is $0$. A state at which an active gradient is $0$ coincides with the target.

\subsection{Estimates at finite width and a numerical example}

Theorems~A and~B are statements about limits. To obtain statements that can be checked by numerical computation, this paper treats three kinds of finiteness: the width of the network is finite, the loss uses finitely many directions, and the numerical minimization is restricted to the finitely many directions in which the singular values of the tangent map are large.

First, solving the two estimates in the sketch of the proof of Theorem~A for $u$ while keeping the width finite, we find (Proposition~\ref{thm:finite}) that a stationary point in the neighborhood of the target satisfies
\[
\delta_\omega^2\ \le\ u(\psi_\omega)\ \le\ \tau\big(\delta_\omega^2\big).
\]
The function $\tau(s)/s$ can be written explicitly, and as $s\to0$ it is at most a constant multiple of a power of $\log(1/s)$. That is, the infidelity of a stationary point in the neighborhood is at least the square of the error of approximating the target by the range of the derivative, and is at most this square multiplied by a constant multiple of a power of a logarithm. Moreover, if the initial centers are drawn independently and the target can be represented as an average of units, then the stationary points in the neighborhood whose centers are displaced from their initial values by at most a given amount satisfy, with high probability, $u(\psi_\omega)=O(1/m)$ up to a logarithmic factor. This proposition is a statement about a single network of finite width, so it can be checked numerically.

Second, for a single mode, we replace the directions of the loss by finitely many equally spaced angles. If the number of angles is at least $\pi$ times the photon-number cutoff, then the loss over finitely many angles is bounded from above and from below by constant multiples of the loss over all directions, up to an exponentially small additive error (Proposition~\ref{prop:fdir}). Therefore an estimate of the same form as in Proposition~\ref{thm:finite} also holds for the stationary points of the loss over finitely many angles (Proposition~\ref{prop:restricted}).

Third, the numerical minimization is carried out only in the directions in which the singular values of the tangent map are large. The upper bound on the infidelity in the estimate above remains valid when stationarity is restricted to the directions in a subspace of the parameter space and the range of the tangent map is replaced by the image of that subspace (the same proposition).

In the numerical example (Section~\ref{sec:numerics}), the target is a cat state with mean photon number $9$. The network cannot represent this target. While increasing the width of the network from $12$ to $28$ in steps of $4$, we minimize, for each of $q=1,2,\infty$, the loss over $240$ equally spaced angles with all the amplitudes and centers as variables. The numerical minimization is carried out only in the directions for which the singular value of the tangent map is larger than $10^{-6}$ times the largest singular value. The infidelity of the point where the minimization stopped at each width is between $1.05$ and $1.63$ times the square of the error of approximating the target by the image of these directions under the tangent map. As the width is increased this error decreases, and the infidelity decreases accordingly. For $q=2,\infty$, the points of width at least $20$ satisfy all the hypotheses of Proposition~\ref{prop:restricted}, except that the stationarity restricted to these directions is a numerical approximation. For $q=1$, the size of the neighborhood is smaller than the infidelities reached in the numerical computation, so Proposition~\ref{prop:restricted} does not apply.

The loss of this paper is one in which the slices of the target are given exactly and all directions are used. We do not treat the error due to finitely many samples. Proposition~\ref{prop:fdir} and Proposition~\ref{prop:restricted}, which concern finitely many directions, are stated for equally spaced angles in a single mode. Proposition~\ref{prop:restricted} says nothing about infidelities smaller than a threshold determined by the number of angles. We do not treat the question of whether the iterates of training converge to a stationary point. Nor do we show that training the centers makes $\delta_\omega$ or $\dist(\phi,\mathcal V(c))$ smaller. About stationary points outside the neighborhood of the target we can say nothing, except in the case where $q=\infty$ and the loss is differentiable. The estimates with the explicit constants are pessimistic, and the infidelities of the points that entered the neighborhood are at least $10$ orders of magnitude smaller than the explicit upper bound (Section~\ref{sec:discussion}). The numerical example concerns a single state of a single mode.

\subsection{Related work}\label{sec:related}

\paragraph{Overparameterization and landscapes.}
The theory of optimization of overparameterized networks includes the analysis of the regime in which the tangent kernel hardly changes during training \citep{jacot,chizatlazy}, a result showing that, in the mean-field limit of infinite width, if the gradient flow converges then its limit is a global minimizer \citep{chizatbach}, and results showing that the loss landscape of shallow networks has no spurious local minima or valleys \citep{soltanolkotabi,venturi}. These works mainly treat smooth losses. The loss of this paper is nonsmooth, the states lie on the unit sphere, and the conclusion at finite width is not about global minimizers but is a quantitative estimate for stationary points in a neighborhood of the target. The statement at infinite width (Theorem~\ref{thm:limit}) is, like that of \citet{chizatbach}, a statement about limits, but it does not assume convergence of the sequence and derives the existence of a convergent subsequence from the upper bound on the exponential moment. Moreover, the statements of this paper assume neither that the tangent kernel does not change during training nor any condition on the eigenvalues of the tangent kernel at infinite width. All that we use about the derivative with respect to the parameters is its range at the stationary point.

\paragraph{Nonsmooth optimization and sharpness.}
Stationary points are defined by the generalized gradient of Clarke \citep{clarke}, and the proofs are also carried out with generalized directional derivatives only. The $\mathcal W_1$ in each direction is the maximum of a family of $C^1$ functions indexed by a compact set, so its generalized directional derivative equals the maximum of the derivatives at the indices that attain the maximum, and it coincides with the one-sided directional derivative. Hence the loss is regular in the sense of Clarke, and for $q<\infty$ the same holds at the points where the loss is positive (Proposition~\ref{prop:lipschitz}). The properties of generalized gradients used in this paper are proved from the definition in Appendix~\ref{sec:clarke}. The property that the excess of the value of a function over its minimum value is at least a positive constant multiple of the distance to the set of minimizers is known as a weak sharp minimum \citep{burkeferris} and is the key to the local linear convergence of subgradient methods \citep{davis}. Lemma~\ref{prop:sharp} shows this property for the distance to the set of minimizers $\{e^{i\alpha}\phi\}$ (the square of this distance is bounded from above and from below by constant multiples of the infidelity), within the range of states whose exponential moment is at most a given bound, and up to a logarithmic factor.

\paragraph{Sliced Wasserstein distances.}
The Wasserstein distances of one-dimensional projections averaged over directions (for example \citealp{bonneel}) or maximized over directions \citep{deshpande}, and their generalization to projections onto subspaces \citep{patycuturi}, have been proposed, and their statistical and topological properties have been studied \citep{nadjahi}. The regularity and the critical points when the squared sliced $\mathcal W_2$ between discrete measures is taken as the loss, and the convergence of stochastic gradient methods to critical points in the sense of Clarke, are studied by \citet{tanguy}. What this paper treats is a quantitative estimate of the error at the stationary points of a model. The slices of this paper are the measurement distributions of the mutually noncommuting operators $X_\theta$, and they are not projections of a probability distribution but projections (the Radon transform) of the Wigner function, which can take negative values \citep{vogelrisken}. Instead of inverting the Radon transform, this paper uses the Weyl operators and the Moyal identity \citep{folland} to obtain that a state is determined by its slices, together with a quantitative estimate of this fact.

\paragraph{Phase retrieval.}
In phase retrieval, it is known that when there are sufficiently many independent Gaussian measurement vectors, the least-squares loss has, with high probability, no spurious local minima \citep{sunquwright}. Phase retrieval can also be regarded as the problem of recovering the weight of a network with a single unit with quadratic activation from its output. For shallow networks with quadratic activation fitted to the output of a network of the same form, the landscape of the loss and the convergence of gradient descent have been studied in the overparameterized regime \citep{saraomannelli,soltanolkotabi}. Like the assumption of the prior work, these results obtain a benign landscape by taking the width large. In this paper the measurements form the fixed family of quadratures, the object is infinite-dimensional, the width is finite, and the conclusion is a quantitative estimate for stationary points in a neighborhood of the target.

\paragraph{Quantum tomography of optical systems.}
Tomography of states of light by homodyne detection began experimentally with the reconstruction of the Wigner function by filtered back-projection \citep{smithey}, which uses the fact that the family of quadrature distributions is the Radon transform of the Wigner function \citep{vogelrisken}. Subsequently, methods were derived that obtain the elements of the density matrix in the photon-number basis directly, without back-projection, as averages of the values that kernels called pattern functions take at the measured values \citep{dariano,leonhardt}. What is commonly used at present is iterative maximum-likelihood estimation, which preserves the positivity of the density matrix \citep{lvovskyml}. The development of these methods is reviewed by \citet{lvovsky}. These methods represent the density matrix under a photon-number cutoff, so if the photon-number cutoff in each mode is $N$, the number of parameters to be estimated is of the order of $(N+1)^{2n}$. Representing the state by a network avoids this growth of the number of parameters. Neural quantum states, which represent the states of many-body systems by restricted Boltzmann machines \citep{carleo}, have been used for the tomography of systems of qubits \citep{torlai}, and for continuous-variable systems measured by homodyne detection a representation by a network has also been trained on the basis of the likelihood of the measurement outcomes \citep{tiunov}. The term neural quantum state comes from these works. The networks of the prior work and of this paper are not restricted Boltzmann machines and represent the wave function in the position representation itself. The loss is also not a likelihood but the $\mathcal W_1$ of the quadrature distributions aggregated over directions.

\subsection{Organization}

Section~\ref{sec:setup} defines the slices, the loss, the networks with Gaussian units, and the stationary points. Section~\ref{sec:results} states the lemmas on the lower bound and the upper bound on the loss and the two main theorems. Section~\ref{sec:numerics} states the quantitative form at finite width and the way to transfer it to the loss over finitely many angles, and presents a numerical example in a single mode. Section~\ref{sec:proofs} proves the statements of Section~\ref{sec:results}, Section~\ref{sec:proofs4} proves the statements of Section~\ref{sec:numerics}, and Section~\ref{sec:discussion} contains the discussion. The list of notation is in Appendix~\ref{sec:notation}.

\section{Setting}\label{sec:setup}

\subsection{Slices}\label{sec:quadslice}

We consider $n\ge1$ modes and put
\begin{equation}\label{eq:hilbert}
\Hs:=L^2(\R^n),\qquad\ip fg:=\int_{\R^n}\overline{f(x)}\,g(x)\,dx.
\end{equation}
The inner product is conjugate linear in the first argument. We write $x_j$ for multiplication by the $j$-th coordinate and $\partial_jf$ for the partial derivative in the sense of distributions, we put $p_jf:=-i\partial_jf$, and we write $\langle x,y\rangle$ for the inner product of $\R^n$ and $|x|$ for the Euclidean norm. The space of Schwartz functions on $\R^n$ is denoted by $\Sc$, and the domain of an operator is denoted by $D(\cdot)$. We define the space $\Qs$ of all $f\in\Hs$ for which $x_jf$ and $\partial_jf$ all belong to $\Hs$, together with the energy $\En(f)$ and the norm $\|f\|_{\Qs}$ of an element $f$ of this space, by
\begin{equation}\label{eq:energy}
\begin{gathered}
\Qs:=\{f\in\Hs:\ x_jf,\ \partial_jf\in\Hs\ (j=1,\dots,n)\},\\
\En(f):=\sum_{j=1}^n\big(\|x_jf\|^2+\|\partial_jf\|^2\big),\qquad\|f\|_{\Qs}:=\big(\|f\|^2+\En(f)\big)^{1/2}.
\end{gathered}
\end{equation}
The quantity $\En(f)^{1/2}$ is a seminorm on $\Qs$. The space $\Qs$ is complete with respect to $\|\cdot\|_{\Qs}$. Indeed, let $(f_k)$ be a Cauchy sequence with respect to $\|\cdot\|_{\Qs}$. Then $f_k$, $x_jf_k$, and $\partial_jf_k$ converge in $\Hs$. Let $f$ be the limit of $f_k$. For $\chi\in\Sc$ we have $\ip\chi{x_jf_k}=\ip{x_j\chi}{f_k}\to\ip{x_j\chi}f$ and $\ip\chi{\partial_jf_k}=-\ip{\partial_j\chi}{f_k}\to-\ip{\partial_j\chi}f$, so the limits of $x_jf_k$ and $\partial_jf_k$ in $\Hs$ are equal, as distributions, to $x_jf$ and $\partial_jf$. Hence $f\in\Qs$ and $\|f_k-f\|_{\Qs}\to0$. In the notation of quantum optics, $\En(f)=\ip f{(2\hat N+n)f}$, where $\hat N$ is the total photon number. We take the target $\phi$ to be a unit vector in $\Qs$. In what follows, unit vectors and the unit sphere are always understood with respect to the norm $\|\cdot\|$ of $\Hs$.

\begin{definition}[Slices]\label{def:slice}
For $(\xi,\varpi)\in\R^n\times\R^n$, we let $X_{\xi,\varpi}$ be the closure of the essentially self-adjoint operator $\sum_j(\xi_jx_j+\varpi_jp_j)$ on $\Sc$ (see the proof of Lemma~\ref{lem:slice}), and we call the unitary operator $V(\xi,\varpi):=e^{iX_{\xi,\varpi}}$ on $\Hs$ the Weyl operator. For a point $\theta=(\xi,\varpi)$ of the unit sphere $\Sph$ of $\R^n\times\R^n$, the operator $X_\theta:=X_{\xi,\varpi}$ is called the quadrature in the direction $\theta$, and $E_\theta$ denotes its spectral measure. For a unit vector $f\in\Hs$, a Borel set $B\subset\R$ and $s\in\R$, we put
\begin{equation}\label{eq:slicedef}
\Pm^\theta_f(B):=\ip f{E_\theta(B)f},\qquad F^\theta_f(s):=\Pm^\theta_f\big((-\infty,s]\big),
\end{equation}
and we call the probability measure $\Pm^\theta_f$ on $\R$ the slice of $f$ in the direction $\theta$, and $F^\theta_f$ its distribution function.
\end{definition}

If $t\ne0$ then $X_{t\xi,t\varpi}=tX_{\xi,\varpi}$, and $X_{0,0}=0$, so $V(t\xi,t\varpi)=e^{itX_{\xi,\varpi}}$ for all $t\in\R$. The measure $\Pm^\theta_f$ is the distribution of the measured values of the quadrature $X_\theta$, which is measured with a passive interferometer and homodyne detection. For $n=1$ and $\theta=(\cos t,\sin t)$, the operator $X_\theta=\cos t\,x+\sin t\,p$ is the rotated quadrature and $\Pm^\theta_f$ is the slice of the prior work.

\begin{lemma}[Moments and continuity of slices, Weyl operators]\label{lem:slice}
First, for $f\in\Qs$ and $(\xi,\varpi)\in\R^n\times\R^n$ we have $f\in D(X_{\xi,\varpi})$ and
\begin{equation}\label{eq:Xxi}
X_{\xi,\varpi}f=\sum_{j=1}^n\big(\xi_j\,x_jf+\varpi_j\,p_jf\big),\qquad\|X_{\xi,\varpi}f\|\le|(\xi,\varpi)|\,\En(f)^{1/2},
\end{equation}
and if $f$ is a unit vector, then for all $\theta\in\Sph$ we have $\int s^2\,d\Pm^\theta_f(s)=\|X_\theta f\|^2\le\En(f)$. Second, for a unit vector $f$ in $\Qs$ and $\theta_0\in\Sph$, we have $\int_{\R}|F^\theta_f-F^{\theta_0}_f|\,ds\to0$ as $\theta\to\theta_0$. Third, for $(\xi,\varpi)\in\R^n\times\R^n$ and $f\in\Hs$, the Weyl operator $V(\xi,\varpi)$ acts by
\begin{equation}\label{eq:weyldef}
(V(\xi,\varpi)f)(x)=e^{i\langle\xi,\varpi\rangle/2}e^{i\langle\xi,x\rangle}f(x+\varpi).
\end{equation}
Moreover, if $f\in\Qs$ then $V(\xi,\varpi)f\in\Qs$, and for $j=1,\dots,n$ we have
\begin{equation}\label{eq:weylcomm}
x_jV(\xi,\varpi)f=V(\xi,\varpi)(x_jf)-\varpi_jV(\xi,\varpi)f,\qquad p_jV(\xi,\varpi)f=\xi_jV(\xi,\varpi)f+V(\xi,\varpi)(p_jf).
\end{equation}
\end{lemma}

\begin{proof}
We first check that $\sum_j(\xi_jx_j+\varpi_jp_j)$ is essentially self-adjoint on $\Sc$ and that, for its closure $X_{\xi,\varpi}$, the operator $V(\xi,\varpi)=e^{iX_{\xi,\varpi}}$ acts by \eqref{eq:weyldef}. For $t\in\R$ and $f\in\Hs$ put
\begin{equation}\label{eq:weyl}
(V_tf)(x):=e^{it^2\langle\xi,\varpi\rangle/2}e^{it\langle\xi,x\rangle}f(x+t\varpi).
\end{equation}
A direct computation gives $V_sV_t=V_{s+t}$, and $V_t$ is a strongly continuous one-parameter unitary group that preserves $\Sc$. For $g\in\Sc$, the quotient $t^{-1}(V_tg-g)$ converges in $\Hs$ to $i\sum_j(\xi_jx_jg+\varpi_jp_jg)$ as $t\to0$ (by the dominated convergence theorem). Hence, for the generator $B$ of $V_t$ given by Stone's theorem, $\Sc$ is contained in $D(B)$, on this space we have $B=\sum_j(\xi_jx_j+\varpi_jp_j)$, and by \citet[Theorem~VIII.11]{reedsimon}, $\Sc$ is a core for $B$. The restriction of a self-adjoint operator to a core is essentially self-adjoint, which gives the claim, with $X_{\xi,\varpi}=B$ and $V(\xi,\varpi)=V_1$.

We prove the first assertion. For $g\in\Sc$, the definition of the distributional derivative gives $\ip{\sum_j(\xi_jx_jg+\varpi_jp_jg)}f=\ip g{\sum_j(\xi_jx_jf+\varpi_jp_jf)}$, and the vector $\sum_j(\xi_jx_jf+\varpi_jp_jf)$ on the right-hand side belongs to $\Hs$ because $f\in\Qs$. Thus $f$ belongs to the domain of the adjoint of the operator $\sum_j(\xi_jx_j+\varpi_jp_j)$ on $\Sc$, and by essential self-adjointness this adjoint equals $X_{\xi,\varpi}$, which gives the equality in \eqref{eq:Xxi}. The inequality follows from the Cauchy--Schwarz inequality. The equality for the moment follows from the functional calculus.

We prove the second assertion. By \eqref{eq:weyldef}, the continuity of translations in $\Hs$, and the dominated convergence theorem, the map $(\xi,\varpi)\mapsto V(\xi,\varpi)f$ into $\Hs$ is continuous. Hence the characteristic function $t\mapsto\int e^{its}\,d\Pm^\theta_f(s)=\ip f{V(t\theta)f}$ of $\Pm^\theta_f$ converges, for each $t$, to the characteristic function of $\Pm^{\theta_0}_f$ as $\theta\to\theta_0$. By L\'evy's continuity theorem, $\Pm^\theta_f$ converges weakly to $\Pm^{\theta_0}_f$, and $F^\theta_f(s)$ converges to $F^{\theta_0}_f(s)$ for all $s$ outside an at most countable set. On the other hand, by the first assertion and Chebyshev's inequality, we have $|F^\theta_f(s)-F^{\theta_0}_f(s)|\le\min\{1,\En(f)/s^2\}$ for all $\theta$ and all $s\ne0$, and the right-hand side is integrable. The assertion follows from the dominated convergence theorem.

We prove the third assertion. The formula \eqref{eq:weyldef} was shown at the beginning of the proof. Let $f\in\Qs$. By \eqref{eq:weyldef} we have, as distributions, $x_jV(\xi,\varpi)f=V(\xi,\varpi)(x_jf)-\varpi_jV(\xi,\varpi)f$ and $\partial_j(V(\xi,\varpi)f)=i\xi_jV(\xi,\varpi)f+V(\xi,\varpi)\partial_jf$. Both right-hand sides belong to $\Hs$, so $V(\xi,\varpi)f\in\Qs$, and multiplying the second identity by $-i$ gives \eqref{eq:weylcomm}.
\end{proof}

\subsection{Loss}\label{sec:loss}

For unit vectors $a,b,\psi$ in $\Qs$ and $\theta\in\Sph$ put
\begin{equation}\label{eq:Wtheta}
\begin{gathered}
\mathcal W_\theta(a,b):=\mathcal W_1\big(\Pm^\theta_a,\Pm^\theta_b\big),\qquad u(a,b):=1-|\ip ab|^2,\\
\mathcal W_\theta(\psi):=\mathcal W_\theta(\psi,\phi),\qquad u(\psi):=u(\psi,\phi).
\end{gathered}
\end{equation}
Here $\mathcal W_1$ is the 1-Wasserstein distance on $\R$. We call $u(a,b)$ the infidelity between $a$ and $b$, and $u(\psi)$ the infidelity of $\psi$. As in these definitions, for notation that takes two states as arguments we omit the second argument when it is the target $\phi$. The first moments of the slices are finite by Lemma~\ref{lem:slice}, so $\mathcal W_\theta(a,b)$ is finite.

We express the distance $\mathcal W_1$ between the slices in each direction in terms of Kantorovich potentials. We put
\begin{equation}\label{eq:potentials}
\Kc:=\{\mu:\R\to\R:\ |\mu(s)-\mu(s')|\le|s-s'|,\ \mu(0)=0\},\qquad\Ac:=\Sph\times\Kc,\qquad M_{\theta,\mu}:=\mu(X_\theta).
\end{equation}
We equip $\Kc$ with the topology of locally uniform convergence, and for $(\theta,\mu)\in\Ac$ the operator $M_{\theta,\mu}$ is defined by the functional calculus. For $(\theta,\mu)\in\Ac$ and a unit vector $\psi$ in $\Qs$ put
\begin{equation}\label{eq:ell}
\ell_{\theta,\mu}(\psi):=\ip\psi{M_{\theta,\mu}\psi}-\ip\phi{M_{\theta,\mu}\phi}.
\end{equation}
That the right-hand side is well defined follows from the next lemma.

\begin{lemma}[Representation by Kantorovich potentials]\label{lem:kant}
Let $\psi$ be a unit vector in $\Qs$. For $(\theta,\mu)\in\Ac$ we have $\Qs\subset D(M_{\theta,\mu})$, for $f\in\Qs$ we have $\|M_{\theta,\mu}f\|\le\En(f)^{1/2}$, and $\ell_{\theta,\mu}(\psi)=\int\mu\,d(\Pm^\theta_\psi-\Pm^\theta_\phi)$. For all $\theta\in\Sph$ we have
\begin{equation}\label{eq:Lpot}
\mathcal W_\theta(\psi)=\sup_{\mu\in\Kc}\ell_{\theta,\mu}(\psi)=\int_{\R}\big|F^\theta_\phi-F^\theta_\psi\big|\,ds,
\end{equation}
and the supremum is attained by the function $s\mapsto\int_0^s\sgn(F^\theta_\phi-F^\theta_\psi)\,dt$. The map $\theta\mapsto\mathcal W_\theta(\psi)$ is continuous.
\end{lemma}

\begin{proof}
Since $|\mu(s)|\le|s|$, the functional calculus and the first assertion of Lemma~\ref{lem:slice} give $\Qs\subset D(M_{\theta,\mu})$ and $\|M_{\theta,\mu}f\|\le\|X_\theta f\|\le\En(f)^{1/2}$, and for a unit vector $f\in\Qs$ we have $\ip f{M_{\theta,\mu}f}=\int\mu\,d\Pm^\theta_f$. For probability measures $P$, $Q$ on $\R$ with finite first moments, we write $F_P$, $F_Q$ for their distribution functions. For a Lipschitz function $\mu$, integrating $\mu(x)=\mu(0)+\int_0^x\mu'$ and using Fubini's theorem gives
\begin{equation}\label{eq:ibp}
\int\mu\,d(P-Q)=\int_{\R}\mu'(s)\big(F_Q(s)-F_P(s)\big)\,ds.
\end{equation}
If $\mu\in\Kc$ then $|\mu'|\le1$, so the right-hand side is at most $\int|F_P-F_Q|$, with equality for $\mu(s)=\int_0^s\sgn(F_Q-F_P)\,dt$. Together with the Kantorovich--Rubinstein duality \citep{villani} this gives
\begin{equation}\label{eq:KR}
\mathcal W_1(P,Q)=\sup_{\mu\in\Kc}\int\mu\,d(P-Q)=\int_{\R}\big|F_P-F_Q\big|\,ds.
\end{equation}
Taking $P=\Pm^\theta_\psi$ and $Q=\Pm^\theta_\phi$ gives \eqref{eq:Lpot}. Let $\theta_0\in\Sph$. By the right-hand side of \eqref{eq:Lpot} and the second assertion of Lemma~\ref{lem:slice}, as $\theta\to\theta_0$ we have $|\mathcal W_\theta(\psi)-\mathcal W_{\theta_0}(\psi)|\le\int|F^\theta_\psi-F^{\theta_0}_\psi|\,ds+\int|F^\theta_\phi-F^{\theta_0}_\phi|\,ds\to0$.
\end{proof}

Lemma~\ref{lem:kant} holds with any unit vector in $\Qs$ as the target. Hence, for unit vectors $a,b$ in $\Qs$, the map $\theta\mapsto\mathcal W_\theta(a,b)$ is continuous and can be aggregated over directions.

\begin{definition}[Loss]\label{def:loss}
Let $\varsigma$ be the rotation-invariant probability measure on $\Sph$ (the normalized surface measure). For unit vectors $a,b,\psi$ in $\Qs$ and $1\le q\le\infty$ we put
\begin{equation}\label{eq:Lq}
\begin{gathered}
\Lc_q(a,b):=\Big(\int_{\Sph}\mathcal W_\theta(a,b)^q\,d\varsigma(\theta)\Big)^{1/q}\quad(q<\infty),\qquad\Lc_\infty(a,b):=\sup_{\theta\in\Sph}\mathcal W_\theta(a,b),\\
\Lc_q(\psi):=\Lc_q(\psi,\phi),
\end{gathered}
\end{equation}
and we call $\Lc_q(\psi)$ the loss. For $f\in\Qs\setminus\{0\}$ we put $\ell_{\theta,\mu}(f):=\ell_{\theta,\mu}(f/\|f\|)$, and we extend $\mathcal W_\theta(f)$ and $\Lc_q(f)$ to $\Qs\setminus\{0\}$ in the same way.
\end{definition}

The loss $\Lc_1$ is the average over directions of the distance $\mathcal W_1$ between the slices (sliced $\mathcal W_1$), $\Lc_\infty$ is the supremum (max-sliced $\mathcal W_1$), and the loss of the prior work is $\Lc_\infty$. If $q\le q'$ then $\Lc_q\le\Lc_{q'}$. By \eqref{eq:Lpot},
\begin{equation}\label{eq:Linf}
\Lc_\infty(\psi)=\sup_{(\theta,\mu)\in\Ac}\ell_{\theta,\mu}(\psi).
\end{equation}
For a unit vector $\psi$ in $\Qs$ and $\theta\in\Sph$, we denote the set of all $\mu$ that attain the supremum in \eqref{eq:Lpot} and the set of all $(\theta,\mu)$ that attain the supremum in \eqref{eq:Linf} by
\begin{equation}\label{eq:active}
\Kc_\theta(\psi):=\{\mu\in\Kc:\ \ell_{\theta,\mu}(\psi)=\mathcal W_\theta(\psi)\},\qquad\Ac(\psi):=\{(\theta,\mu)\in\Ac:\ \ell_{\theta,\mu}(\psi)=\Lc_\infty(\psi)\},
\end{equation}
and we call $\Ac(\psi)$ the active set. The space $\Ac$ is compact, and $\Ac(\psi)$ is a nonempty compact set (Lemma~\ref{lem:loss}).

\subsection{Networks}\label{sec:nets}

\begin{definition}[Networks with Gaussian units]\label{def:rbf}\label{def:model}
Take a width $m\ge1$, $\sigma>0$, and $\varepsilon>0$, and let the parameter space be $\mathcal Z:=\C^m\times(\R^n)^m$. For $y\in\R^n$ and $\omega=(a,c)\in\mathcal Z$ put
\begin{equation}\label{eq:net}
\begin{gathered}
\varphi_y(x):=\exp\Big(-\frac{|x-y|^2}{2\sigma^2}\Big),\qquad\Psi_\omega:=\frac{\varepsilon}{\sqrt m}\sum_{j=1}^ma_j\varphi_{c_j},\\
\psi_\omega:=\frac{\Psi_\omega}{\|\Psi_\omega\|}\qquad(\omega\in\Omega:=\{\omega\in\mathcal Z:\Psi_\omega\ne0\}).
\end{gathered}
\end{equation}
The map $\omega\mapsto\psi_\omega$ is called a network with Gaussian units, and we call $\varphi_{c_j}$ the units, $c_j$ the centers, and $a_j$ the amplitudes. An element of $\Omega$ is called a point of the network. We define the span of the units by
\begin{equation}\label{eq:Vc}
\mathcal V(c):=\spn_{\C}\{\varphi_{c_1},\dots,\varphi_{c_m}\},
\end{equation}
and we write $\Pi_{\mathcal V(c)}$ for the orthogonal projection of $\Hs$ onto $\mathcal V(c)$. We equip $(\R^n)^m$ with the Euclidean norm $\|c\|$. Regarding complex numbers as pairs of real and imaginary parts, we view $\mathcal Z$ as a real vector space of dimension $(2+n)m$ with the Euclidean norm, and write $\zeta_1,\dots,\zeta_{(2+n)m}$ for the real coordinates of an element $\zeta$ of this space. For $\omega\in\mathcal Z$, we define a real-linear map by
\begin{equation}\label{eq:tangent}
\mathscr J(\omega)\zeta:=\sum_{i=1}^{(2+n)m}\zeta_i\,\partial_{\omega_i}\Psi_\omega\qquad(\zeta\in\mathcal Z),
\end{equation}
and we call it the tangent map. The partial derivatives $\partial_{\omega_i}\Psi_\omega\in\Qs$ are taken as limits with respect to $\|\cdot\|_{\Qs}$. The range of the tangent map is the real linear span of the partial derivatives,
\[
\ran\mathscr J(\omega)=\spn_{\R}\big\{\partial_{\omega_1}\Psi_\omega,\dots,\partial_{\omega_{(2+n)m}}\Psi_\omega\big\},
\]
which is a real subspace of $\Qs$ of dimension at most $(2+n)m$.
\end{definition}

This network is a one-hidden-layer network (a radial basis function network) with input in $\R^n$ and $m$ Gaussian units, in which the centers $c_j$ are the weights of the hidden layer and the amplitudes $a_j$ are the weights of the output layer.

\begin{lemma}[Tangent map]\label{lem:tangent}
The map $\omega\mapsto\Psi_\omega$ is a $C^1$ map from $\mathcal Z$ into $\Qs$ (Definition~\ref{def:frechet} in the appendix), all the partial derivatives in Definition~\ref{def:rbf} exist, and the Fr\'echet derivative at $\omega$ is $\mathscr J(\omega)$. In particular $\Omega$ is open.
\end{lemma}

\begin{proof}
For $\omega=(a,c)$ we have $\Psi_\omega=\frac{\varepsilon}{\sqrt m}\sum_{j=1}^ma_j\varphi_{c_j}$. The map $(a_j,f)\mapsto a_jf$ is a continuous real-bilinear map from $\C\times\Qs$ into $\Qs$, and $c_j\mapsto\varphi_{c_j}$ is a $C^1$ map from $\R^n$ into $\Qs$ (the third assertion of Lemma~\ref{lem:gaussunit}). We write $D\varphi_y$ for the Fr\'echet derivative of $y\mapsto\varphi_y$ at $y$. The product rule shows that $\omega\mapsto\Psi_\omega$ is a $C^1$ map from $\mathcal Z$ into $\Qs$ whose Fr\'echet derivative at $\omega$ is, for $\zeta=(b,d)\in\mathcal Z$,
\[
\zeta\ \mapsto\ \frac{\varepsilon}{\sqrt m}\sum_{j=1}^m\big(b_j\varphi_{c_j}+a_jD\varphi_{c_j}[d_j]\big).
\]
For each $i$, the value of this map at the unit vector of the $i$-th real coordinate is the partial derivative $\partial_{\omega_i}\Psi_\omega$. Since this map is real-linear, its value at $\zeta$ equals $\sum_{i=1}^{(2+n)m}\zeta_i\partial_{\omega_i}\Psi_\omega=\mathscr J(\omega)\zeta$. That is, with respect to $\|\cdot\|_{\Qs}$,
\[
\lim_{\zeta\to0}\frac{\|\Psi_{\omega+\zeta}-\Psi_\omega-\mathscr J(\omega)\zeta\|_{\Qs}}{\|\zeta\|}=0,
\]
and $\omega\mapsto\mathscr J(\omega)$ is continuous in the operator norm. Since $\omega\mapsto\Psi_\omega$ is also continuous as a map into $\Hs$, the set $\Omega=\{\omega:\Psi_\omega\ne0\}$ is open.
\end{proof}

The partial derivatives with respect to the real and imaginary parts of $a_j$ are $\frac{\varepsilon}{\sqrt m}\varphi_{c_j}$ and $\frac{\varepsilon}{\sqrt m}i\varphi_{c_j}$, so
\begin{equation}\label{eq:VinJ}
\psi_\omega\in \mathcal V(c)\subset\ran\mathscr J(\omega).
\end{equation}

For $f\in\Hs$ and a nonempty subset $A$ of $\Hs$ we write $\dist(f,A):=\inf_{g\in A}\|f-g\|$. At a point $\omega$ of the network, we define the error of approximating the target by the range of the tangent map by
\begin{equation}\label{eq:deltaomega}
\delta_\omega:=\dist\big(\ip\phi{\psi_\omega}\phi,\ \ran\mathscr J(\omega)\big).
\end{equation}
The range $\ran\mathscr J(\omega)$ is a finite-dimensional real subspace of $\Hs$. By \eqref{eq:VinJ} we have $\mathcal V(c)\subset\ran\mathscr J(\omega)$ and $\C\psi_\omega\subset\ran\mathscr J(\omega)$, so
\begin{equation}\label{eq:deltabounds}
\delta_\omega\le|\ip\phi{\psi_\omega}|\,\dist(\phi,\mathcal V(c)),\qquad\delta_\omega^2\le|\ip\phi{\psi_\omega}|^2\big(1-|\ip\phi{\psi_\omega}|^2\big).
\end{equation}
Besides $\mathcal V(c)$, the range $\ran\mathscr J(\omega)$ contains the real multiples of the partial derivatives $a_j\,\partial_{c_{j,i}}\varphi_{c_j}$ with respect to the $i$-th coordinate $c_{j,i}$ of the center $c_j$.

\subsection{Stationary points}\label{sec:stat}

For a unit vector $\psi$ in $\Qs$ and $(\theta,\mu)\in\Ac$ put
\begin{equation}\label{eq:vdef}
v_{\theta,\mu}(\psi):=\big(M_{\theta,\mu}-\ip\psi{M_{\theta,\mu}\psi}\big)\psi,\qquad\mathcal G(\psi):=\{v_{\theta,\mu}(\psi):(\theta,\mu)\in\Ac(\psi)\}.
\end{equation}
We call $\mathcal G(\psi)$ the set of active gradients, and we write $\overline{\conv}\,\mathcal G(\psi)$ for the closure in $\Hs$ of its convex hull. Since $M_{\theta,\mu}$ is self-adjoint and $\ip\psi{M_{\theta,\mu}\psi}$ is real, we have $\ip\psi{v_{\theta,\mu}(\psi)}=0$. By Lemma~\ref{lem:kant} we have $\|v_{\theta,\mu}(\psi)\|^2=\|M_{\theta,\mu}\psi\|^2-\ip\psi{M_{\theta,\mu}\psi}^2\le\En(\psi)$. The vector $2v_{\theta,\mu}(\psi)$ is the gradient of $\ell_{\theta,\mu}$ on the unit sphere.

For a point $\omega\in\Omega$ of the network and $1\le q\le\infty$ put
\begin{equation}\label{eq:Lsq}
\Ls_q(\omega):=\Lc_q(\psi_\omega).
\end{equation}
We check that the loss is locally Lipschitz.

\begin{proposition}[Local Lipschitz continuity and regularity of the loss]\label{prop:lipschitz}
For unit vectors $\psi,\tilde\psi$ in $\Qs$ we have
\begin{equation}\label{eq:lip}
\sup_{(\theta,\mu)\in\Ac}|\ell_{\theta,\mu}(\psi)-\ell_{\theta,\mu}(\tilde\psi)|\ \le\ \En(\psi)^{1/2}\|\psi-\tilde\psi\|+\En(\psi-\tilde\psi)^{1/2},
\end{equation}
and the right-hand side of this inequality also bounds $|\mathcal W_\theta(\psi)-\mathcal W_\theta(\tilde\psi)|$ for all $\theta\in\Sph$ and $|\Lc_q(\psi)-\Lc_q(\tilde\psi)|$ for all $q\in[1,\infty]$. Consequently, for every point of $\Qs\setminus\{0\}$ there exist a neighborhood with respect to $\|\cdot\|_{\Qs}$ and a constant such that, for all $(\theta,\mu)\in\Ac$ and all $q\in[1,\infty]$, the functions $\ell_{\theta,\mu}$, $\mathcal W_\theta$, and $\Lc_q$ have, on this neighborhood, this constant as a Lipschitz constant with respect to $\|\cdot\|_{\Qs}$. For all $q\in[1,\infty]$, the function $\Ls_q$ of \eqref{eq:Lsq} is locally Lipschitz on $\Omega$. Moreover, $\Lc_\infty$ and $\Ls_\infty$ are regular in the sense of Clarke at every point, and for $1\le q<\infty$ the functions $\Lc_q$ and $\Ls_q$ are regular in the sense of Clarke at the points where their value is positive. That is, the one-sided directional derivative exists in every direction and equals the generalized directional derivative (the definitions are in \eqref{eq:clarkeQ}, \eqref{eq:clarkeZ}, and Appendix~\ref{sec:clarke}).
\end{proposition}

\begin{proof}
Put $M:=M_{\theta,\mu}$. The terms with $\phi$ in \eqref{eq:ell} cancel when we take the difference, and by the symmetry of $M$ we have $\ell_{\theta,\mu}(\psi)-\ell_{\theta,\mu}(\tilde\psi)=\ip{\psi-\tilde\psi}{M\psi}+\ip{\tilde\psi}{M(\psi-\tilde\psi)}$. The bound $\|Mf\|\le\En(f)^{1/2}$ of Lemma~\ref{lem:kant} gives \eqref{eq:lip}. By \eqref{eq:Lpot}, $\mathcal W_\theta$ is the supremum of $\ell_{\theta,\mu}$ over $\mu$, so the absolute value of the difference of the two values of $\mathcal W_\theta$ is at most the left-hand side of \eqref{eq:lip}. Since $\Lc_q$ is the $L^q(\varsigma)$ norm of $\theta\mapsto\mathcal W_\theta$ (the supremum for $q=\infty$), Minkowski's inequality gives $|\Lc_q(\psi)-\Lc_q(\tilde\psi)|\le\sup_\theta|\mathcal W_\theta(\psi)-\mathcal W_\theta(\tilde\psi)|$. The normalization $f\mapsto f/\|f\|$ is a locally Lipschitz map from $\Qs\setminus\{0\}$ into $\Qs$ and $\En(f)^{1/2}\le\|f\|_{\Qs}$, so \eqref{eq:lip} gives the assertion on $\Qs\setminus\{0\}$. The map $\omega\mapsto\Psi_\omega$ is a $C^1$ map from $\Omega$ into $\Qs\setminus\{0\}$ (Lemma~\ref{lem:tangent}) and $\Ls_q(\omega)=\Lc_q(\Psi_\omega)$, so $\Ls_q$ is locally Lipschitz on $\Omega$ by Lemma~\ref{lem:clarkechain} in the appendix. The regularity is proved in the fourth assertion of Lemma~\ref{lem:stat} and in Lemma~\ref{lem:netstat}.
\end{proof}

By Proposition~\ref{prop:lipschitz}, for all $\theta\in\Sph$ and all $q\in[1,\infty]$, the functions $\mathcal W_\theta$ and $\Lc_q$ are locally Lipschitz on the open subset $\Qs\setminus\{0\}$ of the real Banach space $\Qs$ (completeness was shown in Section~\ref{sec:quadslice}). We write $\Qs^*$ for the set of all real-linear functionals on $\Qs$ that are continuous with respect to $\|\cdot\|_{\Qs}$. For $\Lc_q$ at $f\in\Qs\setminus\{0\}$, we define Clarke's generalized directional derivative in the direction $h\in\Qs$ and Clarke's generalized gradient \citep[\S~2.1]{clarke} by
\begin{equation}\label{eq:clarkeQ}
\begin{gathered}
D^\circ_h\Lc_q(f):=\limsup_{f'\to f,\ t\downarrow0}\frac{\Lc_q(f'+th)-\Lc_q(f')}{t},\\
\partial^\circ\Lc_q(f):=\big\{x^*\in\Qs^*:\ x^*(h)\le D^\circ_h\Lc_q(f)\ \ (h\in\Qs)\big\}.
\end{gathered}
\end{equation}
Here $f'\to f$ means convergence with respect to $\|\cdot\|_{\Qs}$. For $\mathcal W_\theta$ they are defined in the same way, by replacing $\Lc_q$ with $\mathcal W_\theta$ in \eqref{eq:clarkeQ}. The set $\partial^\circ\Lc_q(f)$ is nonempty and $D^\circ_h\Lc_q(f)=\max_{x^*\in\partial^\circ\Lc_q(f)}x^*(h)$ (Lemma~\ref{lem:clarkebasic} in the appendix). The properties of generalized gradients used in this paper are proved from the definitions in Appendix~\ref{sec:clarke}, for locally Lipschitz functions on open subsets of real Banach spaces. For a unit vector $\psi$ in $\Qs$, we say that $\Lc_\infty$ is differentiable at $\psi$ if $\partial^\circ\Lc_\infty(\psi)$ consists of a single point. By \citet[Proposition~2.2.4]{clarke}, this condition is equivalent to $\Lc_\infty$, as a function on $\Qs\setminus\{0\}$, being strictly differentiable at $\psi$ \citep[\S~2.2]{clarke}.

\begin{definition}[Stationary points]\label{def:stat}
Let $1\le q\le\infty$. Since $\mathcal Z$ is finite-dimensional, all real-linear functionals on it are continuous. We write $\mathcal Z^*$ for the set of all of them. For the function $\Ls_q$ of \eqref{eq:Lsq} at a point $\omega\in\Omega$ of the network, we define the generalized directional derivative in the direction $\zeta\in\mathcal Z$ and the generalized gradient by
\begin{equation}\label{eq:clarkeZ}
\begin{gathered}
D^\circ_\zeta\Ls_q(\omega):=\limsup_{\omega'\to\omega,\ t\downarrow0}\frac{\Ls_q(\omega'+t\zeta)-\Ls_q(\omega')}{t},\\
\partial^\circ\Ls_q(\omega):=\big\{z^*\in\mathcal Z^*:\ z^*(\zeta)\le D^\circ_\zeta\Ls_q(\omega)\ \ (\zeta\in\mathcal Z)\big\}.
\end{gathered}
\end{equation}
Since $\Ls_q$ is locally Lipschitz on $\Omega$ (Proposition~\ref{prop:lipschitz}), $D^\circ_\zeta\Ls_q(\omega)$ is finite. If $D^\circ_\zeta\Ls_q(\omega)\ge0$ for all $\zeta\in\mathcal Z$, we call $\omega$ a stationary point of the network for $\Lc_q$. This condition is equivalent to $0\in\partial^\circ\Ls_q(\omega)$.
\end{definition}

\section{Main results}\label{sec:results}

For a unit vector $\psi$ in $\Qs$, the infidelity of \eqref{eq:Wtheta} is $u(\psi)=1-|\ip\phi\psi|^2$. When $u(\psi)>0$, we define $g_\psi$ and $\Gamma(\psi)$ by
\begin{equation}\label{eq:udef}
g_\psi:=\frac{\psi-\ip\phi\psi\,\phi}{\sqrt{u(\psi)}},\qquad\Gamma(\psi):=\En(\phi)^{1/2}+\En(g_\psi)^{1/2}.
\end{equation}
The vector $g_\psi$ is the normalized component of $\psi$ orthogonal to $\phi$, and we call it the error direction of $\psi$. The vector $g_\psi$ is a unit vector in $\Qs$, and $\ip\phi{g_\psi}=0$.

The main results are statements about states whose exponential moment of the photon number is at most a given bound. Let $\N_0:=\{0,1,2,\dots\}$. Let $h_k$ ($k\in\N_0^n$) be the orthonormal basis of $\Hs$ consisting of the products $h_k(x)=\prod_jh_{k_j}(x_j)$ of Hermite functions of one variable, and write $|k|:=k_1+\dots+k_n$. For $\beta>0$ and $f\in\Hs$ we put
\begin{equation}\label{eq:Mbeta}
\Mo_\beta(f):=\sum_{k\in\N_0^n}e^{\beta|k|}\,|\ip{h_k}f|^2\ \in[0,\infty],
\end{equation}
and call this the exponential moment of the photon number of $f$. In the notation of quantum optics, $\Mo_\beta(f)=\ip f{e^{\beta\hat N}f}$. If $\Mo_\beta(f)<\infty$ then $f\in\Qs$ (Lemma~\ref{lem:expmom}).

\paragraph{Constants.}
We collect the constants that appear in the statements. For $1\le q\le\infty$ put
\begin{equation}\label{eq:exponents}
r:=\min\{q,2\},\qquad\gamma:=\frac{2n+r}{2r},\qquad\kappa:=\big((2n+r)\,2^{\,n+r-3}\,(n-1)!\big)^{1/r}.
\end{equation}
For $\beta>0$, $\Lambda\ge1$, and $s\in(0,1]$ put
\begin{equation}\label{eq:consts}
\begin{gathered}
E_*(s):=n+\frac2\beta\log\frac{4\Lambda}{s},\qquad\Gamma_*(s):=\En(\phi)^{1/2}+E_*(s)^{1/2},\\
\kappa_*(s):=\frac{\kappa}{4}\big(4(\En(\phi)+E_*(s))\big)^{-\gamma},
\end{gathered}
\end{equation}
and
\begin{equation}\label{eq:tau}
\tau(s):=16\,s\,\frac{\Gamma_*(s)^2}{\kappa_*(s)^2},\qquad\bar s:=\min\{1,4\Lambda e^{-(1+2\gamma)}\}.
\end{equation}
The function $\tau$ is continuous and nondecreasing on $(0,\bar s]$, and $\tau(s)\to0$ as $s\downarrow0$ (Lemma~\ref{lem:tau}). Hence the set of all $s\in(0,\bar s]$ that satisfy $\tau(s)\le1$ and the set of all $s\in(0,\bar s]$ that satisfy $16\,\tau(s)\le1$ are both nonempty, and each has a maximum. We write these maxima as
\begin{equation}\label{eq:ubar}
\bar u:=\max\{s\in(0,\bar s]:\tau(s)\le1\},\qquad\bar u':=\max\{s\in(0,\bar s]:16\,\tau(s)\le1\}.
\end{equation}
The roles of these constants are as follows.
\begin{itemize}
\item Quantities that appear in the assumptions. The constants $\beta$ and $\Lambda$ appear in the assumption $\Mo_\beta(f)\le\Lambda$ on the exponential moment of a state $f$.
\item Quantities determined by the number of modes $n$ and $q$ alone. These are $r$, $\gamma$, and $\kappa$. For $q\ge2$ we have $\gamma=(n+1)/2$ and $\kappa=(2^n(n+1)(n-1)!)^{1/2}$, which do not depend on $q$, and for $q=1$ we have $\gamma=(2n+1)/2$ and $\kappa=(2n+1)\,2^{n-2}(n-1)!$. The constants in the statements below depend on $q$ through $\gamma$ and $\kappa$.
\item An upper bound on the energy of the error direction, and the coefficients determined by it. When $\Mo_\beta(\phi)\le\Lambda$, $\Mo_\beta(\psi)\le\Lambda$, and $u(\psi)>0$, the quantity $E_*(u(\psi))$ is an upper bound on the energy of the error direction $g_\psi$ (Lemma~\ref{lem:expmom}). The quantity $\Gamma_*(s)$ is obtained from $\Gamma(\psi)$ of \eqref{eq:udef} by replacing $\En(g_\psi)$ with $E_*(s)$. The quantity $\kappa_*(s)$ is the coefficient in the lower bound on the loss (the second assertion of Lemma~\ref{prop:sharp}).
\item The size of the neighborhood of the target. The function $\tau$ appears in the upper bound on the infidelity of a stationary point of finite width (Proposition~\ref{thm:finite}). The constant $\bar s$ is the right endpoint of the interval $(0,\bar s]$ on which Lemma~\ref{lem:tau} shows that $\tau$ is nondecreasing. The constants $\bar u$ and $\bar u'$ are thresholds for the infidelity, and the statements concern states whose infidelity is at most these thresholds. We use $\bar u$ for the loss over all directions and $\bar u'$ for the loss over finitely many directions in Section~\ref{sec:numerics}.
\end{itemize}

\begin{restatable}[\restatetitle{Lower bound on the loss}]{lemma}{restsharp}\label{prop:sharp}
Let $1\le q\le\infty$, let $\psi$ be a unit vector in $\Qs$, and let $u:=u(\psi)$. First, for $r$, $\gamma$, and $\kappa$ of \eqref{eq:exponents},
\[
\Lc_q(\psi)\ \ge\ \kappa\,u^{1/r+\gamma}\big(4(\En(\psi)+\En(\phi))\big)^{-\gamma},
\]
and in particular $\Lc_q(\psi)=0$ if and only if $\psi=e^{i\alpha}\phi$ for some $\alpha$. Second, let $\beta>0$ and $\Lambda\ge1$, and assume $\Mo_\beta(\phi)\le\Lambda$. For $\Gamma_*$ and $\kappa_*$ of \eqref{eq:consts}, if $\psi$ satisfies $\Mo_\beta(\psi)\le\Lambda$, $u>0$, and $\sqrt u\le\kappa_*(u)/\Gamma_*(u)$, then $\Lc_q(\psi)\ge\kappa_*(u)\sqrt u$.
\end{restatable}

We call the inequality of the second assertion the sharpness of the loss and $\kappa_*(u)$ the sharpness coefficient. As $u\to0$ we have $E_*(u)\sim\frac2\beta\log\frac1u$, so
\[
\kappa_*(u)\sim\frac\kappa4\Big(\frac8\beta\log\frac1u\Big)^{-\gamma},\qquad\Gamma_*(u)\sim\Big(\frac2\beta\log\frac1u\Big)^{1/2},
\]
where $\sim$ means that the ratio converges to $1$. Hence, as $u\to0$, the ratio $\kappa_*(u)/\Gamma_*(u)$ is asymptotic to a positive constant multiple of $(\log(1/u))^{-(1+2\gamma)/2}$. Since $\sqrt u$ is a positive power of $u$, the condition $\sqrt u\le\kappa_*(u)/\Gamma_*(u)$ holds if $u$ is sufficiently small. For $q\ge2$, the constants are the same for the root mean square over directions ($q=2$) and for the supremum ($q=\infty$). For $q<2$, the constant $\gamma$ is larger than its value for $q\ge2$, and it is largest for the average ($q=1$).

\begin{restatable}[\restatetitle{Upper bound on the loss at a stationary point}]{lemma}{restonepoint}\label{prop:onepoint}
Let $\psi$ be a unit vector in $\Qs$ that satisfies $u:=u(\psi)>0$, and put $z:=\ip\phi\psi$. Let $\Gamma(\psi)$ be as in \eqref{eq:udef}. First, for all $(\theta,\mu)\in\Ac$ we have $\|v_{\theta,\mu}(\psi)\|\le\Gamma(\psi)$ and
\begin{equation}\label{eq:graddomre}
\ell_{\theta,\mu}(\psi)\ \le\ -2\Re\ip{v_{\theta,\mu}(\psi)}{z\phi}+2\Gamma(\psi)\,u.
\end{equation}
Second, let $1\le q\le\infty$, let the point $\omega$ of the network be a stationary point for $\Lc_q$, and let $\psi=\psi_\omega$. Then, for $\delta_\omega$ of \eqref{eq:deltaomega},
\begin{equation}\label{eq:Lupper}
\Lc_q(\psi)\ \le\ 2\Gamma(\psi)\big(\delta_\omega+u\big).
\end{equation}
\end{restatable}

The second assertion is the key to the argument of this paper. At a stationary point, the generalized gradient of the loss with respect to the state contains an element that vanishes on the range of the tangent map (Lemma~\ref{lem:netstat}). Applying this element to the element of the range closest to $\ip\phi\psi\phi$ gives \eqref{eq:Lupper}. In the upper bound \eqref{eq:Lupper}, the term with $\delta_\omega$ converges to $0$ as the width tends to infinity, provided that the sequence of initial centers is dense in $\R^n$ and the sum of the squared displacements of the centers from their initial values is bounded. The remaining term $2\Gamma(\psi)\,u$ is linear in $u$, and when $u$ is sufficiently small, the lower bound $\kappa_*(u)\sqrt u$ of Lemma~\ref{prop:sharp} is larger than this term. Hence the infidelity of the limit is $0$ if it is sufficiently small. This statement is Theorem~\ref{thm:limit}.

\begin{definition}[Bounded sequences of stationary points]\label{def:seq}
Let the sequence of initial centers $c_1,c_2,\dots\in\R^n$ be such that $\{c_j:j\ge1\}$ is dense in $\R^n$, and let the initial centers of the network of width $m$ be its first $m$ terms $(c_1,\dots,c_m)$. Fix $\sigma$ and $\varepsilon$. Let $1\le q\le\infty$, $\beta>0$, and $\Lambda\ge1$. Suppose that there exist $R\ge0$ and positive integers $m_k$ ($k\ge1$) with $m_k\to\infty$ such that, for each $k$, $\omega_k$ is a stationary point of the network of width $m_k$ for $\Lc_q$, and the centers $c^{(k)}$ of $\omega_k$ and the state $\psi_{\omega_k}$ satisfy
\begin{equation}\label{eq:seqbound}
\|c^{(k)}-(c_1,\dots,c_{m_k})\|\le R,\qquad\Mo_\beta(\psi_{\omega_k})\le\Lambda.
\end{equation}
Then $(\omega_k)_{k\ge1}$ is called a bounded sequence of stationary points for $\Lc_q$. A limit with respect to $\|\cdot\|_{\Qs}$ of a subsequence of $\psi_{\omega_k}$ is called a limit point of this sequence. When the centers of a point of the network are written alongside the initial centers, we write them as $c^{(k)}$, as here, or as $c'$.
\end{definition}

A sequence of independent random variables distributed according to a Borel probability measure whose support is all of $\R^n$ satisfies, with probability $1$, the condition on the initial centers in Definition~\ref{def:seq} (Lemma~\ref{lem:approx}). In an implementation, it suffices to draw the initial centers in this way.

For a bounded sequence of stationary points, every subsequence of $\psi_{\omega_k}$ has a subsequence that converges with respect to $\|\cdot\|_{\Qs}$, and the limit points are unit vectors in $\Qs$ (Lemma~\ref{lem:expmom}).

\begin{restatable}[\restatetitle{Limits near the target}]{mainthm}{restlimit}\label{thm:limit}
Let $1\le q\le\infty$, $\beta>0$, and $\Lambda\ge1$, and assume $\Mo_\beta(\phi)\le\Lambda$. With $\beta$ and $\Lambda$ in Definition~\ref{def:seq} taken to be these values, every limit point $\psi_*$ of a bounded sequence of stationary points for $\Lc_q$ satisfies $u(\psi_*)=0$ or satisfies $\Lc_q(\psi_*)\le2\,\Gamma(\psi_*)\,u(\psi_*)$ with $\Gamma$ of \eqref{eq:udef}. In particular, if $u(\psi_*)\le\bar u$ with $\bar u$ of \eqref{eq:ubar}, then $\psi_*=e^{i\alpha}\phi$ for some $\alpha$.
\end{restatable}

\begin{restatable}[\restatetitle{Landscape for $q=\infty$}]{mainthm}{restlandscape}\label{thm:landscape}
Every limit point $\psi_*$ of a bounded sequence of stationary points for $\Lc_\infty$ satisfies $0\in\overline{\conv}\,\mathcal G(\psi_*)$. If moreover $\Lc_\infty$ is differentiable at $\psi_*$ (more generally, if $0\in\mathcal G(\psi_*)$), then $\psi_*=e^{i\alpha}\phi$ for some $\alpha$.
\end{restatable}

Theorem~\ref{thm:limit} is obtained by comparing the limit as $\delta_\omega\to0$ of the upper bound of Lemma~\ref{prop:onepoint} with the lower bound of Lemma~\ref{prop:sharp}, and it holds for all $q$. The constant $\bar u$ is determined by $q$, $\beta$, $\Lambda$, $\En(\phi)$, and $n$ alone. Theorem~\ref{thm:landscape} gives the conclusion of the landscape theorem of the prior work, without the overparameterization assumption, for limits of stationary points of finite width. Theorem~\ref{thm:landscape} is not a statement restricted to a neighborhood of the target, and it makes no assumption on the exponential moment of the target. For $q<\infty$, the method of this paper cannot exclude the possibility that the gradients in the individual directions are nonzero while their integral over the directions is zero, so a statement of this form is not obtained.

\section{Quantitative form at finite width and a numerical example}\label{sec:numerics}

In this section we state the form obtained by solving the lower bound of Lemma~\ref{prop:sharp} and the upper bound of Lemma~\ref{prop:onepoint} for $u$ while keeping $\delta_\omega>0$ (Proposition~\ref{thm:finite}). Next, we transfer this estimate to the loss over finitely many directions (Proposition~\ref{prop:fdir}). Stationarity is stated in a form restricted to the directions in a subspace of the parameter space (Proposition~\ref{prop:restricted}). We then take one target with a single mode, compute stationary points numerically while increasing the width of the network from $12$ to $28$, and compare their infidelity with the square of the error of approximating the target by the image, under the tangent map, of the directions used in the minimization. The computations use double-precision floating-point numbers and are not certified by interval arithmetic. The proofs of the propositions and the corollary of this section are given in Section~\ref{sec:proofs4}.

\begin{restatable}[\restatetitle{Quantitative form at finite width}]{proposition}{restfinite}\label{thm:finite}
Let $1\le q\le\infty$, $\beta>0$, and $\Lambda\ge1$, and assume $\Mo_\beta(\phi)\le\Lambda$. For each point $\omega$ of the network, we write $c'$ for its centers, and we write $\psi:=\psi_\omega$ and $u:=u(\psi)$. For $\tau$ of \eqref{eq:tau} and $\bar u$ of \eqref{eq:ubar}, the following hold. First, $\delta_\omega$ of \eqref{eq:deltaomega} satisfies $\delta_\omega^2\le\dist(\phi,\mathcal V(c'))^2\le u$. Moreover, if $\omega$ is a stationary point for $\Lc_q$ and satisfies $\Mo_\beta(\psi)\le\Lambda$ and $0<u\le\bar u$, then $\delta_\omega>0$ and
\[
\delta_\omega^2\ \le\ u\ \le\ \tau\big(\delta_\omega^2\big)\ \le\ \tau\big(\dist(\phi,\mathcal V(c'))^2\big).
\]
Second, let $\nu$ be a Borel probability measure on $\R^n$. Suppose that the target can be written as a continuous superposition of units with some $\varrho\in L^2(\nu)$, that is,
\begin{equation}\label{eq:superpos}
\phi=\int_{\R^n}\varrho(y)\varphi_y\,\nu(dy),
\end{equation}
where the integral is a Bochner integral, and that the components $c_1,\dots,c_m$ of the initial centers $c=(c_1,\dots,c_m)\in(\R^n)^m$ are independent random variables with distribution $\nu$. For $\eta\in(0,1]$ and $R\ge0$ put
\begin{equation}\label{eq:deltam}
\delta_m^2:=\frac{2}{m\eta}\,\|\varrho\|_{L^2(\nu)}^2\,(\pi\sigma^2)^{n/2}\Big(1+\frac{R}{\sqrt2\,\sigma}\Big)^2.
\end{equation}
Then, if $\delta_m^2\le\bar u$, on an event of probability at least $1-\eta$, every stationary point $\omega$ of the network for $\Lc_q$ whose centers $c'$ satisfy $\|c'-c\|\le R$ and which satisfies $\Mo_\beta(\psi)\le\Lambda$ and $0<u\le\bar u$ also satisfies $u\le\tau(\delta_m^2)$.
\end{restatable}

The first assertion of Proposition~\ref{thm:finite} states that a point in the neighborhood that satisfies $u(\psi_\omega)>\tau(\delta_\omega^2)$ is not a stationary point. Here $\delta_\omega$ is the error of approximating the target by the range of the tangent map, which also includes the directions that move the centers. The neighborhood $\{\psi:0<u(\psi)\le\bar u,\ \Mo_\beta(\psi)\le\Lambda\}$ is determined by $q$, $\beta$, $\Lambda$, $\En(\phi)$, and $n$ alone and depends neither on the width nor on the centers. The quantity $\Mo_\beta(\psi_\omega)$ is determined by the parameters of the network alone, and it is finite for all $\omega\in\Omega$ if $e^\beta|\sigma^2-1|<\sigma^2+1$ (Lemma~\ref{lem:gaussunit}). If the centers $c_1,c_2,\dots$ are drawn independently from a distribution whose support is all of $\R^n$, then with probability $1$ the distance $\dist(\phi,\mathcal V(c_1,\dots,c_m))$ is nonincreasing in $m$ and converges to $0$ as $m\to\infty$ (Lemma~\ref{lem:approx}).

\paragraph{Finitely many directions.}
Only finitely many directions can be measured in an experiment, so the numerical example uses the loss over finitely many equally spaced angles for a single mode. Let $n=1$, and for a positive integer $K$, a unit vector $\psi$ in $\Qs$, and a point $\omega\in\Omega$ of the network put
\begin{equation}\label{eq:LqK}
\begin{gathered}
\theta_k:=\big(\cos(k\pi/K),\sin(k\pi/K)\big)\quad(k=0,\dots,K-1),\\
\Lc_{q,K}(\psi):=\Big(\frac1K\sum_{k=0}^{K-1}\mathcal W_{\theta_k}(\psi)^q\Big)^{1/q}\quad(q<\infty),\qquad\Lc_{\infty,K}(\psi):=\max_{0\le k<K}\mathcal W_{\theta_k}(\psi),\\
\Ls_{q,K}(\omega):=\Lc_{q,K}(\psi_\omega).
\end{gathered}
\end{equation}
As in Definition~\ref{def:loss}, for $f\in\Qs\setminus\{0\}$ we put $\Lc_{q,K}(f):=\Lc_{q,K}(f/\|f\|)$. By Proposition~\ref{prop:lipschitz}, in a neighborhood of each point of $\Qs\setminus\{0\}$ the functions $\mathcal W_{\theta_k}$ have a Lipschitz constant that does not depend on $k$, so by Minkowski's inequality $\Lc_{q,K}$ is locally Lipschitz on $\Qs\setminus\{0\}$. The generalized directional derivative and the generalized gradient of $\Lc_{q,K}$ are defined by replacing $\Lc_q$ with $\Lc_{q,K}$ in \eqref{eq:clarkeQ}. The quantity $D^\circ_\zeta\Ls_{q,K}(\omega)$ and stationary points for $\Lc_{q,K}$ are defined by replacing $\Ls_q$ with $\Ls_{q,K}$ and $\Lc_q$ with $\Lc_{q,K}$ in \eqref{eq:clarkeZ} and Definition~\ref{def:stat}. To transfer Proposition~\ref{thm:finite} to the stationary points of this loss, we use the following two propositions and a corollary (the proofs are in Section~\ref{sec:prooffdir}). For $\beta>0$, $\Lambda\ge1$, and a positive integer $N$ put
\begin{equation}\label{eq:epsN}
\epsilon_N:=8\sqrt\Lambda\,e^{-\beta(N+1)/2}\Big(3\Big(1+\frac2\beta\log\Lambda\Big)^{1/2}+(2N+3)^{1/2}\Big).
\end{equation}

\begin{restatable}[\restatetitle{Finitely many directions}]{proposition}{restfdir}\label{prop:fdir}
Let $n=1$, $1\le q\le\infty$, $\beta>0$, and $\Lambda\ge1$. Let the positive integers $K$ and $N$ satisfy $K\ge\pi N$ and $\beta(N+1)\ge\max\{1,\log(2\Lambda)\}$, and let $\epsilon_N$ be as in \eqref{eq:epsN}. If a unit vector $\psi$ in $\Qs$ and the target $\phi$ satisfy $\Mo_\beta(\psi)\le\Lambda$ and $\Mo_\beta(\phi)\le\Lambda$, then
\[
\tfrac12\,\Lc_q(\psi)-\epsilon_N\ \le\ \Lc_{q,K}(\psi)\ \le\ \tfrac32\,\Lc_q(\psi)+\epsilon_N.
\]
\end{restatable}

In a numerical minimization, sometimes stationarity can be checked only with respect to some of the directions of the parameter space. For a real subspace $\mathcal Z_0$ of $\mathcal Z$ and a point $\omega$ of the network, we therefore call the condition that $D^\circ_\zeta\Ls_{q,K}(\omega)\ge0$ for all $\zeta\in\mathcal Z_0$ stationarity at $\omega$ restricted to the directions in $\mathcal Z_0$, and we state stationarity in this form.

\begin{restatable}[\restatetitle{Stationarity restricted to a subspace}]{proposition}{restrestricted}\label{prop:restricted}
Let $n$, $q$, $\beta$, $\Lambda$, $K$, and $N$ be as in Proposition~\ref{prop:fdir}, and assume $\Mo_\beta(\phi)\le\Lambda$. Let $\kappa_*$, $\tau$, $\bar u'$, and $\epsilon_N$ be as in \eqref{eq:consts}, \eqref{eq:tau}, \eqref{eq:ubar} and \eqref{eq:epsN}. Let $\omega$ be a point of the network, and write $\psi:=\psi_\omega$ and $u:=u(\psi)$. Let $\mathcal Z_0$ be a real subspace of the parameter space $\mathcal Z$, and put
\begin{equation}\label{eq:deltatilde}
\tilde\delta_\omega:=\dist\big(\ip\phi{\psi_\omega}\phi,\ \mathscr J(\omega)\mathcal Z_0\big).
\end{equation}
If $D^\circ_\zeta\Ls_{q,K}(\omega)\ge0$ for all $\zeta\in\mathcal Z_0$, and if $\Mo_\beta(\psi)\le\Lambda$, $0<u\le\bar u'$, and $\epsilon_N\le\frac14\kappa_*(u)\sqrt u$ hold, then $\tilde\delta_\omega>0$ and $u\le16\,\tau(\tilde\delta_\omega^2)$. In particular, when $\mathcal Z_0=\mathcal Z$, the hypothesis on stationarity says that $\omega$ is a stationary point for $\Lc_{q,K}$, the quantity $\tilde\delta_\omega$ equals $\delta_\omega$ of \eqref{eq:deltaomega}, and $\delta_\omega^2\le u\le16\,\tau(\delta_\omega^2)$.
\end{restatable}

Since $\epsilon_N$ is exponentially small in $N$, the number of angles needed to satisfy the condition $\epsilon_N\le\frac14\kappa_*(u)\sqrt u$ at a point of infidelity $u$ is of the order of $\frac1\beta\log\frac\Lambda u$. For small $u$ that do not satisfy this condition, Proposition~\ref{prop:restricted} says nothing. The proof of Proposition~\ref{prop:fdir} relies on the fact that, for a pair of states with photon number at most $N$, the difference of the distribution functions of the slices is a trigonometric polynomial of degree at most $N$ in the angle, and on the interpolation formula of M.~Riesz.

\paragraph{Setting.}
The target is the cat state obtained from a coherent state by the time evolution $e^{-i\pi\hat N^2/2}$ under the Kerr effect,
\[
\phi\ \propto\ e^{-i\pi/4}e^{-(x-3\sqrt2)^2/2}+e^{i\pi/4}e^{-(x+3\sqrt2)^2/2}.
\]
Its photon number distribution is the Poisson distribution with mean $9$, so $\En(\phi)=19$ and $\Mo_\beta(\phi)=\exp(9(e^\beta-1))$. The units of the network are the Gaussian functions $\varphi_y(x)=e^{-(x-y)^2}$ ($y\in\R$) with $\sigma^2=1/2$. The target is a sum of Gaussian functions with $\sigma^2=1$, so it cannot be represented by the network at any width. On the other hand,
\[
\int_{\R}e^{-(x-y)^2}e^{-(y\mp3\sqrt2)^2}\,dy=\sqrt{\frac\pi2}\,e^{-(x\mp3\sqrt2)^2/2},
\]
so the target can be written as a superposition of units of the form \eqref{eq:superpos}. If we take $\nu$ to be the standard normal distribution, then $\varrho(y)$ is a constant multiple of $e^{y^2/2}(e^{-i\pi/4}e^{-(y-3\sqrt2)^2}+e^{i\pi/4}e^{-(y+3\sqrt2)^2})$, and $|\varrho(y)|^2e^{-y^2/2}$ is at most a constant multiple of $e^{-3y^2/2+12\sqrt2|y|}$, so $\varrho\in L^2(\nu)$. Hence the target satisfies those hypotheses of the second assertion of Proposition~\ref{thm:finite} that concern the target. The loss is $\Lc_{q,240}$ ($q=1,2,\infty$), the loss over $K=240$ equally spaced angles. To compute the loss, we represent states by their coefficients with respect to the Hermite functions of photon number at most $119$ and compute the distribution functions of the slices on a grid of spacing $0.02$ on the interval $[-20,20]$. The value $\mathcal W_{\theta_k}$ is obtained by interpolating the difference of the distribution functions linearly between grid points and computing the integral of its absolute value exactly.

\paragraph{Training with increasing width.}
We start from width $12$ and increase the width in steps of $4$ up to $28$. At width $12$, we first place the centers equally spaced on the interval $[-6,6]$ and, keeping the centers fixed at these positions, minimize $\Lc_{q,240}$ using the coefficients with respect to an orthonormal basis of $\mathcal V(c)$ as coordinates, and we take the resulting point as the starting point. At each width, we minimize $\Lc_{q,240}$ with respect to all $3m$ real parameters, namely the real and imaginary parts of the amplitudes and the centers. For $q=1,2$ we use a quasi-Newton method, and for $q=\infty$ we use sequential quadratic programming in the form of minimizing $t$ subject to $\max_k\mathcal W_{\theta_k}\le t$. We regard the tangent map $\mathscr J(\omega)$ as a real-linear map into $\Hs$ from the parameter space $\mathcal Z$, whose coordinates are the real and imaginary parts of the amplitudes and the centers, and we denote the singular values of this map by $s_1\ge s_2\ge\cdots$. Let $\mathcal Z_0$ be the subspace of $\mathcal Z$ spanned by the right singular vectors whose singular values are larger than $10^{-6}s_1$. As coordinates for the minimization we take elements of $\mathcal Z_0$, scaled so that their images under $\mathscr J(\omega)$ are orthonormal. If a direction with a small singular value is included in the coordinates, the scale of that coordinate grows in proportion to $1/s_j$, and the change of the loss caused by a step in that direction can no longer be distinguished from the rounding error of double precision, so we exclude these directions. When there are two units with nearby centers or a unit with a small amplitude, $\mathscr J(\omega)$ has directions with small singular values. We repeat the minimization, recomputing the basis each time, until the relative decrease of the loss falls below $10^{-9}$ or no step that decreases the loss can be found. For $q=2$ at width $20$, however, we stopped at the upper limit of $40$ on the number of recomputations. For $q=\infty$ at width $20$, we stopped at the point where the singular value decomposition no longer converged. At the last point of each width, the number of right singular vectors not contained in $\mathcal Z_0$ is $2$ out of $36$ at width $12$, and $17$ to $19$ out of $84$ at width $28$.

When the width is increased, we add $4$ units with amplitude $10^{-10}$ to the last point. The centers to be added are chosen one at a time. Let $c$ denote the centers at that moment. Among the points $y$ of the grid of spacing $0.02$ on the interval $[-8,8]$ that are at distance at least $0.05$ from the existing centers, the one that maximizes $|\ip{\varphi_y}{\phi-\Pi_{\mathcal V(c)}\phi}|/\|\varphi_y\|$ is taken as the center to be added. Since the added units have amplitude $10^{-10}$, the norm of the change of the state is at most $10^{-9}$, but $\mathscr J(\omega)\mathcal Z_0$ becomes larger by the directions of the amplitudes of the added units.

\paragraph{Quantities compared.}
The minimization is carried out only in the directions in $\mathcal Z_0$, so what can be checked at a point where the minimization stopped is stationarity restricted to the directions in $\mathcal Z_0$. We therefore compare the infidelity with the square of $\tilde\delta_\omega$ of Proposition~\ref{prop:restricted}, that is, with the square of the quantity \eqref{eq:deltatilde} in which $\mathcal Z_0$ is the subspace defined above. Since $\mathscr J(\omega)\mathcal Z_0\subset\ran\mathscr J(\omega)$, we have $\delta_\omega\le\tilde\delta_\omega$. At the points of width $12$ the two agree to the digits shown in the table, but the larger the width, the smaller $\delta_\omega^2$ is in comparison, and at width $28$ it is smaller by about $3$ orders of magnitude (Table~\ref{tab:stair}).

Stationarity is checked by the following quantities. For $q=1,2$, we compare the norm of the gradient of the loss with respect to the coordinates on $\mathcal Z_0$ with the maximum over $k$ of the norms of the gradients of the individual $\mathcal W_{\theta_k}$ (called the reference below). For $q=\infty$, we take the angles at which $\mathcal W_{\theta_k}$ is at least $1-10^{-6}$ times its maximum over $k$, compute the minimum norm of a convex combination of the gradients of $\mathcal W_{\theta_k}$ at these angles, taken with respect to the coordinates on $\mathcal Z_0$, and compare it with the same reference. Because of rounding errors, one cannot decide whether $\mathcal W_{\theta_k}$ coincides exactly with this maximum, so we allowed a relative difference of $10^{-6}$.

\begin{figure}[tbp]
\centering
\includegraphics[width=0.8\textwidth]{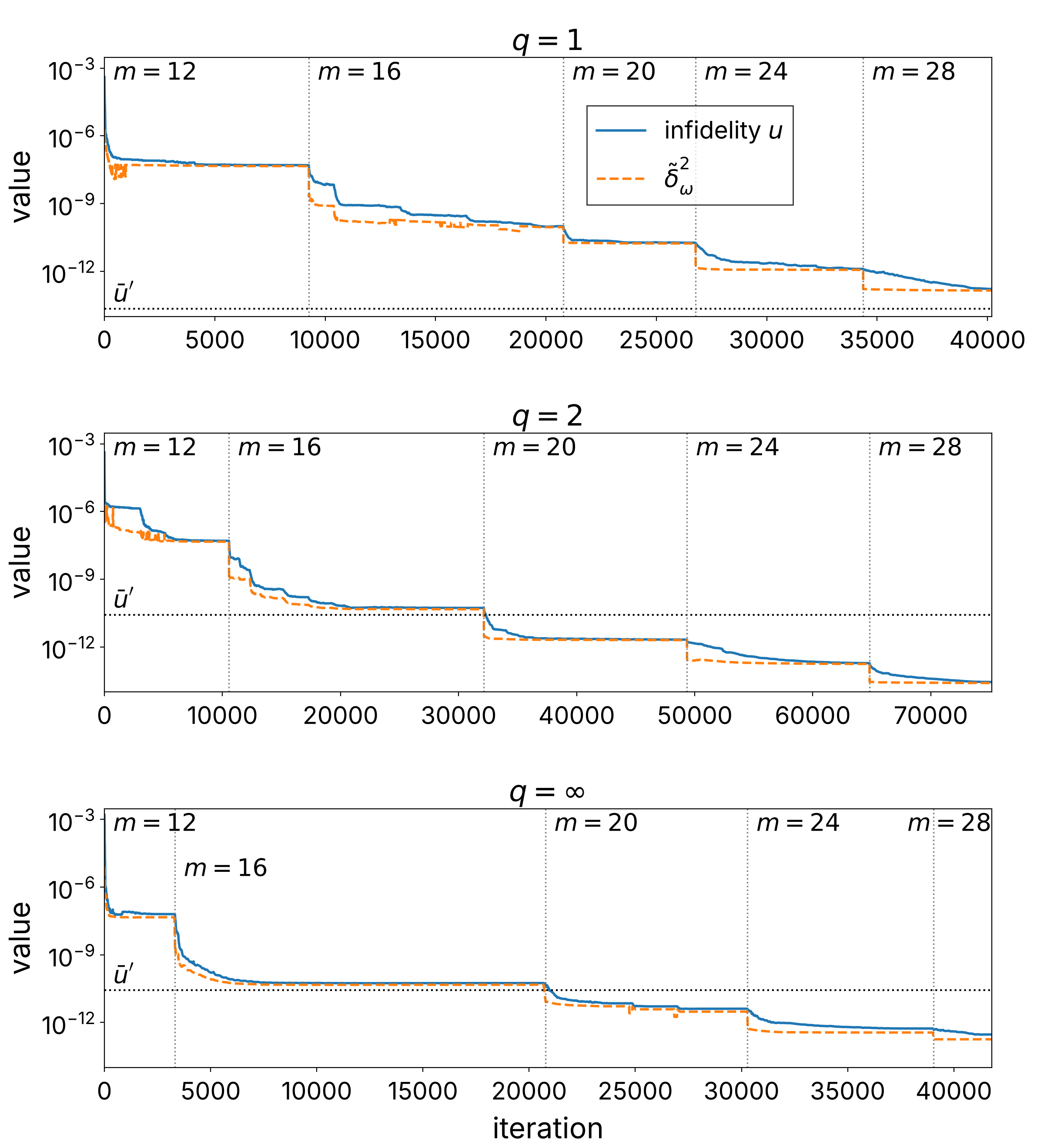}
\caption{Infidelity $u$ (solid line) and $\tilde\delta_\omega^2$ (dashed line) when $\Lc_{q,240}$ is minimized with respect to all amplitudes and centers while the width is increased from $12$ to $28$ in steps of $4$. The vertical dotted lines mark the iterations at which the width was increased, and the horizontal dotted line marks $\bar u'$ ($\beta=1.05$, $\Lambda=1.5\times10^9$). The horizontal axis is the number of iterations of the quasi-Newton method ($q=1,2$) or of sequential quadratic programming ($q=\infty$), counted cumulatively across the recomputations of the basis and across the widths. For $q=\infty$, only the iterations at which the loss improved on its previous minimum are plotted. For $q=1$, $\bar u'$ lies below the infidelities reached numerically, and Proposition~\ref{prop:restricted} does not apply.}
\label{fig:stair}
\end{figure}

\paragraph{Results.}
Figure~\ref{fig:stair} shows the infidelity $u$ and $\tilde\delta_\omega^2$ along the iterations of the training, and Table~\ref{tab:stair} shows the values at the last point of each width. For every $q$, at the end of each width the infidelity stops near $\tilde\delta_\omega^2$, and the ratio $u/\tilde\delta_\omega^2$ is between $1.05$ and $1.17$ for $q=1,2$ and between $1.18$ and $1.63$ for $q=\infty$. Immediately after the width is increased, $\tilde\delta_\omega^2$ drops and the ratio is then between $3$ and $20$. In the subsequent minimization the infidelity decreases and again stops near $\tilde\delta_\omega^2$. The infidelity at the last point of each width is between $5.0\times10^{-8}$ and $6.3\times10^{-8}$ at width $12$, and between $2.8\times10^{-14}$ and $2.9\times10^{-13}$ at width $28$. The stationarity quantity at the last point of each width is between $1.8\times10^{-6}$ and $5.9\times10^{-3}$ times the reference.

\begin{table}[t]
\centering
\caption{Values at the last point of each width (Kerr cat, $\sigma^2=1/2$, $K=240$). The quantity $\delta_\omega^2$ was computed with the image of the directions whose singular values are larger than $10^{-11}$ times $s_1$. The ratio $u/\tilde\delta_\omega^2$ was computed from the values before rounding. The column ``stationarity'' is the quantity defined in the text divided by the reference.}
\label{tab:stair}
\small\setlength{\tabcolsep}{6pt}
\begin{tabular}{ccccccc}
\toprule
$q$ & $m$ & $u$ & $\tilde\delta_\omega^2$ & $u/\tilde\delta_\omega^2$ & $\delta_\omega^2$ & stationarity\\
\midrule
$1$ & $12$ & $5.06\times10^{-8}$ & $4.58\times10^{-8}$ & $1.11$ & $4.6\times10^{-8}$ & $3.9\times10^{-5}$\\
$1$ & $16$ & $9.95\times10^{-11}$ & $8.99\times10^{-11}$ & $1.11$ & $5.4\times10^{-11}$ & $8.1\times10^{-5}$\\
$1$ & $20$ & $1.84\times10^{-11}$ & $1.71\times10^{-11}$ & $1.08$ & $9.7\times10^{-13}$ & $3.1\times10^{-4}$\\
$1$ & $24$ & $1.28\times10^{-12}$ & $1.16\times10^{-12}$ & $1.10$ & $5.0\times10^{-15}$ & $5.1\times10^{-4}$\\
$1$ & $28$ & $1.65\times10^{-13}$ & $1.41\times10^{-13}$ & $1.17$ & $6.3\times10^{-17}$ & $5.9\times10^{-3}$\\
\midrule
$2$ & $12$ & $5.02\times10^{-8}$ & $4.58\times10^{-8}$ & $1.10$ & $4.6\times10^{-8}$ & $1.8\times10^{-6}$\\
$2$ & $16$ & $5.35\times10^{-11}$ & $4.66\times10^{-11}$ & $1.15$ & $4.7\times10^{-11}$ & $1.0\times10^{-3}$\\
$2$ & $20$ & $2.09\times10^{-12}$ & $2.00\times10^{-12}$ & $1.05$ & $3.1\times10^{-13}$ & $5.2\times10^{-3}$\\
$2$ & $24$ & $1.91\times10^{-13}$ & $1.76\times10^{-13}$ & $1.09$ & $2.5\times10^{-15}$ & $1.9\times10^{-3}$\\
$2$ & $28$ & $2.77\times10^{-14}$ & $2.52\times10^{-14}$ & $1.10$ & $2.1\times10^{-17}$ & $2.6\times10^{-3}$\\
\midrule
$\infty$ & $12$ & $6.33\times10^{-8}$ & $4.60\times10^{-8}$ & $1.38$ & $4.6\times10^{-8}$ & $2.7\times10^{-5}$\\
$\infty$ & $16$ & $5.58\times10^{-11}$ & $4.71\times10^{-11}$ & $1.18$ & $4.7\times10^{-11}$ & $1.4\times10^{-4}$\\
$\infty$ & $20$ & $4.08\times10^{-12}$ & $3.03\times10^{-12}$ & $1.35$ & $6.4\times10^{-13}$ & $2.1\times10^{-4}$\\
$\infty$ & $24$ & $5.42\times10^{-13}$ & $3.59\times10^{-13}$ & $1.51$ & $5.6\times10^{-15}$ & $1.3\times10^{-3}$\\
$\infty$ & $28$ & $2.94\times10^{-13}$ & $1.80\times10^{-13}$ & $1.63$ & $5.4\times10^{-17}$ & $3.3\times10^{-4}$\\
\bottomrule
\end{tabular}
\end{table}

With $\beta=1.05$ and $\Lambda=1.5\times10^9$, $\bar u'$ is $2.7\times10^{-11}$ for $q=2,\infty$ and $2.2\times10^{-14}$ for $q=1$. For $q=2,\infty$, the infidelities of the points of widths $20$, $24$, and $28$ are at most $\bar u'$, and the infidelities of the points of width $16$ ($5.35\times10^{-11}$ and $5.58\times10^{-11}$) are about $2$ times $\bar u'$. We verify the hypotheses of Proposition~\ref{prop:restricted} with $\beta=1.05$, $\Lambda=1.5\times10^9$, and $N=76$. Since $e^\beta|\sigma^2-1|/(\sigma^2+1)=0.95$, the exponential moments of the units are finite by Lemma~\ref{lem:gaussunit}. Computing the Hermite coefficients of the units by the three-term recurrence up to photon number $399$, we find that $\Mo_\beta$ of the states at the $15$ points of Table~\ref{tab:stair} is between $2.8\times10^7$ and $1.45\times10^9$, and that $\Mo_\beta(\phi)=1.8\times10^7$ for the target. We have $K=240\ge\pi N=238.8$ and $\beta(N+1)=80.9\ge\log(2\Lambda)=21.8$, and $\epsilon_N=2.7\times10^{-11}$. At the six points whose infidelity is at most $\bar u'$, the quantity $\frac14\kappa_*(u)\sqrt u$ is between $4.3\times10^{-11}$ and $5.6\times10^{-10}$, in every case at least $\epsilon_N$. Hence these six points satisfy all the hypotheses of Proposition~\ref{prop:restricted}, except that stationarity restricted to the directions in $\mathcal Z_0$ is a numerical approximation. The upper bound $16\,\tau(\tilde\delta_\omega^2)$ of the proposition is between $1.3\times10^{-3}$ and $1.3\times10^{-1}$ at these six points, which is $2.9\times10^{10}$ to $4.7\times10^{10}$ times the observed infidelity. This factor is the value of $16\,\tau(s)/s$ at $s=\tilde\delta_\omega^2$ (between $4.2\times10^{10}$ and $5.2\times10^{10}$) divided by the ratio $u/\tilde\delta_\omega^2$.

For $q=1$, the infidelities of all the points are larger than $\bar u'$, and Proposition~\ref{prop:restricted} does not apply. The reason why $\bar u'$ for $q=1$ is $3$ orders of magnitude smaller than $\bar u'$ for $q=2,\infty$ is that the estimates of the constants are cruder for $q<2$ (Section~\ref{sec:discussion}). Nevertheless, the infidelities of the points for $q=1$ are at most $1.17$ times $\tilde\delta_\omega^2$, and they decrease with the width in the same way as for $q=2,\infty$.

\section{Proofs}\label{sec:proofs}

For the functional calculus of self-adjoint operators and Stone's theorem we follow \citet[\S~VIII.3, \S~VIII.4]{reedsimon}.

\subsection{Hermite expansions and exponential moments}\label{sec:prooffock}

The symbols $h_k$ ($k\in\N_0^n$) and $|k|$ denote, as defined just before \eqref{eq:Mbeta}, the orthonormal basis of $\Hs$ and the sum $k_1+\dots+k_n$. For $j=1,\dots,n$ and $N\in\N_0$ we put
\begin{equation}\label{eq:ladder}
\hat a_j:=\frac{x_j+\partial_j}{\sqrt2},\qquad\hat a_j^*:=\frac{x_j-\partial_j}{\sqrt2},\qquad\mathcal D_N:=\spn_{\C}\{h_k:|k|=N\},\qquad\mathcal D:=\spn_{\C}\{h_k:k\in\N_0^n\}.
\end{equation}
We call $\hat a_j$ the annihilation operator and $\hat a_j^*$ the creation operator, and we let them act on distributions, as $x_j$ and $\partial_j$ do. The space $\mathcal D$ is the set of all finite linear combinations of the Hermite functions $h_k$, and $\mathcal D_N$ is the set of all linear combinations of those $h_k$ with $|k|=N$.

\begin{lemma}[Hermite expansion]\label{lem:fock}
A vector $f\in\Hs$ belongs to $\Qs$ if and only if $\sum_k(2|k|+n)|\ip{h_k}f|^2<\infty$, and in that case
\begin{equation}\label{eq:fock}
\En(f)=\sum_k(2|k|+n)|\ip{h_k}f|^2.
\end{equation}
In particular $\En(f)\ge n\|f\|^2$.
\end{lemma}

\begin{proof}
Put $c_k:=\ip{h_k}f$ and let $e_j$ be the $j$-th unit vector of $\N_0^n$. For a distribution $T$ and $g\in\Sc$ we write $\ip gT:=T(\overline g)$. For $\hat a_j$ and $\hat a_j^*$ of \eqref{eq:ladder}, the Hermite functions satisfy $\hat a_j^*h_k=\sqrt{k_j+1}\,h_{k+e_j}$, and they satisfy $\hat a_jh_k=\sqrt{k_j}\,h_{k-e_j}$ if $k_j\ge1$ and $\hat a_jh_k=0$ if $k_j=0$. Moreover, for $g\in\Sc$, the definition of the derivative of a distribution gives $\ip g{\hat a_jf}=\ip{\hat a_j^*g}f$ and $\ip g{\hat a_j^*f}=\ip{\hat a_jg}f$. Taking $g=h_k$, we get $\ip{h_k}{\hat a_jf}=\sqrt{k_j+1}\,c_{k+e_j}$ for all $k\in\N_0^n$, and we get $\ip{h_k}{\hat a_j^*f}=\sqrt{k_j}\,c_{k-e_j}$ if $k_j\ge1$ and $\ip{h_k}{\hat a_j^*f}=0$ if $k_j=0$.

Let $f\in\Qs$. Since $\hat a_jf$ and $\hat a_j^*f$ belong to $\Hs$, the completeness of $\{h_k\}$ allows us to use Parseval's identity, and $\|\hat a_jf\|^2=\sum_{k\in\N_0^n}(k_j+1)|c_{k+e_j}|^2$. Since $k\mapsto k+e_j$ is a bijection from $\N_0^n$ onto $\{k\in\N_0^n:k_j\ge1\}$, the right-hand side equals $\sum_{k\in\N_0^n,\,k_j\ge1}k_j|c_k|^2=\sum_{k\in\N_0^n}k_j|c_k|^2$. By Parseval's identity and the same reindexing as above, $\|\hat a_j^*f\|^2=\sum_{k\in\N_0^n,\,k_j\ge1}k_j|c_{k-e_j}|^2=\sum_{k\in\N_0^n}(k_j+1)|c_k|^2$. By the parallelogram law, $\|\hat a_jf\|^2+\|\hat a_j^*f\|^2=\|x_jf\|^2+\|\partial_jf\|^2$, so summing over $j$ gives \eqref{eq:fock}.

Conversely, assume $\sum_k(2|k|+n)|c_k|^2<\infty$. By the same reindexing as above, $\sum_{k\in\N_0^n}(k_j+1)|c_{k+e_j}|^2$ and $\sum_{k\in\N_0^n,\,k_j\ge1}k_j|c_{k-e_j}|^2$ are finite, so $y_j:=\sum_{k\in\N_0^n}\sqrt{k_j+1}\,c_{k+e_j}h_k$ and $z_j:=\sum_{k\in\N_0^n,\,k_j\ge1}\sqrt{k_j}\,c_{k-e_j}h_k$ converge in $\Hs$. For $g\in\Sc$, the quantity $\ip{h_k}{\hat a_j^*g}=\ip{\hat a_jh_k}g$ equals $\sqrt{k_j}\,\ip{h_{k-e_j}}g$ if $k_j\ge1$ and $0$ if $k_j=0$, so Parseval's identity and the reindexing give
\[
\ip g{\hat a_jf}=\ip{\hat a_j^*g}f=\sum_{k\in\N_0^n,\,k_j\ge1}\sqrt{k_j}\,\overline{\ip{h_{k-e_j}}g}\,c_k=\sum_{k\in\N_0^n}\overline{\ip{h_k}g}\,\sqrt{k_j+1}\,c_{k+e_j}=\ip g{y_j},
\]
and hence $\hat a_jf=y_j$ as distributions. The same computation with the roles of $\hat a_j$ and $\hat a_j^*$ interchanged gives $\hat a_j^*f=z_j$ as distributions. Hence $x_jf=(y_j+z_j)/\sqrt2$ and $\partial_jf=(y_j-z_j)/\sqrt2$ belong to $\Hs$, and $f\in\Qs$. By \eqref{eq:fock} and Parseval's identity, $\En(f)\ge n\|f\|^2$.
\end{proof}

\begin{lemma}[Change of the direction of the quadrature]\label{lem:rot}
Let $\theta=(\xi,\varpi)\in\Sph$, and for $j=1,\dots,n$ take $r_j\ge0$ and $t_j\in\R$ such that $\xi_j+i\varpi_j=r_je^{it_j}$. Then $r:=(r_1,\dots,r_n)$ is a unit vector of $\R^n$. Let $\mathbf e_1:=(1,0,\dots,0)$, and let $O$ be an orthogonal matrix of order $n$ with $O\mathbf e_1=r$. Define unitary operators $R$ and $G$ on $\Hs$ by
\begin{equation}\label{eq:rotdef}
(Rf)(x):=f(O^{\mathsf T}x),\qquad Gh_k:=e^{i(t_1k_1+\dots+t_nk_n)}h_k\quad(k\in\N_0^n),
\end{equation}
and put $U:=GR$. Then $E_\theta(B)=U\,\mathbf 1_B(x_1)\,U^*$ for all Borel sets $B\subset\R$. Here $\mathbf 1_B(x_1)$ is the operator of multiplication by the function $x\mapsto\mathbf 1_B(x_1)$. Consequently, the slice $\Pm^\theta_f$ of a unit vector $f\in\Hs$ is the marginal distribution in the first coordinate of the probability measure $|(U^*f)(x)|^2dx$, it is absolutely continuous with respect to Lebesgue measure, and $F^\theta_f$ is continuous.
\end{lemma}

\begin{proof}
The operator $R$ is a change of variables by an orthogonal transformation, so it is unitary. The operator $G$ maps each element $h_k$ of the orthonormal basis to a multiple of it by a constant of modulus $1$, so it is unitary and maps $\mathcal D$ into itself. Let $e_j$ be the $j$-th unit vector of $\N_0^n$. From $\hat a_j^*h_k=\sqrt{k_j+1}\,h_{k+e_j}$, from $\hat a_jh_k=\sqrt{k_j}\,h_{k-e_j}$ for $k_j\ge1$, and from $\hat a_jh_k=0$ for $k_j=0$, we have on $\mathcal D$
\[
G\,\hat a_j\,G^*=e^{-it_j}\hat a_j,\qquad G\,\hat a_j^*\,G^*=e^{it_j}\hat a_j^*.
\]
Since $x_j=(\hat a_j+\hat a_j^*)/\sqrt2$ and $p_j=-i(\hat a_j-\hat a_j^*)/\sqrt2$, we have on $\mathcal D$
\begin{equation}\label{eq:frft}
G\,x_j\,G^*=\frac{e^{-it_j}\hat a_j+e^{it_j}\hat a_j^*}{\sqrt2}=\cos t_j\,x_j+\sin t_j\,p_j.
\end{equation}
We write $\langle r,x\rangle$ for the operator of multiplication by the function $x\mapsto\langle r,x\rangle$. This operator is self-adjoint and maps $\mathcal D$ into itself. Multiplying \eqref{eq:frft} by $r_j$ and summing over $j$, we obtain from $r_j\cos t_j=\xi_j$, $r_j\sin t_j=\varpi_j$, and \eqref{eq:Xxi} that $G\,\langle r,x\rangle\,G^*=\sum_j(\xi_jx_j+\varpi_jp_j)=X_\theta$ on $\mathcal D$.

We show that $\mathcal D$ is a core for $X_\theta$. For $g\in\Sc$ put $g_N:=\sum_{|k|\le N}\ip{h_k}g\,h_k\in\mathcal D$. By Lemma~\ref{lem:fock} we have $\En(g-g_N)\to0$, and by \eqref{eq:Xxi} we have $\|X_\theta(g-g_N)\|\le\En(g-g_N)^{1/2}\to0$. Hence the closure of the restriction of $X_\theta$ to $\mathcal D$ contains the restriction of $X_\theta$ to $\Sc$. Since $\Sc$ is a core for $X_\theta$ (see the proof of Lemma~\ref{lem:slice}), $\mathcal D$ is also a core for $X_\theta$. The self-adjoint operator $G\,\langle r,x\rangle\,G^*$ coincides with $X_\theta$ on the core $\mathcal D$, so it is an extension of $X_\theta$, and since $X_\theta$ is also self-adjoint, $G\,\langle r,x\rangle\,G^*=X_\theta$. Hence $E_\theta(B)=G\,\mathbf 1_B(\langle r,x\rangle)\,G^*$. Since $(O^{\mathsf T}x)_1=\langle O\mathbf e_1,x\rangle=\langle r,x\rangle$, the operator $R\,\mathbf 1_B(x_1)\,R^*$ is multiplication by the function $x\mapsto\mathbf 1_B(\langle r,x\rangle)$, and we obtain $E_\theta(B)=U\,\mathbf 1_B(x_1)\,U^*$. The latter assertions of the lemma follow from $\Pm^\theta_f(B)=\ip{U^*f}{\mathbf 1_B(x_1)U^*f}=\int_{\R^n}\mathbf 1_B(x_1)\,|(U^*f)(x)|^2dx$ and Fubini's theorem.
\end{proof}

When $n=1$ and $\theta=(\cos t,\sin t)$, the operator $R$ is the identity and $U^*=G^*$ maps $h_k$ to $e^{-itk}h_k$. This is the fractional Fourier transform $\mathcal F_t$ of angle $t$, and the density of the slice is $|\mathcal F_tf|^2$. For $0<t<\pi$, Mehler's formula gives, for $f\in\Sc$,
\[
(\mathcal F_tf)(x)=\frac{e^{it/2}}{\sqrt{2\pi i\sin t}}\int_{\R}\exp\Big(\frac{i}{2\sin t}\big((x^2+y^2)\cos t-2xy\big)\Big)f(y)\,dy
\]
\citep{folland}. For general $n$, the operator $G^*$ is the product of the fractional Fourier transforms of angle $t_j$ in the individual variables.

\begin{lemma}[Exponential moments]\label{lem:expmom}
Let $\beta>0$ and $\Lambda\ge1$. First, if a unit vector $f\in\Hs$ satisfies $\Mo_\beta(f)\le\Lambda$, then $f\in\Qs$ and $\En(f)\le n+\frac2\beta\log\Lambda$. The set of all unit vectors $f\in\Hs$ satisfying $\Mo_\beta(f)\le\Lambda$ is compact with respect to $\|\cdot\|_{\Qs}$. Second, assume $\Mo_\beta(\phi)\le\Lambda$. If a unit vector $\psi$ satisfies $\Mo_\beta(\psi)\le\Lambda$ and $u:=u(\psi)>0$, then $\En(g_\psi)\le E_*(u)$ for $E_*$ of \eqref{eq:consts}.
\end{lemma}

\begin{proof}
Since $\Mo_\beta^{1/2}$ is a weighted $\ell^2$ norm, it satisfies the triangle inequality. For a unit vector $f$ put $p_k:=|\ip{h_k}f|^2$. Then $\sum_kp_k=1$, and Jensen's inequality gives $\exp(\beta\sum_kp_k|k|)\le\sum_kp_ke^{\beta|k|}=\Mo_\beta(f)$, that is,
\begin{equation}\label{eq:jensen}
\sum_kp_k|k|\ \le\ \frac1\beta\log\Mo_\beta(f).
\end{equation}
If $\Mo_\beta(f)\le\Lambda$, then by Lemma~\ref{lem:fock} we have $f\in\Qs$ and $\En(f)=2\sum_kp_k|k|+n\le n+\frac2\beta\log\Lambda$.

We prove the compactness. Take a sequence $f_j$ of unit vectors satisfying $\Mo_\beta(f_j)\le\Lambda$. Passing to a subsequence, we may assume that $f_j$ converges weakly to some $f\in\Hs$. Since $\ip{h_k}{f_j}\to\ip{h_k}f$ for each $k$, Fatou's lemma gives $\Mo_\beta(f)\le\Lambda$. By \eqref{eq:fock}, $\|f_j-f\|_{\Qs}^2=\sum_k(2|k|+n+1)|\ip{h_k}{f_j-f}|^2$. For $d\ge0$, the sum of the terms with $|k|>d$ of the series on the right-hand side is at most $\Mo_\beta(f_j-f)\sup_{t>d}(2t+n+1)e^{-\beta t}\le4\Lambda\sup_{t>d}(2t+n+1)e^{-\beta t}$, and the sum of the terms with $|k|\le d$ is a finite sum and therefore converges to $0$ as $j\to\infty$. Letting $d\to\infty$, we obtain $\|f_j-f\|_{\Qs}\to0$. In particular $\|f\|=1$.

We prove the second assertion. By \eqref{eq:udef}, $\sqrt u\,g_\psi=\psi-\ip\phi\psi\phi$. Hence the triangle inequality and the homogeneity of $\Mo_\beta^{1/2}$, together with the bound $|\ip\phi\psi|\le1$ from the Cauchy--Schwarz inequality, give $\sqrt u\,\Mo_\beta(g_\psi)^{1/2}\le\Mo_\beta(\psi)^{1/2}+|\ip\phi\psi|\,\Mo_\beta(\phi)^{1/2}\le2\Lambda^{1/2}$, that is, $\Mo_\beta(g_\psi)\le4\Lambda/u$. Since $g_\psi$ is a unit vector, \eqref{eq:jensen} and \eqref{eq:fock} give $\En(g_\psi)\le n+\frac2\beta\log\frac{4\Lambda}{u}=E_*(u)$.
\end{proof}

\subsection{Networks}\label{sec:proofnet}

\begin{lemma}[Gaussian units]\label{lem:gaussunit}
First, for all $c,c'\in\R^n$ we have $\|\varphi_c\|^2=(\pi\sigma^2)^{n/2}$ and $\|\varphi_c-\varphi_{c'}\|\le|c-c'|\,\|\varphi_0\|/(\sqrt2\,\sigma)$. Second, for every dense subset $D$ of $\R^n$, the subspace spanned by $\{\varphi_c:c\in D\}$ is dense in $\Hs$. Third, $c\mapsto\varphi_c$ is a $C^1$ map from $\R^n$ into $\Qs$. Fourth, if $\beta>0$ satisfies $e^\beta|\sigma^2-1|<\sigma^2+1$, then $\Mo_\beta(\varphi_c)<\infty$ for all $c\in\R^n$ and $\Mo_\beta(\psi_\omega)<\infty$ for all $\omega\in\Omega$.
\end{lemma}

\begin{proof}
The equality in the first assertion follows from the fact that translations do not change the norm of $\Hs$, the fact that $\varphi_0$ is a product of functions of the individual coordinates, and the one-variable Gaussian integral $\int e^{-t^2/\sigma^2}dt=(\pi\sigma^2)^{1/2}$. Let $v:=c'-c$. Since $\frac{d}{dt}\varphi_0(x-c-tv)=-\langle v,\nabla\varphi_0(x-c-tv)\rangle$, the fundamental theorem of calculus gives
\[
\varphi_{c'}(x)-\varphi_c(x)=-\int_0^1\langle v,\nabla\varphi_0(x-c-tv)\rangle\,dt.
\]
The integrand on the right-hand side is the translate by $c+tv$ of the function $g_v(x):=\langle v,\nabla\varphi_0(x)\rangle=-\sigma^{-2}\langle v,x\rangle\varphi_0(x)$. Since $\varphi_0$ is invariant under rotations, the one-variable Gaussian integral $\int t^2e^{-t^2/\sigma^2}dt=\frac{\sigma^2}2(\pi\sigma^2)^{1/2}$ shows that the norm of $g_v$ in $\Hs$ satisfies $\|g_v\|^2=\sigma^{-4}|v|^2\int x_1^2\varphi_0(x)^2dx=\frac{|v|^2}{2\sigma^2}\|\varphi_0\|^2$. Since translations do not change the norm of $\Hs$, Minkowski's integral inequality gives $\|\varphi_{c'}-\varphi_c\|\le\int_0^1\|g_v\|\,dt=\|g_v\|$, and we obtain the inequality in the first assertion.

We prove the second assertion. Suppose that $f\in\Hs$ satisfies $\ip{\varphi_c}f=0$ for all $c\in\R^n$. Since $\varphi_0$ is real-valued and $\varphi_0(-x)=\varphi_0(x)$, we have $\ip{\varphi_c}f=\int\varphi_0(c-x)f(x)\,dx=(f*\varphi_0)(c)$, and the Fourier transform of $f*\varphi_0\in\Hs$ is $(2\pi)^{n/2}\hat f\hat\varphi_0$. Since $\hat\varphi_0$ is a Gaussian function that is positive everywhere, $\hat f=0$, that is, $f=0$. Hence the subspace spanned by $\{\varphi_c:c\in\R^n\}$ is dense in $\Hs$. By the inequality in the first assertion, $c\mapsto\varphi_c$ is a continuous map into $\Hs$, so the closure of the subspace spanned by $\{\varphi_c:c\in D\}$ contains $\varphi_c$ for all $c\in\R^n$. Therefore this closure equals $\Hs$.

We prove the third assertion. For $g\in\Hs$ put $T_cg:=g(\cdot-c)$. Then $T_c$ is a unitary operator on $\Hs$, and $c\mapsto T_cg$ is a continuous map into $\Hs$ (strong continuity of translations). Let $\chi\in\Sc$. We first show that $c\mapsto T_c\chi$ is a continuous map into $\Qs$. Since $x_jT_c\chi=T_c(x_j\chi)+c_jT_c\chi$ and $\partial_jT_c\chi=T_c\partial_j\chi$, we have
\[
x_j(T_{c'}\chi-T_c\chi)=(T_{c'}-T_c)(x_j\chi)+c'_jT_{c'}\chi-c_jT_c\chi,\qquad\partial_j(T_{c'}\chi-T_c\chi)=(T_{c'}-T_c)\partial_j\chi,
\]
and since $\chi$, $x_j\chi$, and $\partial_j\chi$ belong to $\Hs$, the strong continuity of translations gives $\|T_{c'}\chi-T_c\chi\|_{\Qs}\to0$ as $c'\to c$. Next, let $e_i$ be the $i$-th unit vector of $\R^n$. For each $x$ we have $\frac{d}{ds}\chi(x-c-se_i)=-(\partial_i\chi)(x-c-se_i)$, so by the fundamental theorem of calculus, for each $x$,
\[
(T_{c+te_i}\chi)(x)-(T_c\chi)(x)=-\int_0^t(T_{c+se_i}\partial_i\chi)(x)\,ds.
\]
Since $\partial_i\chi\in\Sc$, the map $s\mapsto T_{c+se_i}\partial_i\chi$ is a continuous map into $\Qs$, and since $\Qs$ is complete (Section~\ref{sec:quadslice}), its Bochner integral $\int_0^tT_{c+se_i}\partial_i\chi\,ds$ in $\Qs$ exists. Since bounded linear functionals commute with the Bochner integral, for all $g\in\Sc$ the inner product of $g$ with this Bochner integral is $\int_0^t\ip g{T_{c+se_i}\partial_i\chi}\,ds$. By Fubini's theorem, this value equals the inner product of $g$ with the function $x\mapsto\int_0^t(T_{c+se_i}\partial_i\chi)(x)\,ds$. Since $\Sc$ is dense in $\Hs$, the identity above, obtained from the fundamental theorem of calculus, holds for all $c\in\R^n$ and $t\in\R$ as an identity between elements of $\Qs$. For $d\in\R^n$ and $i=0,\dots,n$ put $d^{(i)}:=(d_1,\dots,d_i,0,\dots,0)$. Replacing $c$ by $c+d^{(i-1)}$ and $t$ by $d_i$ in this identity and summing over $i=1,\dots,n$, we obtain
\[
T_{c+d}\chi-T_c\chi+\sum_{i=1}^nd_i\,T_c\partial_i\chi=-\sum_{i=1}^n\int_0^{d_i}\big(T_{c+d^{(i-1)}+se_i}-T_c\big)\partial_i\chi\,ds.
\]
The norm $\|\cdot\|_{\Qs}$ of the right-hand side is at most $\sum_{i=1}^n|d_i|\,\sup_{|c'-c|\le|d|}\big\|(T_{c'}-T_c)\partial_i\chi\big\|_{\Qs}$. Since $c'\mapsto T_{c'}\partial_i\chi$ is a continuous map into $\Qs$, this supremum converges to $0$ as $d\to0$, and the norm $\|\cdot\|_{\Qs}$ of the right-hand side divided by $|d|$ also converges to $0$ as $d\to0$. Hence $c\mapsto T_c\chi$ is Fr\'echet differentiable at each $c$, and its Fr\'echet derivative is $d\mapsto-\sum_{i=1}^nd_i\,T_c\partial_i\chi$. Since $c\mapsto T_c\partial_i\chi$ is a continuous map into $\Qs$, this Fr\'echet derivative is continuous in $c$ in the operator norm, and $c\mapsto T_c\chi$ is a $C^1$ map into $\Qs$. Taking $\chi=\varphi_0$ gives the third assertion.

We prove the fourth assertion. Since $h_k$ and $\varphi_c$ are both products of functions of the individual coordinates, $\Mo_\beta(\varphi_c)$ is the product, over the coordinate factors of $\varphi_c$, of the values of $\Mo_\beta$ in the case $n=1$. Hence we may assume $n=1$. Put $b_k:=\ip{h_k}{\varphi_c}$. Then $\Mo_\beta(\varphi_c)=\sum_{k\ge0}e^{\beta k}|b_k|^2$. By the generating function of the Hermite polynomials, for $w\in\C$ we have $\sum_kh_k(x)\,w^k/\sqrt{k!}=\pi^{-1/4}\exp\left[-\frac{x^2}{2}+\sqrt2\,xw-\frac{w^2}{2}\right]$, and the left-hand side converges in $\Hs$. Multiplying both sides by $\varphi_c$, integrating, and evaluating the Gaussian integral, we obtain, for some $A>0$,
\[
F(w):=\sum_{k\ge0}\frac{b_k}{\sqrt{k!}}\,w^k=A\exp\left[\alpha_2 w^2+\alpha_1 w\right],\qquad\alpha_2:=\frac{\sigma^2-1}{2(\sigma^2+1)},\qquad\alpha_1:=\frac{\sqrt2\,c}{\sigma^2+1}.
\]
If $\alpha_2=0$, comparing the coefficients of the series defining $F$ with the Taylor coefficients of the right-hand side gives $b_k=A\alpha_1^k/\sqrt{k!}$, and $\Mo_\beta(\varphi_c)=A^2\sum_k(e^\beta\alpha_1^2)^k/k!=A^2\exp\left[e^\beta\alpha_1^2\right]$. Now let $\alpha_2\ne0$. The function $F$ is entire and $b_k/\sqrt{k!}$ is its $k$-th Taylor coefficient, so by Cauchy's integral formula, for all $t>0$,
\[
\frac{b_k}{\sqrt{k!}}=\frac1{2\pi i}\oint_{|w|=t}\frac{F(w)}{w^{k+1}}\,dw.
\]
The contour has length $2\pi t$, and on it $|w^{k+1}|=t^{k+1}$ and $|F(w)|\le A\exp\left[|\alpha_2|t^2+|\alpha_1|t\right]$, so
\[
\frac{|b_k|}{\sqrt{k!}}\ \le\ \frac{1}{t^k}\max_{|w|=t}|F(w)|\ \le\ \frac{A}{t^k}\exp\left[|\alpha_2|t^2+|\alpha_1|t\right].
\]
For $k\ge1$ take $t=\sqrt{\frac{k}{2|\alpha_2|}}$. Then $\frac{1}{t^{2k}}=\left(\frac{2|\alpha_2|}{k}\right)^k$ and $\exp\left[2|\alpha_2|t^2\right]=e^k$, so
\[
|b_k|^2\ \le\ A^2\,k!\,\left(\frac{2|\alpha_2|e}{k}\right)^k\exp\left[2|\alpha_1|\sqrt{\frac{k}{2|\alpha_2|}}\right],
\]
and $k!\le e\sqrt k\,\left(\frac ke\right)^k$ gives
\[
|b_k|^2\ \le\ A^2e\sqrt k\,(2|\alpha_2|)^k\exp\left[2|\alpha_1|\sqrt{\frac{k}{2|\alpha_2|}}\right].
\]
Since $2|\alpha_2|=\frac{|\sigma^2-1|}{\sigma^2+1}$, the hypothesis gives $\gamma_0:=e^\beta\cdot2|\alpha_2|<1$, and
\[
\Mo_\beta(\varphi_c)\ \le\ |b_0|^2+A^2e\sum_{k\ge1}\sqrt k\,\gamma_0^k\exp\left[2|\alpha_1|\sqrt{\frac{k}{2|\alpha_2|}}\right].
\]
The $k$-th root of the $k$-th term of the series on the right-hand side is $k^{1/(2k)}\,\gamma_0\exp\left[2|\alpha_1|\left(2|\alpha_2|k\right)^{-1/2}\right]$, which converges to $\gamma_0<1$ as $k\to\infty$, so the series converges by Cauchy's root test, and $\Mo_\beta(\varphi_c)<\infty$. Since $\Mo_\beta^{1/2}$ satisfies the triangle inequality and $\Psi_\omega$ is a finite linear combination of the $\varphi_{c_j}$, we have $\Mo_\beta(\psi_\omega)=\Mo_\beta(\Psi_\omega)/\|\Psi_\omega\|^2<\infty$.
\end{proof}

For a sequence $c_1,c_2,\dots\in\R^n$ we put $\mathcal V(c_1,c_2,\dots):=\spn_{\C}\{\varphi_{c_j}:j\ge1\}$. This is equal to the union $\bigcup_{m\ge1}\mathcal V(c_1,\dots,c_m)$ of the subspaces of \eqref{eq:Vc}.

\begin{lemma}[Approximation by the span of the units]\label{lem:approx}
First, let $c_1,c_2,\dots\in\R^n$ be a sequence such that $\{c_j:j\ge1\}$ is dense in $\R^n$. Then, for all $f\in\Hs$, $\dist(f,\mathcal V(c_1,\dots,c_m))$ is nonincreasing in $m$ and converges to $0$ as $m\to\infty$. Moreover, let $R\ge0$, and suppose that $m_k\to\infty$ and $c^{(k)}\in(\R^n)^{m_k}$ satisfy $\|c^{(k)}-(c_1,\dots,c_{m_k})\|\le R$. Then $\dist(f,\mathcal V(c^{(k)}))\to0$ for all $f\in\Hs$. Furthermore, a sequence of independent random variables distributed according to a Borel probability measure whose support is all of $\R^n$ satisfies, with probability $1$, the above condition on the sequence. Second, let $\nu$ be a Borel probability measure on $\R^n$, and let $c_1,c_2,\dots$ be independent random variables distributed according to $\nu$. Assume that the target $\phi$ can be written as in \eqref{eq:superpos} with some $\varrho\in L^2(\nu)$ (the integral is a Bochner integral in $\Hs$), and for $m\ge1$ put $c:=(c_1,\dots,c_m)\in(\R^n)^m$. Then, for all $\eta\in(0,1]$, there is an event of probability at least $1-\eta$ on which, for all $c'\in(\R^n)^m$, we have
\[
\dist\big(\phi,\mathcal V(c')\big)\ \le\ \sqrt{\frac{2}{m\eta}}\ \|\varrho\|_{L^2(\nu)}\,\|\varphi_0\|\,\Big(1+\frac{\|c'-c\|}{\sqrt2\,\sigma}\Big).
\]
\end{lemma}

\begin{proof}
We prove the first assertion. By hypothesis, the set $\{c_j:j\ge1\}$ is dense in $\R^n$, so by the second assertion of Lemma~\ref{lem:gaussunit}, $\mathcal V(c_1,c_2,\dots)=\bigcup_m\mathcal V(c_1,\dots,c_m)$ is dense in $\Hs$. Since the sequence of subspaces $\mathcal V(c_1,\dots,c_m)$ ($m\ge1$) is increasing, for all $f\in\Hs$, $\dist(f,\mathcal V(c_1,\dots,c_m))$ is nonincreasing in $m$ and converges to $0$ as $m\to\infty$.

We prove the assertion for the case where the centers are moved. We first show that, for all $j\ge1$, $\dist(\varphi_{c_j},\mathcal V(c^{(k)}))\to0$ as $k\to\infty$. Fix $\epsilon>0$. The open ball of radius $\epsilon$ centered at $c_j$ contains infinitely many pairwise disjoint nonempty open sets, and by the hypothesis on the sequence each of them contains a term of the sequence. Hence $I:=\{i\ge1:|c_i-c_j|<\epsilon\}$ is an infinite set, and $I\cap\{1,\dots,m_k\}$ contains more than $R^2/\epsilon^2$ elements if $k$ is sufficiently large. Since $\sum_i|c^{(k)}_i-c_i|^2\le R^2$, we have $|c^{(k)}_i-c_i|\le\epsilon$ for at least one element $i$ of $I\cap\{1,\dots,m_k\}$, and therefore $|c^{(k)}_i-c_j|\le2\epsilon$. From $\varphi_{c^{(k)}_i}\in \mathcal V(c^{(k)})$ and the first assertion of Lemma~\ref{lem:gaussunit}, we get $\dist(\varphi_{c_j},\mathcal V(c^{(k)}))\le2\epsilon\|\varphi_0\|/(\sqrt2\,\sigma)$ for sufficiently large $k$. Since $\epsilon$ is arbitrary, $\dist(\varphi_{c_j},\mathcal V(c^{(k)}))\to0$. Next, since $\mathcal V(c^{(k)})$ is a subspace, for $g=\sum_{j\le l}b_j\varphi_{c_j}\in\mathcal V(c_1,c_2,\dots)$ we have $\dist(g,\mathcal V(c^{(k)}))\le\sum_{j\le l}|b_j|\dist(\varphi_{c_j},\mathcal V(c^{(k)}))\to0$. For $f\in\Hs$ and $g\in\mathcal V(c_1,c_2,\dots)$ we have $\dist(f,\mathcal V(c^{(k)}))\le\|f-g\|+\dist(g,\mathcal V(c^{(k)}))$, and $\mathcal V(c_1,c_2,\dots)$ is dense in $\Hs$, so $\dist(f,\mathcal V(c^{(k)}))\to0$.

We prove the assertion that holds with probability $1$. We write $\Pr$ for the probability and $\mathbb E$ for the expectation, and let $c_1,c_2,\dots$ be a sequence of independent random variables distributed according to a Borel probability measure $\nu$ whose support is all of $\R^n$. Let $\mathcal B$ be the set of all open balls of $\R^n$ whose centers have rational coordinates and whose radii are rational. Then $\mathcal B$ is a countable set, and every nonempty open subset of $\R^n$ contains an element of $\mathcal B$. Hence it suffices to show that, with probability $1$, every $B\in\mathcal B$ contains a term of the sequence. For $B\in\mathcal B$ put
\[
\mathcal F_B:=\{\text{for all }j\ge1\text{, }c_j\notin B\}.
\]
Since the support of $\nu$ is all of $\R^n$, we have $\nu(B)>0$, and since the $c_j$ are independent, for all $j_1\ge1$,
\[
\Pr[\mathcal F_B]\ \le\ \Pr\big[c_1\notin B,\dots,c_{j_1}\notin B\big]=\big(1-\nu(B)\big)^{j_1}.
\]
Letting $j_1\to\infty$, we obtain $\Pr[\mathcal F_B]=0$. Since $\mathcal B$ is countable, $\Pr[\bigcup_{B\in\mathcal B}\mathcal F_B]=0$.

We prove the second assertion. Take a Borel measurable representative of $\varrho$. The map $y\mapsto\varrho(y)\varphi_y$ is strongly measurable and $\int_{\R^n}|\varrho(y)|\,\|\varphi_y\|\,\nu(dy)\le\|\varphi_0\|\|\varrho\|_{L^2(\nu)}$, so it is Bochner integrable. Put $Y_j:=\varrho(c_j)\varphi_{c_j}$. Then the $Y_j$ are independent and identically distributed. The mean of $Y_j$ is $\phi$ by \eqref{eq:superpos}, and $\mathbb E\|Y_j\|^2=\|\varphi_0\|^2\|\varrho\|^2_{L^2(\nu)}$. By independence, the expectations of the cross terms are $0$, so
\[
\mathbb E\Big\|\phi-\frac1m\sum_jY_j\Big\|^2=\frac1m\big(\mathbb E\|Y_1\|^2-\|\phi\|^2\big)\le\frac{\|\varphi_0\|^2\|\varrho\|^2_{L^2(\nu)}}{m}.
\]
Moreover, $A_m:=\frac1m\sum_j|\varrho(c_j)|^2$ satisfies $\mathbb EA_m=\|\varrho\|^2_{L^2(\nu)}$. Applying Markov's inequality to the nonnegative random variables $\|\phi-\frac1m\sum_jY_j\|^2$ and $A_m$, we see that the complements of the following two events both have probability at most $\eta/2$:
\[
\mathcal E_1:=\Big\{\Big\|\phi-\frac1m\sum_jY_j\Big\|^2\le\frac{2}{m\eta}\|\varphi_0\|^2\|\varrho\|^2_{L^2(\nu)}\Big\},\qquad\mathcal E_2:=\Big\{A_m\le\frac2\eta\|\varrho\|^2_{L^2(\nu)}\Big\}.
\]
Hence $\Pr[\mathcal E_1\cap\mathcal E_2]\ge1-\eta$. On $\mathcal E_1\cap\mathcal E_2$, take an arbitrary $c'\in(\R^n)^m$ and put $f:=\frac1m\sum_j\varrho(c_j)\varphi_{c'_j}\in \mathcal V(c')$. Then
\[
\phi-f=\Big(\phi-\frac1m\sum_jY_j\Big)+\frac1m\sum_j\varrho(c_j)\big(\varphi_{c_j}-\varphi_{c'_j}\big).
\]
By the first assertion of Lemma~\ref{lem:gaussunit} and the Cauchy--Schwarz inequality, the norm of the second term is at most $\frac1m\sum_j|\varrho(c_j)|\,|c_j-c'_j|\,\frac{\|\varphi_0\|}{\sqrt2\,\sigma}\le A_m^{1/2}\frac{\|c'-c\|}{\sqrt m}\frac{\|\varphi_0\|}{\sqrt2\,\sigma}$. Combining this bound with the bound on the norm of the first term on $\mathcal E_1$ and the bound on $A_m$ on $\mathcal E_2$, we see that, on the event $\mathcal E_1\cap\mathcal E_2$ of probability at least $1-\eta$, the inequality of the assertion holds for all $c'\in(\R^n)^m$.
\end{proof}

\subsection{The active set and generalized gradients}\label{sec:proofquad}\label{sec:proofstat}

In this subsection we compute the generalized directional derivatives of $\Lc_q$ (Lemma~\ref{lem:stat}) and state a necessary condition satisfied by the stationary points of the network in terms of the generalized gradient with respect to the state (Lemma~\ref{lem:netstat}). The facts about generalized gradients that we use are the four lemmas of Appendix~\ref{sec:clarke}. We use Lemma~\ref{lem:clarkebasic} for the basic properties of generalized gradients, Lemma~\ref{lem:clarkemax} on maxima for $\mathcal W_\theta$ in each direction and for $\Lc_\infty$, Lemma~\ref{lem:clarkeint} on aggregation by the $L^q$ norm for $\Lc_q$ with $q<\infty$, and Lemma~\ref{lem:clarkechain} on composition with a $C^1$ map for the pullback to the parameters of the network. Among the hypotheses of these lemmas, we first verify continuity and compactness, in Lemma~\ref{lem:loss}.

\begin{lemma}[Topology of the active set]\label{lem:loss}
If $\Qs$ is given the topology defined by $\|\cdot\|_{\Qs}$ and $\Hs$ the weak topology, then $(f,\theta,\mu)\mapsto M_{\theta,\mu}f$ is a continuous map from $\Qs\times\Ac$ into $\Hs$, and $\Ac$ is a compact metrizable space. Consequently, for unit vectors $\psi$ in $\Qs$ and $(\theta,\mu)\in\Ac$, the value $\ell_{\theta,\mu}(\psi)$ is continuous in $(\psi,\theta,\mu)$, the vector $v_{\theta,\mu}(\psi)$ is weakly continuous and locally bounded in $(\psi,\theta,\mu)$, and $\Ac(\psi)$ is a nonempty compact set.
\end{lemma}

\begin{proof}
By \eqref{eq:Xxi}, for $f\in\Qs$ the map $(\xi,\varpi)\mapsto X_{\xi,\varpi}f$ is linear, and $\|X_\theta f-X_{\theta_0}f\|\le|\theta-\theta_0|\,\En(f)^{1/2}$ for $\theta,\theta_0\in\Sph$.

We first show that, for every bounded continuous function $b:\R\to\C$ and every $g\in\Hs$, we have $b(X_\theta)g\to b(X_{\theta_0})g$ as $\theta\to\theta_0$. Since $\Sph$ is a metric space, it suffices to prove this for sequences converging to $\theta_0$. By the third assertion of Lemma~\ref{lem:slice}, the Weyl operator $V(\xi,\varpi)$ maps $\Qs$ into $\Qs$. Since $\Qs$ is dense in $\Hs$, and since $\Qs\subset D(X_\theta)$ and $e^{itX_\theta}=V(t\theta)$, the space $\Qs$ is a core for every $X_\theta$ by \citet[Theorem~VIII.11]{reedsimon}. For every element $f$ of $\Qs$ we have $X_\theta f\to X_{\theta_0}f$ as $\theta\to\theta_0$, so $X_\theta$ converges to $X_{\theta_0}$ in the strong resolvent sense by \citet[Theorem~VIII.25(a)]{reedsimon}, and we obtain the assertion by \citet[Theorem~VIII.20(b)]{reedsimon}.

Next, we fix $f\in\Qs$ and prove the weak continuity of $(\theta,\mu)\mapsto M_{\theta,\mu}f$. Since $\|M_{\theta,\mu}f\|\le\En(f)^{1/2}$ and $\Qs$ is dense in $\Hs$, it suffices to show that $(\theta,\mu)\mapsto\ip g{M_{\theta,\mu}f}$ is continuous for every $g\in\Qs$. Let $A\ge1$. If a Borel function $w$ vanishes for $|s|\le A$ and satisfies $|w(s)|\le2|s|$ for all $s$, then Chebyshev's inequality $\|E_\theta(\{s:|s|>A\})g\|^2=\ip g{E_\theta(\{s:|s|>A\})g}\le A^{-2}\|X_\theta g\|^2\le A^{-2}\En(g)$ and $\|w(X_\theta)f\|\le2\|X_\theta f\|\le2\En(f)^{1/2}$ give, for all $\theta\in\Sph$,
\begin{equation}\label{eq:tailpair}
|\ip g{w(X_\theta)f}|=|\ip{E_\theta(\{s:|s|>A\})g}{w(X_\theta)f}|\le\frac{2\En(g)^{1/2}\En(f)^{1/2}}{A}.
\end{equation}
Fix $(\theta_0,\mu_0)\in\Ac$, and let $b$ be the continuous function that equals $\mu_0$ on $[-A,A]$, is affine on $[A,2A]$ and on $[-2A,-A]$, and vanishes for $|s|\ge2A$. The function $\mu_0-b$ vanishes for $|s|\le A$, and $|\mu_0(s)-b(s)|\le2|s|$. For $(\theta,\mu)\in\Ac$,
\[
M_{\theta,\mu}f-M_{\theta_0,\mu_0}f=(\mu-\mu_0)(X_\theta)f+(\mu_0-b)(X_\theta)f+\big(b(X_\theta)f-b(X_{\theta_0})f\big)+(b-\mu_0)(X_{\theta_0})f.
\]
We estimate the inner product of $g$ with each of the four terms on the right-hand side. For the first term, we split $\mu-\mu_0$ into the part on $|s|\le A$ and the part on $|s|>A$. The absolute value of the inner product with $g$ is then at most $\sup_{|s|\le A}|\mu-\mu_0|\,\|g\|\,\|f\|$ for the former part, and at most $2\En(g)^{1/2}\En(f)^{1/2}/A$ for the latter part by $|\mu-\mu_0|\le2|s|$ and \eqref{eq:tailpair}. The absolute values of the inner products of $g$ with the second term and with the fourth term are also each at most $2\En(g)^{1/2}\En(f)^{1/2}/A$ by \eqref{eq:tailpair}. The inner product of $g$ with the third term converges to $0$ as $\theta\to\theta_0$, by the convergence shown in the previous paragraph. If $\mu\to\mu_0$ in $\Kc$, then $\mu$ converges to $\mu_0$ uniformly on $[-A,A]$, so
\[
\limsup_{(\theta,\mu)\to(\theta_0,\mu_0)}\big|\ip g{M_{\theta,\mu}f-M_{\theta_0,\mu_0}f}\big|\le\frac{6\En(g)^{1/2}\En(f)^{1/2}}{A}.
\]
Since $A$ is arbitrary, $(\theta,\mu)\mapsto\ip g{M_{\theta,\mu}f}$ is continuous at $(\theta_0,\mu_0)$. The weak continuity jointly in $f$ and $(\theta,\mu)$ follows from this result and $\|M_{\theta,\mu}(\tilde f-f)\|\le\En(\tilde f-f)^{1/2}\le\|\tilde f-f\|_{\Qs}$.

By the Arzel\`a--Ascoli theorem, $\Kc$ is compact with respect to the metric $\sum_k2^{-k}\min\{1,\sup_{|s|\le k}|\mu-\tilde\mu|\}$ of locally uniform convergence, so $\Ac$ is a compact metrizable space. If $\psi_k\to\psi$ with respect to $\|\cdot\|_{\Qs}$ and $(\theta_k,\mu_k)\to(\theta,\mu)$, then $M_{\theta_k,\mu_k}\psi_k$ converges weakly to $M_{\theta,\mu}\psi$ and $\psi_k$ converges to $\psi$ in the norm of $\Hs$, so $\ip{\psi_k}{M_{\theta_k,\mu_k}\psi_k}\to\ip\psi{M_{\theta,\mu}\psi}$. The weak convergence of $M_{\theta_k,\mu_k}\psi_k$ and this convergence give the continuity of $\ell_{\theta,\mu}(\psi)$ and the weak continuity of $v_{\theta,\mu}(\psi)$, and the local boundedness follows from $\|v_{\theta,\mu}(\psi)\|^2\le\En(\psi)$. The set $\Ac(\psi)$ is the set of maximizers of the continuous function $(\theta,\mu)\mapsto\ell_{\theta,\mu}(\psi)$ on the compact set $\Ac$, so it is a nonempty compact set.
\end{proof}

\begin{lemma}[Generalized directional derivatives on $\Qs$]\label{lem:stat}
Let $\psi$ be a unit vector in $\Qs$ and let $h\in\Qs$. First, for every $\theta\in\Sph$,
\begin{equation}\label{eq:clarkeW}
D^\circ_h\mathcal W_\theta(\psi)=\max_{\mu\in\Kc_\theta(\psi)}2\Re\ip{v_{\theta,\mu}(\psi)}h.
\end{equation}
Second,
\begin{equation}\label{eq:clarkeinf}
D^\circ_h\Lc_\infty(\psi)=\max_{(\theta,\mu)\in\Ac(\psi)}2\Re\ip{v_{\theta,\mu}(\psi)}h,
\end{equation}
and in particular $\Lc_\infty$ is differentiable at $\psi$ if and only if the set of active gradients $\mathcal G(\psi)$ in \eqref{eq:vdef} consists of a single point. Third, let $1\le q<\infty$ and $\Lc_q(\psi)>0$, and for $\theta\in\Sph$ put
\begin{equation}\label{eq:rho}
\rho_\theta:=\Big(\frac{\mathcal W_\theta(\psi)}{\Lc_q(\psi)}\Big)^{q-1}\quad(q>1),\qquad\rho_\theta:=1\quad(q=1).
\end{equation}
Then $\rho_\theta\ge0$, $\int\rho_\theta\,d\varsigma\le1$, $\int\rho_\theta\mathcal W_\theta(\psi)\,d\varsigma=\Lc_q(\psi)$, and
\begin{equation}\label{eq:ddq}
D^\circ_h\Lc_q(\psi)=\int_{\Sph}\rho_\theta\,\max_{\mu\in\Kc_\theta(\psi)}2\Re\ip{v_{\theta,\mu}(\psi)}h\,d\varsigma(\theta).
\end{equation}
Fourth, in the three formulas above, the generalized directional derivative on the left-hand side equals the one-sided directional derivative. That is, $\mathcal W_\theta$ and $\Lc_\infty$ are regular in the sense of Clarke at $\psi$, and for $1\le q<\infty$ the function $\Lc_q$ is regular at $\psi$ if $\Lc_q(\psi)>0$.
\end{lemma}

\begin{proof}
Let $(\theta,\mu)\in\Ac$. By Lemma~\ref{lem:kant}, $|\ip{g_1}{M_{\theta,\mu}g_2}|\le\|g_1\|_{\Qs}\|g_2\|_{\Qs}$ for $g_1,g_2\in\Qs$, and $M_{\theta,\mu}$ is symmetric on $\Qs$, so $\ell_{\theta,\mu}(f)=\ip f{M_{\theta,\mu}f}/\|f\|^2-\ip\phi{M_{\theta,\mu}\phi}$ is $C^1$ on $\Qs\setminus\{0\}$, and its G\^ateaux derivative $D\ell_{\theta,\mu}(f)\in\Qs^*$ at $f$ is
\[
D\ell_{\theta,\mu}(f)[h]:=\lim_{t\to0}\frac{\ell_{\theta,\mu}(f+th)-\ell_{\theta,\mu}(f)}{t}=\frac{2}{\|f\|}\Re\ip{v_{\theta,\mu}\Big(\frac{f}{\|f\|}\Big)}{h}\qquad(h\in\Qs).
\]
At a unit vector $\psi$ we have $D\ell_{\theta,\mu}(\psi)[h]=2\Re\ip{v_{\theta,\mu}(\psi)}h$.

To prove the first and second assertions, we apply Lemma~\ref{lem:clarkemax} in the appendix with $X=\Qs$, $U=\Qs\setminus\{0\}$, and the family of functions $\ell_{\theta,\mu}$. The index space is $T=\{\theta\}\times\Kc$ for the first assertion and $T=\Ac$ for the second assertion. The space $\Ac$ is a compact metrizable space (Lemma~\ref{lem:loss}), and $\{\theta\}\times\Kc$ is a closed subset of it, so in both cases $T$ is a compact metrizable space. The normalization $f\mapsto f/\|f\|$ is a continuous map from $\Qs\setminus\{0\}$ into $\Qs$, so by Lemma~\ref{lem:loss} the map $(\theta,\mu,f)\mapsto\ell_{\theta,\mu}(f)$ is continuous, and by the weak continuity of $v_{\theta,\mu}$ the map $(\theta,\mu,f)\mapsto D\ell_{\theta,\mu}(f)[h]$ is also continuous for each $h\in\Qs$. Every point of $\Qs\setminus\{0\}$ has a neighborhood on which the functions $\ell_{\theta,\mu}$ have a Lipschitz constant that does not depend on $(\theta,\mu)$ (Proposition~\ref{prop:lipschitz}). The maximum over $T$ is $\mathcal W_\theta$ by \eqref{eq:Lpot} when $T=\{\theta\}\times\Kc$, and it is $\Lc_\infty$ by \eqref{eq:Linf} when $T=\Ac$. The sets of maximizers at $\psi$ are $\{\theta\}\times\Kc_\theta(\psi)$ and $\Ac(\psi)$, respectively, so we obtain \eqref{eq:clarkeW} and \eqref{eq:clarkeinf}. Moreover, by Lemma~\ref{lem:clarkemax} in the case $T=\{\theta\}\times\Kc$, the one-sided directional derivative $D^+_h\mathcal W_\theta(\psi)$ exists and equals $D^\circ_h\mathcal W_\theta(\psi)$.

We prove the assertion on differentiability. Suppose that $\mathcal G(\psi)=\{v\}$. By \eqref{eq:clarkeinf}, every $x^*\in\partial^\circ\Lc_\infty(\psi)$ satisfies $x^*(h)\le2\Re\ip vh$ for all $h\in\Qs$. Replacing $h$ by $-h$ in this inequality gives the reverse inequality, so $x^*(h)=2\Re\ip vh$. Since $\partial^\circ\Lc_\infty(\psi)$ is nonempty (Lemma~\ref{lem:clarkebasic}), $\partial^\circ\Lc_\infty(\psi)$ consists of a single point. Conversely, suppose that $\partial^\circ\Lc_\infty(\psi)=\{x^*\}$. By the third assertion of Lemma~\ref{lem:clarkebasic}, $D^\circ_h\Lc_\infty(\psi)=x^*(h)$ for all $h\in\Qs$. Hence, by \eqref{eq:clarkeinf}, every $v\in\mathcal G(\psi)$ satisfies $2\Re\ip vh\le x^*(h)$ for all $h\in\Qs$. Replacing $h$ by $-h$ in this inequality gives the reverse inequality, so equality holds. Therefore $\Re\ip{v-v'}h=0$ for $v,v'\in\mathcal G(\psi)$ and all $h\in\Qs$. Combining this identity with the one obtained from it by replacing $h$ by $ih$, and using that $\Qs$ is dense in $\Hs$, we get $v=v'$.

We prove the third assertion. The inequality $\int\rho_\theta\,d\varsigma\le1$ follows from H\"older's inequality, and the identity $\int\rho_\theta\mathcal W_\theta(\psi)\,d\varsigma=\Lc_q(\psi)$ follows from \eqref{eq:rho} and \eqref{eq:Lq}. We apply Lemma~\ref{lem:clarkeint} in the appendix with $S=\Sph$, $w_\theta=\mathcal W_\theta$, and $x=\psi$. The first hypothesis of Lemma~\ref{lem:clarkeint} follows from the continuity of $\theta\mapsto\mathcal W_\theta(f)$ for every $f\in\Qs\setminus\{0\}$ (Lemma~\ref{lem:kant}). We verify the second hypothesis. On some neighborhood of $\psi$, the functions $\mathcal W_\theta$ have a Lipschitz constant that does not depend on $\theta$ (Proposition~\ref{prop:lipschitz}). This neighborhood can be taken to be bounded. Moreover, $0\le\mathcal W_\theta(\psi)\le\En(\psi)^{1/2}+\En(\phi)^{1/2}$, so, since the Lipschitz constant does not depend on $\theta$, the functions $\mathcal W_\theta$ have an upper bound on that neighborhood that does not depend on $\theta$. The third hypothesis was shown at the end of the paragraph in which the first and second assertions were proved. Since $G=\Lc_q$, Lemma~\ref{lem:clarkeint} and \eqref{eq:clarkeW} give \eqref{eq:ddq}.

The fourth assertion is contained in the conclusions of Lemma~\ref{lem:clarkemax} and Lemma~\ref{lem:clarkeint}.
\end{proof}

\begin{lemma}[Stationary points of the network]\label{lem:netstat}
Let $1\le q\le\infty$. If a point $\omega$ of the network is a stationary point for $\Lc_q$, then there exists $x^*\in\partial^\circ\Lc_q(\psi_\omega)$ such that $x^*(h)=0$ for all $h\in\ran\mathscr J(\omega)$. Moreover, $\Ls_\infty$ is regular in the sense of Clarke at every point of the network, and for $1\le q<\infty$ the function $\Ls_q$ is regular at every point $\omega$ of the network with $\Ls_q(\omega)>0$.
\end{lemma}

\begin{proof}
The map $\omega\mapsto\Psi_\omega$ is a $C^1$ map from $\Omega$ into $\Qs\setminus\{0\}$, and its Fr\'echet derivative at $\omega$ is $\mathscr J(\omega)$ (Lemma~\ref{lem:tangent}). The function $\Lc_q$ is locally Lipschitz on $\Qs\setminus\{0\}$ (Proposition~\ref{prop:lipschitz}), and $\Ls_q(\omega)=\Lc_q(\Psi_\omega)$. We apply Lemma~\ref{lem:clarkechain} in the appendix with $Z=\mathcal Z$, $\Omega_0=\Omega$, $X=\Qs$, $U=\Qs\setminus\{0\}$, $F(\omega)=\Psi_\omega$, and $g=\Lc_q$. Since $DF(\omega)=\mathscr J(\omega)$, if $\omega$ is a stationary point, then there exists $y^*\in\partial^\circ\Lc_q(\Psi_\omega)$ that vanishes on $\ran\mathscr J(\omega)$.

Next, we express the generalized gradient of $\Lc_q$ at $\Psi_\omega$ in terms of the generalized gradient at $\psi_\omega$. Put $\lambda:=\|\Psi_\omega\|$. Since $\Lc_q(\lambda f)=\Lc_q(f)$ for all $f\in\Qs\setminus\{0\}$, the change of variable $f'=\lambda f$ gives, for $h\in\Qs$,
\[
\limsup_{f'\to\Psi_\omega,\ t\downarrow0}\frac{\Lc_q(f'+th)-\Lc_q(f')}{t}
=\limsup_{f\to\psi_\omega,\ t\downarrow0}\frac{\Lc_q\big(f+\frac t\lambda h\big)-\Lc_q(f)}{t}
=\frac1\lambda\limsup_{f\to\psi_\omega,\ s\downarrow0}\frac{\Lc_q(f+sh)-\Lc_q(f)}{s}.
\]
The generalized gradient of $\Lc_q$ at $\Psi_\omega$ is the set of all elements of $\Qs^*$ whose value at every $h\in\Qs$ is at most the limit superior on the left-hand side, so $\partial^\circ\Lc_q(\Psi_\omega)=\frac1\lambda\,\partial^\circ\Lc_q(\psi_\omega)$. Hence $x^*:=\lambda\,y^*$ belongs to $\partial^\circ\Lc_q(\psi_\omega)$ and vanishes on $\ran\mathscr J(\omega)$.

We prove the assertion on regularity. Since $\Lc_q$ is invariant under multiplication by positive constants, the fourth assertion of Lemma~\ref{lem:stat} shows that $\Lc_q$ is also regular at $\Psi_\omega$ (for $q<\infty$, when $\Ls_q(\omega)>0$). Since $\Ls_q(\omega)=\Lc_q(\Psi_\omega)$, the assertion on equality in Lemma~\ref{lem:clarkechain} shows that $\Ls_q$ is regular at $\omega$.
\end{proof}

\subsection{Separation of slices and sharpness}\label{sec:proofsharp}

\begin{lemma}[Separation of the slices of two states]\label{lem:separate}
Let $1\le q\le\infty$, and let $r$, $\gamma$, and $\kappa$ be as in \eqref{eq:exponents}. For all unit vectors $a,b$ in $\Qs$ and for the infidelity $u(a,b)$ between $a$ and $b$ defined in \eqref{eq:Wtheta}, we have
\[
\Lc_q(a,b)\ \ge\ \kappa\,u(a,b)^{1/r+\gamma}\big(4(\En(a)+\En(b))\big)^{-\gamma}.
\]
\end{lemma}

\begin{proof}
If $u(a,b)=0$, then the right-hand side of the assertion is $0$ and the assertion holds, so we assume $u(a,b)>0$. The right-hand side of the assertion depends on $q$ only through $r$, and $r\le q$. Since $\varsigma$ is a probability measure and the $L^q(\varsigma)$ norm is nondecreasing in $q$, we have $\Lc_r(a,b)\le\Lc_q(a,b)$, and we may assume $q=r\in[1,2]$. For $(\xi,\varpi)\in\R^n\times\R^n$ put $Z(\xi,\varpi):=\ip a{V(\xi,\varpi)a}-\ip b{V(\xi,\varpi)b}$, and for $\theta\in\Sph$ let $Z^\sharp(\theta):=\sup_{k>0}|Z(k\theta)|/k\in[0,\infty]$. As a supremum of continuous functions, $Z^\sharp$ is lower semicontinuous, and in particular measurable. The proof consists of three steps.

In the first step, we show that $\mathcal W_\theta(a,b)\ge Z^\sharp(\theta)$ for all $\theta$. Take $\theta\in\Sph$ and $k>0$. Since $V(k\theta)=e^{ikX_\theta}$, the functional calculus gives $\ip f{V(k\theta)f}=\int e^{iks}\,d\Pm^\theta_f(s)$ for every unit vector $f$, and
\[
Z(k\theta)=\int_{\R}e^{iks}\,d\big(\Pm^\theta_a-\Pm^\theta_b\big)(s).
\]
Choose $\alpha\in\R$ such that $e^{-i\alpha}Z(k\theta)=|Z(k\theta)|$, and put $\mu(s):=\cos(ks-\alpha)/k=\Re(e^{-i\alpha}e^{iks})/k$. Since $|\mu'|\le1$, we have $\mu-\mu(0)\in\Kc$, and the integral with respect to $\Pm^\theta_a-\Pm^\theta_b$ does not change when a constant is subtracted from $\mu$, so \eqref{eq:KR} gives
\[
\mathcal W_\theta(a,b)\ \ge\ \int\mu\,d\big(\Pm^\theta_a-\Pm^\theta_b\big)=\frac1k\Re\big(e^{-i\alpha}Z(k\theta)\big)=\frac{|Z(k\theta)|}{k}.
\]
Since $k>0$ is arbitrary, $\mathcal W_\theta(a,b)\ge Z^\sharp(\theta)$.

In the second step, we estimate the integral of $|Z|^2$ and its second moment. For $f,g\in\Hs$ and $(\xi,\varpi)\in\R^n\times\R^n$ put
\begin{equation}\label{eq:Ydef}
Y_{g,f}(\xi,\varpi):=\ip g{V(\xi,\varpi)f}.
\end{equation}
Then $Z=Y_{a,a}-Y_{b,b}$. Applying the Moyal identity \eqref{eq:moyal} (Lemma~\ref{lem:moyal} in the appendix) to each term of $|Z|^2=|Y_{a,a}|^2+|Y_{b,b}|^2-2\Re\big(Y_{a,a}\overline{Y_{b,b}}\big)$ gives
\begin{equation}\label{eq:Zl2}
\int_{\R^n\times\R^n}|Z|^2\,d\xi\,d\varpi=(2\pi)^n\big(\|a\|^4+\|b\|^4-2|\ip ab|^2\big)=2(2\pi)^n\,u(a,b).
\end{equation}
Next, let $f\in\Qs$ be a unit vector. The commutation relations \eqref{eq:weylcomm} of the Weyl operators and the symmetry of $x_j$ and $p_j$ on $\Qs$ give
\[
\varpi_jY_{f,f}=Y_{f,x_jf}-Y_{x_jf,f},\qquad \xi_jY_{f,f}=Y_{p_jf,f}-Y_{f,p_jf}.
\]
Since $x_j$ is symmetric, $\ip f{x_jf}$ is real, and the Moyal identity gives
\begin{align*}
\int_{\R^n\times\R^n}\varpi_j^2\,|Y_{f,f}|^2\,d\xi\,d\varpi&=\int_{\R^n\times\R^n}\big|Y_{f,x_jf}-Y_{x_jf,f}\big|^2\,d\xi\,d\varpi\\
&=2(2\pi)^n\big(\|x_jf\|^2-\ip f{x_jf}^2\big)\le2(2\pi)^n\|x_jf\|^2.
\end{align*}
For $\xi_j$ and $p_j$ as well, an inequality of the same form follows from the formula for $\xi_jY_{f,f}$ and the Moyal identity, and summing over $j$ gives $\int(|\xi|^2+|\varpi|^2)|Y_{f,f}|^2\,d\xi\,d\varpi\le2(2\pi)^n\En(f)$. Since $|Z|^2\le2|Y_{a,a}|^2+2|Y_{b,b}|^2$,
\begin{equation}\label{eq:Zmoment}
\int_{\R^n\times\R^n}(|\xi|^2+|\varpi|^2)|Z|^2\,d\xi\,d\varpi\le4(2\pi)^n\big(\En(a)+\En(b)\big).
\end{equation}

In the third step, we estimate the $L^r(\varsigma)$ norm of $Z^\sharp$ from below. Put $r_0^2:=4(\En(a)+\En(b))/u(a,b)$ (we have $\En(a)>0$ by Lemma~\ref{lem:fock}). By \eqref{eq:Zmoment},
\[
\int_{|(\xi,\varpi)|>r_0}|Z|^2\,d\xi\,d\varpi\le\frac1{r_0^2}\int(|\xi|^2+|\varpi|^2)|Z|^2\,d\xi\,d\varpi\le\frac{4(2\pi)^n(\En(a)+\En(b))}{r_0^2}=(2\pi)^n\,u(a,b),
\]
so together with \eqref{eq:Zl2} we get $\int_{|(\xi,\varpi)|\le r_0}|Z|^2\,d\xi\,d\varpi\ge(2\pi)^n\,u(a,b)$. On the other hand, $|Z|\le2$ and the definition of $Z^\sharp$ give $|Z(k\theta)|^2\le2^{2-r}|Z(k\theta)|^r\le2^{2-r}(k\,Z^\sharp(\theta))^r$, so in polar coordinates
\begin{align*}
\int_{|(\xi,\varpi)|\le r_0}|Z|^2\,d\xi\,d\varpi&\le\frac{2\pi^n}{(n-1)!}\,2^{2-r}\int_{\Sph}Z^\sharp(\theta)^r\,d\varsigma(\theta)\int_0^{r_0}k^{2n-1+r}\,dk\\
&=\frac{2^{3-r}\pi^nr_0^{2n+r}}{(2n+r)\,(n-1)!}\,\|Z^\sharp\|_{L^r(\varsigma)}^r,
\end{align*}
where the area of $\Sph$ is $2\pi^n/(n-1)!$. Combining this upper bound on $\int_{|(\xi,\varpi)|\le r_0}|Z|^2\,d\xi\,d\varpi$ with the lower bound above gives
\[
\|Z^\sharp\|_{L^r(\varsigma)}^r\ \ge\ (2n+r)\,2^{n+r-3}(n-1)!\,u(a,b)\,r_0^{-(2n+r)}=\kappa^r\,u(a,b)\,r_0^{-(2n+r)},
\]
and substituting the definition of $r_0$ gives $\|Z^\sharp\|_{L^r(\varsigma)}\ge\kappa\,u(a,b)^{1/r}r_0^{-2\gamma}=\kappa\,u(a,b)^{1/r+\gamma}\big(4(\En(a)+\En(b))\big)^{-\gamma}$. Together with the first step, this gives the assertion.
\end{proof}

\recall{\restsharp}

\begin{proof}
The inequality of the first assertion is obtained by applying Lemma~\ref{lem:separate} with $a=\psi$ and $b=\phi$. If $\Lc_q(\psi)=0$, then $u=0$ by this inequality, and by the condition for equality in the Cauchy--Schwarz inequality we have $\psi=e^{i\alpha}\phi$ for some $\alpha$. Conversely, if $\psi=e^{i\alpha}\phi$ for some $\alpha$, then $\Pm^\theta_\psi=\Pm^\theta_\phi$ for all $\theta$, so $\Lc_q(\psi)=0$.

We prove the second assertion. For $h\in\Qs$ satisfying $\|h\|=1$ and $\ip\phi h=0$, put $a:=(\phi+h)/\sqrt2$ and $b:=(\phi-h)/\sqrt2$. Since $\ip\phi h=0$, the vectors $a,b$ are unit vectors in $\Qs$ and $\ip ab=\frac12(\|\phi\|^2-\|h\|^2-\ip\phi h+\ip h\phi)=0$. For $(\theta,\mu)\in\Ac$, the self-adjointness of $M_{\theta,\mu}$ gives $\ip a{M_{\theta,\mu}a}-\ip b{M_{\theta,\mu}b}=2\Re\ip\phi{M_{\theta,\mu}h}$, so $\Re\ip\phi{M_{\theta,\mu}h}=\frac12(\int\mu\,d\Pm^\theta_a-\int\mu\,d\Pm^\theta_b)$. Taking the supremum over the elements $\mu$ of $\Kc$, we obtain from \eqref{eq:KR}, for all $\theta$,
\begin{equation}\label{eq:Nh}
\sup_{\mu\in\Kc}2\Re\ip\phi{M_{\theta,\mu}h}=\mathcal W_\theta(a,b).
\end{equation}
Since $\ip ab=0$, we have $u(a,b)=1$, and applying Lemma~\ref{lem:separate} to these $a,b$ and substituting $\En(a)+\En(b)=\frac12(\En(\phi+h)+\En(\phi-h))=\En(\phi)+\En(h)$ (the parallelogram law), we obtain
\begin{equation}\label{eq:Nhq}
\Lc_q(a,b)\ \ge\ \kappa\,\big(4(\En(\phi)+\En(h))\big)^{-\gamma}.
\end{equation}
Moreover, $\En(\phi)\ge n$ by Lemma~\ref{lem:fock}. From $r\le2$ and $(n+1)(n-1)!\le2n^n$ we get $\kappa^r\le(n+1)2^n(n-1)!\le2^{n+1}n^n$, hence $\kappa^r(4n)^{-r\gamma}\le2^{1-n-r}\le2^{-r}$, and for all $s\in(0,1]$ we have $\kappa_*(s)\le\frac{\kappa}{4}(4n)^{-\gamma}\le\frac18$ and $\Gamma_*(s)\ge1$.

Suppose that $\psi$ satisfies the hypotheses $\Mo_\beta(\psi)\le\Lambda$, $u>0$, and $\sqrt u\le\kappa_*(u)/\Gamma_*(u)$. By these hypotheses and by $\kappa_*(u)\le\frac18$ and $\Gamma_*(u)\ge1$, which were shown in the previous paragraph, we have $\sqrt u\le1/8$, so $u\le1/64$. With $z:=\ip\phi\psi$ and $g:=g_\psi$, we have $\psi=z\phi+\sqrt u\,g$ and $|z|=\sqrt{1-u}>0$. By the second assertion of Lemma~\ref{lem:expmom}, $\En(g)\le E_*(u)$. The vector $h:=(\overline z/|z|)\,g$ is a unit vector orthogonal to $\phi$, and $\En(h)=\En(g)\le E_*(u)$. For $(\theta,\mu)\in\Ac$ put $M:=M_{\theta,\mu}$. Since $M$ is symmetric on $\Qs$, the identity $\overline z\ip\phi{Mg}=|z|\ip\phi{Mh}$ gives
\[
\begin{aligned}
\ell_{\theta,\mu}(\psi)&=-u\,\ip\phi{M\phi}+2\sqrt{u(1-u)}\,\Re\ip\phi{Mh}+u\,\ip g{Mg}\\
&\ge2\sqrt{u(1-u)}\,\Re\ip\phi{Mh}-u\,\Gamma_*(u).
\end{aligned}
\]
The inequality in the second line is obtained as follows. By $\|Mf\|\le\En(f)^{1/2}$ and the Cauchy--Schwarz inequality, $|\ip f{Mf}|\le\En(f)^{1/2}$ for every unit vector $f\in\Qs$. Hence $\ip\phi{M\phi}\le\En(\phi)^{1/2}$ and $\ip g{Mg}\ge-\En(g)^{1/2}$, and $\En(g)\le E_*(u)$ gives $\ip g{Mg}\ge-E_*(u)^{1/2}$. Fixing $\theta$ and taking the supremum over the elements $\mu$ of $\Kc$, we obtain $\mathcal W_\theta(\psi)\ge\sqrt{u(1-u)}\,\mathcal W_\theta(a,b)-u\,\Gamma_*(u)$ from \eqref{eq:Nh} for the vectors $a,b$ defined from this $h$. Since $\mathcal W_\theta(\psi)\ge0$ and $\varsigma$ is a probability measure, the triangle inequality in $L^q(\varsigma)$ and \eqref{eq:Nhq} give $\Lc_q(\psi)\ge\sqrt{u(1-u)}\,\kappa\,\big(4(\En(\phi)+E_*(u))\big)^{-\gamma}-u\,\Gamma_*(u)=4\sqrt{u(1-u)}\,\kappa_*(u)-u\,\Gamma_*(u)$. From $u\le1/64$ we get $\sqrt{1-u}\ge1/\sqrt2$, and $\sqrt u\,\Gamma_*(u)\le\kappa_*(u)$ by hypothesis, so $\Lc_q(\psi)\ge\sqrt u\,\big(2\sqrt2-1\big)\kappa_*(u)\ge\kappa_*(u)\sqrt u$.
\end{proof}

\subsection{Upper bound on the loss at a stationary point}\label{sec:proofonepoint}

\recall{\restonepoint}

\begin{proof}
We prove the first assertion. With $g:=g_\psi$ we have $\psi=z\phi+\sqrt u\,g$, $|z|^2=1-u$, and $\ip\phi g=0$. Take $(\theta,\mu)\in\Ac$ and put $M:=M_{\theta,\mu}$, $\lambda:=\ip\psi{M\psi}$, and $v_0:=v_{\theta,\mu}(\psi)=(M-\lambda)\psi$. Since $M$ is symmetric on $\Qs$, substituting $\psi=z\phi+\sqrt u\,g$ gives
\begin{align*}
\lambda&=(1-u)\ip\phi{M\phi}+2\sqrt u\,\Re\big(\overline z\ip\phi{Mg}\big)+u\ip g{Mg},\\
\Re\ip{v_0}{z\phi}&=\Re\ip{M\psi}{z\phi}-(1-u)\lambda=(1-u)\ip\phi{M\phi}+\sqrt u\,\Re\big(\overline z\ip\phi{Mg}\big)-(1-u)\lambda.
\end{align*}
Eliminating $\Re(\overline z\ip\phi{Mg})$ from the two identities and using $\ell_{\theta,\mu}(\psi)=\lambda-\ip\phi{M\phi}$, we obtain
\begin{equation}\label{eq:quadexp}
-2\Re\ip{v_0}{z\phi}=\ell_{\theta,\mu}(\psi)+u\,\big(\ip\phi{M\phi}+\ip g{Mg}-2\lambda\big).
\end{equation}
We bound the second term on the right-hand side of \eqref{eq:quadexp}. The triangle inequality for $\En^{1/2}$ gives $\En(\psi)^{1/2}\le\Gamma(\psi)$, and $\|Mf\|\le\En(f)^{1/2}$ gives $|\ip f{Mf}|\le\En(f)^{1/2}$ for every unit vector $f\in\Qs$. In particular $|\lambda|\le\Gamma(\psi)$. If $z=0$, then $\psi=g$ and $\lambda=\ip g{Mg}$, so $|\ip\phi{M\phi}+\ip g{Mg}-2\lambda|=|\ip\phi{M\phi}-\ip g{Mg}|\le\Gamma(\psi)$. If $z\ne0$, put $e:=\sqrt u\,(z/|z|)\phi-|z|\,g$. Then $\|e\|=1$ and $\ip\psi e=0$, so $\{\phi,g\}$ and $\{\psi,e\}$ are both orthonormal bases of the same two-dimensional subspace. The sum of $\ip f{Mf}$ over the elements $f$ of an orthonormal basis of this subspace does not depend on the choice of the basis, so $\ip\phi{M\phi}+\ip g{Mg}=\lambda+\ip e{Me}$. Since $\En(e)^{1/2}\le\Gamma(\psi)$, we have $|\ip e{Me}-\lambda|\le2\Gamma(\psi)$. In both cases the absolute value of the second term on the right-hand side of \eqref{eq:quadexp} is at most $2\Gamma(\psi)u$, and we obtain \eqref{eq:graddomre}. Moreover, $\|v_0\|^2=\|M\psi\|^2-\lambda^2\le\En(\psi)$ gives $\|v_0\|\le\Gamma(\psi)$.

We prove the second assertion. Since $\ran\mathscr J(\omega)$ is a finite-dimensional real subspace of $\Hs$, there is an element $y_0$ of $\ran\mathscr J(\omega)$ with $\|z\phi-y_0\|=\delta_\omega$. By \eqref{eq:graddomre} and $\|v_{\theta,\mu}(\psi)\|\le\Gamma(\psi)$, for all $(\theta,\mu)\in\Ac$ we have
\begin{equation}\label{eq:pertheta}
\ell_{\theta,\mu}(\psi)\ \le\ -2\Re\ip{v_{\theta,\mu}(\psi)}{y_0}+2\Gamma(\psi)(\delta_\omega+u).
\end{equation}
Put $C:=2\Gamma(\psi)(\delta_\omega+u)$. By Lemma~\ref{lem:netstat} there is $x^*\in\partial^\circ\Lc_q(\psi)$ with $x^*(y_0)=0$, so the definition of the generalized gradient gives $0=x^*(y_0)\le D^\circ_{y_0}\Lc_q(\psi)$. In the case $q=\infty$, by \eqref{eq:clarkeinf} there is $(\theta,\mu)\in\Ac(\psi)$ with $D^\circ_{y_0}\Lc_\infty(\psi)=2\Re\ip{v_{\theta,\mu}(\psi)}{y_0}$. For this $(\theta,\mu)$ we have $\ell_{\theta,\mu}(\psi)=\Lc_\infty(\psi)$, so \eqref{eq:pertheta} gives $\Lc_\infty(\psi)\le-D^\circ_{y_0}\Lc_\infty(\psi)+C\le C$. In the case $q<\infty$, since $u>0$, the first assertion of Lemma~\ref{prop:sharp} gives $\Lc_q(\psi)>0$. For $\theta\in\Sph$ and $\mu\in\Kc_\theta(\psi)$ we have $\ell_{\theta,\mu}(\psi)=\mathcal W_\theta(\psi)$, so \eqref{eq:pertheta} gives $2\Re\ip{v_{\theta,\mu}(\psi)}{y_0}\le C-\mathcal W_\theta(\psi)$. By the third assertion of Lemma~\ref{lem:stat}, the function $\rho_\theta$ in \eqref{eq:rho} satisfies $\rho_\theta\ge0$, $\int\rho_\theta\,d\varsigma\le1$, and $\int\rho_\theta\mathcal W_\theta(\psi)\,d\varsigma=\Lc_q(\psi)$, so \eqref{eq:ddq} gives
\[
0\ \le\ D^\circ_{y_0}\Lc_q(\psi)\ \le\ \int_{\Sph}\rho_\theta\big(C-\mathcal W_\theta(\psi)\big)\,d\varsigma(\theta)=C\int_{\Sph}\rho_\theta\,d\varsigma-\Lc_q(\psi)\ \le\ C-\Lc_q(\psi),
\]
and hence $\Lc_q(\psi)\le C$.
\end{proof}

\subsection{Proofs of the main theorems}\label{sec:prooftheorems}\label{sec:prooflimit}

\begin{lemma}[Properties of the constants]\label{lem:tau}
For $\tau$ and $\bar s$ of \eqref{eq:tau}, the quotient $\tau(s)/s$ is nonincreasing on $(0,1]$. The function $\tau$ is continuous on $(0,1]$, satisfies $\tau(s)\to0$ as $s\downarrow0$, and is nondecreasing on $(0,\bar s]$.
\end{lemma}

\begin{proof}
By \eqref{eq:consts} and \eqref{eq:tau}, for $s\in(0,1]$ we have
\[
\frac{\tau(s)}{s}=\frac{2^8}{\kappa^2}\,\big(\En(\phi)^{1/2}+E_*(s)^{1/2}\big)^2\big(4(\En(\phi)+E_*(s))\big)^{2\gamma}.
\]
Since $E_*$ is decreasing in $s$ and the right-hand side is increasing in $E_*(s)$, the quotient $\tau(s)/s$ is nonincreasing. Since the right-hand side is continuous in $s$, the function $\tau$ is continuous on $(0,1]$. Moreover, when $s$ is sufficiently small, the right-hand side is at most a constant multiple of some power of $\log(1/s)$, so $\tau(s)\to0$ as $s\downarrow0$. Since $\frac{d}{ds}E_*(s)=-\frac{2}{\beta s}$, for $s\in(0,1)$ we have
\[
\frac{d}{ds}\log\tau(s)=\frac1s-\frac{2}{\beta s}\Big(\frac{E_*(s)^{-1/2}}{\En(\phi)^{1/2}+E_*(s)^{1/2}}+\frac{2\gamma}{\En(\phi)+E_*(s)}\Big)\ \ge\ \frac1s\Big(1-\frac{2(1+2\gamma)}{\beta E_*(s)}\Big).
\]
If $s\le4\Lambda e^{-(1+2\gamma)}$, then $E_*(s)\ge\frac2\beta\log\frac{4\Lambda}{s}\ge\frac{2(1+2\gamma)}{\beta}$, so the rightmost expression is at least $0$, and $\tau$ is nondecreasing on $(0,\bar s]$.
\end{proof}

\recall{\restlimit}

\begin{proof}
The limit point $\psi_*$ is the limit with respect to $\|\cdot\|_{\Qs}$ of a subsequence of $\psi_{\omega_k}$, and a subsequence of a bounded sequence of stationary points is again a bounded sequence of stationary points, so we may assume that $\psi_k:=\psi_{\omega_k}$ converges to $\psi_*$ with respect to $\|\cdot\|_{\Qs}$. By the compactness in the first assertion of Lemma~\ref{lem:expmom}, $\psi_*$ is a unit vector in $\Qs$ that satisfies $\Mo_\beta(\psi_*)\le\Lambda$. By the same compactness, every subsequence of $\psi_{\omega_k}$ has a subsequence that converges with respect to $\|\cdot\|_{\Qs}$, so limit points exist. Put $u:=u(\psi_*)$. If $u=0$, then the condition for equality in the Cauchy--Schwarz inequality gives $\psi_*=e^{i\alpha}\phi$ for some $\alpha$, and the assertions of the theorem hold. We therefore assume $u>0$.

By the condition on the initial centers in Definition~\ref{def:seq} and the first assertion of Lemma~\ref{lem:approx}, we have $\dist(\phi,\mathcal V(c^{(k)}))\to0$. The first inequality of \eqref{eq:deltabounds} gives $\delta_{\omega_k}\le|\ip\phi{\psi_k}|\,\dist(\phi,\mathcal V(c^{(k)}))\le\dist(\phi,\mathcal V(c^{(k)}))$, so $\delta_{\omega_k}\to0$.

Since $u_k:=u(\psi_k)$ converges to $u$, we have $u_k>0$ for all sufficiently large $k$, and \eqref{eq:Lupper} in the second assertion of Lemma~\ref{prop:onepoint} gives $\Lc_q(\psi_k)\le2\Gamma(\psi_k)(\delta_{\omega_k}+u_k)$. Since $\Lc_q$ is continuous with respect to $\|\cdot\|_{\Qs}$ (Proposition~\ref{prop:lipschitz}), we have $\Lc_q(\psi_k)\to\Lc_q(\psi_*)$. Moreover, $g_{\psi_k}$ converges to $g_{\psi_*}$ with respect to $\|\cdot\|_{\Qs}$, so $\Gamma(\psi_k)\to\Gamma(\psi_*)$. Hence, letting $k\to\infty$, we obtain $\Lc_q(\psi_*)\le2\Gamma(\psi_*)\,u$.

We prove the last assertion. Suppose $0<u\le\bar u$. By the second assertion of Lemma~\ref{lem:expmom}, $\Gamma(\psi_*)\le\Gamma_*(u)$. By \eqref{eq:ubar}, $\bar u\le\bar s$ and $\tau(\bar u)\le1$. Since $\tau$ is nondecreasing on $(0,\bar s]$ (Lemma~\ref{lem:tau}), we have $\tau(u)\le\tau(\bar u)\le1$. By \eqref{eq:tau}, the inequality $\tau(u)\le1$ is equivalent to $\sqrt u\le\kappa_*(u)/(4\Gamma_*(u))$, and in particular $\sqrt u\le\kappa_*(u)/\Gamma_*(u)$. Since $\Mo_\beta(\psi_*)\le\Lambda$, the hypotheses of the second assertion of Lemma~\ref{prop:sharp} hold, and its conclusion together with the inequalities obtained above gives $\kappa_*(u)\sqrt u\le\Lc_q(\psi_*)\le2\Gamma_*(u)\,u$. This inequality yields $\tau(u)=16\,\Gamma_*(u)^2u/\kappa_*(u)^2\ge4$, which contradicts $\tau(u)\le1$. Hence $u>\bar u$ whenever $u>0$, and we obtain the last assertion.
\end{proof}

\begin{proposition}[Zeros of active gradients]\label{prop:zerograd}
If a unit vector $\psi$ in $\Qs$ satisfies $0\in\mathcal G(\psi)$, then $\psi=e^{i\alpha}\phi$ for some $\alpha$. In particular, if $0\in\overline{\conv}\,\mathcal G(\psi)$ and $\Lc_\infty$ is differentiable at $\psi$, then $\psi=e^{i\alpha}\phi$ for some $\alpha$.
\end{proposition}

\begin{proof}
Let $(\theta,\mu)\in\Ac(\psi)$ satisfy $v_{\theta,\mu}(\psi)=0$, and put $P:=\Pm^\theta_\psi$, $Q:=\Pm^\theta_\phi$, $G:=F_Q-F_P$, and $S:=\operatorname{supp}P$. By Lemma~\ref{lem:rot}, $P$ and $Q$ are absolutely continuous with respect to Lebesgue measure, and $F_P$, $F_Q$, and $G$ are continuous. First, \eqref{eq:ibp} gives $\ell_{\theta,\mu}(\psi)=\int\mu'G$. Since $(\theta,\mu)$ belongs to the active set $\Ac(\psi)$, \eqref{eq:Linf} and \eqref{eq:Lpot} give $\int\mu'G=\Lc_\infty(\psi)\ge\int|G|$, and together with $|\mu'G|\le|G|$ this shows that $\mu'G=|G|$ almost everywhere with respect to Lebesgue measure. Next, put $\lambda:=\ip\psi{M_{\theta,\mu}\psi}$. The functional calculus gives $0=\|v_{\theta,\mu}(\psi)\|^2=\int|\mu-\lambda|^2dP$, so $\mu=\lambda$ almost everywhere with respect to $P$, and by the continuity of $\mu$ we have $\mu\equiv\lambda$ on $S$. With respect to Lebesgue measure, almost every point of $S$ is a density point of $S$ (the Lebesgue density theorem) and, in addition, a point at which $\mu$ is differentiable (Rademacher's theorem). Such a point is the limit of a sequence of points of $S$ different from it, and the difference quotients of $\mu$ along this sequence are $0$, so $\mu'=0$ at that point. Together with $\mu'G=|G|$, this shows that $G=0$ at almost every point of $S$ with respect to Lebesgue measure. Furthermore, since $P$ is absolutely continuous and concentrated on $S$, the intersection of $S$ with every neighborhood of each point of $S$ has positive Lebesgue measure, and by the continuity of $G$ we have $G=0$ on $S$. For a connected component $I$ of $\R\setminus S$, we have $P(I)=0$, so $F_P$ is constant on $I$. A finite endpoint of $I$ belongs to $S$, so $F_Q=F_P$ at that endpoint, and when an endpoint is $\pm\infty$, the limit of $F_Q$ and the limit of $F_P$ at that endpoint are equal. Since $F_Q$ is monotone, $F_Q=F_P$ on $I$. This shows that $G$ vanishes identically. Hence $P=Q$ and $\Lc_\infty(\psi)=\ell_{\theta,\mu}(\psi)=\int\mu'G=0$, and by the first assertion of Lemma~\ref{prop:sharp} we have $\psi=e^{i\alpha}\phi$ for some $\alpha$. We prove the second half of the statement. If $\Lc_\infty$ is differentiable at $\psi$, then $\mathcal G(\psi)$ consists of a single point by the second assertion of Lemma~\ref{lem:stat}, so $\overline{\conv}\,\mathcal G(\psi)=\mathcal G(\psi)$, and the claim reduces to the first half.
\end{proof}

\recall{\restlandscape}

\begin{proof}
The limit point $\psi_*$ is the limit with respect to $\|\cdot\|_{\Qs}$ of a subsequence of $\psi_{\omega_k}$, and a subsequence of a bounded sequence of stationary points is again a bounded sequence of stationary points, so we may assume that $\psi_k:=\psi_{\omega_k}$ converges to $\psi_*$ with respect to $\|\cdot\|_{\Qs}$. By the first assertion of Lemma~\ref{lem:expmom}, $E_{\max}:=\sup_k\En(\psi_k)<\infty$. By Lemma~\ref{lem:netstat}, for each $k$ there is $x_k^*\in\partial^\circ\Lc_\infty(\psi_k)$ that vanishes on $\ran\mathscr J(\omega_k)$. For $h\in\Hs$ put
\[
\chi_k(h):=\max_{(\theta,\mu)\in\Ac(\psi_k)}\Re\ip h{v_{\theta,\mu}(\psi_k)},\qquad\chi_*(h):=\max_{(\theta,\mu)\in\Ac(\psi_*)}\Re\ip h{v_{\theta,\mu}(\psi_*)}.
\]
The maxima are attained by Lemma~\ref{lem:loss}. Since $\|v_{\theta,\mu}(\psi_k)\|\le\En(\psi_k)^{1/2}\le E_{\max}^{1/2}$, we have $|\chi_k(h)|\le E_{\max}^{1/2}\|h\|$. By \eqref{eq:clarkeinf}, for all $h\in\Qs$ we have $x_k^*(h)\le D^\circ_h\Lc_\infty(\psi_k)=2\chi_k(h)\le2E_{\max}^{1/2}\|h\|$. Combining this inequality with the one obtained from it by replacing $h$ with $-h$, we get $|x_k^*(h)|\le2E_{\max}^{1/2}\|h\|$. Since $\Qs$ is dense in $\Hs$, the functional $x_k^*$ extends to a continuous real-linear functional on $\Hs$. Regarding $\Hs$ as a real Hilbert space with the real inner product $\Re\ip\cdot\cdot$ and applying the Riesz representation theorem, we obtain $v_k\in\Hs$ such that $x_k^*(h)=2\Re\ip h{v_k}$ for all $h\in\Qs$, and $\|v_k\|\le E_{\max}^{1/2}$. For all $h\in\Qs$ we have $\Re\ip h{v_k}\le\chi_k(h)$, both sides of this inequality are continuous in $h$ on $\Hs$, and $\Qs$ is dense in $\Hs$, so for all $h\in\Hs$ we have
\[
\Re\ip h{v_k}\ \le\ \chi_k(h).
\]
By \eqref{eq:VinJ} we have $\mathcal V(c^{(k)})\subset\ran\mathscr J(\omega_k)$, and $x_k^*$ vanishes on $\ran\mathscr J(\omega_k)$, so $\Re\ip h{v_k}=0$ for all $h\in \mathcal V(c^{(k)})$. Since $\mathcal V(c^{(k)})$ is a complex subspace, combining this identity with the one obtained from it by replacing $h$ with $ih$ gives $\ip h{v_k}=0$. Passing to a subsequence, we may assume that $v_k$ converges weakly to some $v_*\in\Hs$. Next we show $v_*\in\overline{\conv}\,\mathcal G(\psi_*)$. Fix $h\in\Hs$. We show $\limsup_k\chi_k(h)\le\chi_*(h)$. Take $(\theta_k,\mu_k)\in\Ac(\psi_k)$ attaining the maximum in the definition of $\chi_k(h)$. Passing to a subsequence, we may assume that $\chi_k(h)\to\limsup_k\chi_k(h)$ and that $(\theta_k,\mu_k)\to(\hat\theta,\hat\mu)$ for some point $(\hat\theta,\hat\mu)$ of $\Ac$. In the identity $\ell_{\theta_k,\mu_k}(\psi_k)=\Lc_\infty(\psi_k)$, the left-hand side converges to $\ell_{\hat\theta,\hat\mu}(\psi_*)$ and the right-hand side to $\Lc_\infty(\psi_*)$ (Lemma~\ref{lem:loss}), so $(\hat\theta,\hat\mu)\in\Ac(\psi_*)$. Since $v_{\theta_k,\mu_k}(\psi_k)$ converges weakly to $v_{\hat\theta,\hat\mu}(\psi_*)$, we have $\lim_k\chi_k(h)=\Re\ip h{v_{\hat\theta,\hat\mu}(\psi_*)}\le\chi_*(h)$. Taking the upper limit of both sides of $\Re\ip h{v_k}\le\chi_k(h)$, we obtain $\Re\ip h{v_*}\le\chi_*(h)$ from $\limsup_k\chi_k(h)\le\chi_*(h)$. Since $h$ is arbitrary, the Hahn--Banach separation theorem gives $v_*\in\overline{\conv}\,\mathcal G(\psi_*)$. Finally we show $v_*=0$. By the condition on the initial centers in Definition~\ref{def:seq} and the first assertion of Lemma~\ref{lem:approx}, we have $\dist(\varphi_{c_j},\mathcal V(c^{(k)}))\to0$ for all $j\ge1$. Let $y_k$ be the point of $\mathcal V(c^{(k)})$ closest to $\varphi_{c_j}$. Then $|\ip{\varphi_{c_j}}{v_k}|=|\ip{\varphi_{c_j}-y_k}{v_k}|\le\dist(\varphi_{c_j},\mathcal V(c^{(k)}))\,E_{\max}^{1/2}\to0$, and since $v_k$ converges weakly to $v_*$, we get $\ip{\varphi_{c_j}}{v_*}=0$. By the first assertion of Lemma~\ref{lem:approx}, $\mathcal V(c_1,c_2,\dots)$ is dense in $\Hs$, so $v_*=0$, and hence $0\in\overline{\conv}\,\mathcal G(\psi_*)$. The second half of the statement of the theorem follows from Proposition~\ref{prop:zerograd}.
\end{proof}

\section{Proofs for the quantitative form at finite width}\label{sec:proofs4}

In this section we prove the propositions and the corollary of Section~\ref{sec:numerics}.

\subsection{The loss over all directions}\label{sec:prooffinite}

\recall{\restfinite}

\begin{proof}
We first prove the part $\delta_\omega^2\le\dist(\phi,\mathcal V(c'))^2\le u$ of the first assertion. By \eqref{eq:VinJ}, $\psi$ is a unit vector in $\mathcal V(c')$. For the orthogonal projection $\Pi_{\mathcal V(c')}$ we have $1-u=|\ip\phi\psi|^2=|\ip{\Pi_{\mathcal V(c')}\phi}\psi|^2\le\|\Pi_{\mathcal V(c')}\phi\|^2=1-\dist(\phi,\mathcal V(c'))^2$, so $\dist(\phi,\mathcal V(c'))^2\le u$. Moreover, the first inequality of \eqref{eq:deltabounds} gives $\delta_\omega\le|\ip\phi\psi|\,\dist(\phi,\mathcal V(c'))\le\dist(\phi,\mathcal V(c'))$.

Suppose that $\omega$ is a stationary point for $\Lc_q$ and satisfies $\Mo_\beta(\psi)\le\Lambda$ and $0<u\le\bar u$. By the second assertion of Lemma~\ref{lem:expmom}, $\En(g_\psi)\le E_*(u)$, so $\Gamma(\psi)\le\Gamma_*(u)$. Since $0<u\le\bar u\le \bar s$ and $\tau$ is nondecreasing on $(0,\bar s]$ (Lemma~\ref{lem:tau}), together with \eqref{eq:ubar} we get $\tau(u)\le\tau(\bar u)\le1$. Since $\tau(u)=16\,u\,\Gamma_*(u)^2/\kappa_*(u)^2$, the inequality $\tau(u)\le1$ is equivalent to $16\,u\,\Gamma_*(u)^2\le\kappa_*(u)^2$, that is, to $\sqrt u\le\kappa_*(u)/(4\Gamma_*(u))$, and in particular $\sqrt u\le\kappa_*(u)/\Gamma_*(u)$. Hence the hypotheses of the second assertion of Lemma~\ref{prop:sharp} hold, and combining its conclusion with \eqref{eq:Lupper} and $\Gamma(\psi)\le\Gamma_*(u)$ we get $\kappa_*(u)\sqrt u\le\Lc_q(\psi)\le2\Gamma_*(u)(\delta_\omega+u)$. Multiplying both sides of $\sqrt u\le\kappa_*(u)/(4\Gamma_*(u))$ by $2\Gamma_*(u)\sqrt u$ gives $2\Gamma_*(u)\,u\le\frac{\kappa_*(u)}{2}\sqrt u$, so together with the inequality above we get $\frac{\kappa_*(u)}{2}\sqrt u\le2\Gamma_*(u)\,\delta_\omega$, that is, $u\le16\,\Gamma_*(u)^2\delta_\omega^2/\kappa_*(u)^2=(\tau(u)/u)\,\delta_\omega^2$. Since $u>0$, we have $\delta_\omega>0$, and from $\delta_\omega^2\le u$ and the fact that $\tau(s)/s$ is nonincreasing (Lemma~\ref{lem:tau}) we obtain $u\le(\tau(\delta_\omega^2)/\delta_\omega^2)\,\delta_\omega^2=\tau(\delta_\omega^2)$. Since $0<\delta_\omega^2\le\dist(\phi,\mathcal V(c'))^2\le u\le\bar u\le \bar s$ and $\tau$ is nondecreasing on $(0,\bar s]$, we have $\tau(\delta_\omega^2)\le\tau(\dist(\phi,\mathcal V(c'))^2)$.

We prove the second assertion. On the event of the second assertion of Lemma~\ref{lem:approx}, if $\|c'-c\|\le R$, then $\|\varphi_0\|^2=(\pi\sigma^2)^{n/2}$ and \eqref{eq:deltam} give $\dist(\phi,\mathcal V(c'))^2\le\delta_m^2$. Since $\delta_m^2\le\bar u\le \bar s$, the first assertion and the fact that $\tau$ is nondecreasing on $(0,\bar s]$ give $u\le\tau(\dist(\phi,\mathcal V(c'))^2)\le\tau(\delta_m^2)$.
\end{proof}

\subsection{Finitely many directions}\label{sec:prooffdir}

In this subsection we set $n=1$. For $N\in\N_0$ put
\begin{equation}\label{eq:HN}
\Hs_N:=\spn_{\C}\{h_k:k\le N\},
\end{equation}
and write $\Pi_N$ for the orthogonal projection of $\Hs$ onto $\Hs_N$. For unit vectors $a,b$ in $\Qs$, we write $\Lc_{q,K}(a,b)$ for the quantity obtained by replacing $\mathcal W_{\theta_k}(\psi)$ with $\mathcal W_{\theta_k}(a,b)$ in the definition of $\Lc_{q,K}(\psi)$ in \eqref{eq:LqK}.

\begin{lemma}[Equally spaced angles]\label{lem:equi}
Let $N$ and $K$ be positive integers, and let $a,b$ be unit vectors in the space $\Hs_N$ of \eqref{eq:HN}. For all $q\in[1,\infty]$,
\[
\big|\Lc_{q,K}(a,b)-\Lc_q(a,b)\big|\ \le\ \frac{\pi N}{2K}\,\Lc_q(a,b).
\]
\end{lemma}

\begin{proof}
Put $\theta(t):=(\cos t,\sin t)$. Let $U_t$ be the unitary operator that maps $h_k$ to $e^{-ikt}h_k$ for all $k$. On the space $\mathcal D$ of \eqref{eq:ladder} we have $U_t^*\hat a_1U_t=e^{-it}\hat a_1$ and $X_{\theta(t)}=(e^{-it}\hat a_1+e^{it}\hat a_1^*)/\sqrt2$, so $U_t^*X_{\theta(0)}U_t=X_{\theta(t)}$ on $\mathcal D$. The partial sums of the Hermite expansion of an element of $\Sc$ converge with respect to $\|\cdot\|_{\Qs}$, so by \eqref{eq:Xxi} the operator $X_{\theta(t)}$ is the closure of the restriction of $X_{\theta(t)}$ to $\mathcal D$. The operator $U_t^*X_{\theta(0)}U_t$ is self-adjoint and coincides with $X_{\theta(t)}$ on $\mathcal D$, so it is an extension of $X_{\theta(t)}$. Since $X_{\theta(t)}$ is also self-adjoint, $U_t^*X_{\theta(0)}U_t$ and $X_{\theta(t)}$ are equal. Since $X_{\theta(0)}$ is multiplication by $x$, the Hermite expansions of $a$ and $b$ and the functional calculus give
\[
\Delta(s,t):=F^{\theta(t)}_a(s)-F^{\theta(t)}_b(s)=\sum_{k,l\le N}\Big(\overline{\ip{h_k}a}\,\ip{h_l}a-\overline{\ip{h_k}b}\,\ip{h_l}b\Big)\,e^{i(k-l)t}\int_{-\infty}^sh_k(x)\,h_l(x)\,dx.
\]
For each $s$, the function $t\mapsto\Delta(s,t)$ is a real-valued trigonometric polynomial of degree at most $N$. By the interpolation formula of M.~Riesz \citep{riesz} (see also \citealp[Chapter X, formula (3.11)]{zygmund}), with $\vartheta_j:=(2j-1)\pi/(2N)$ and $w_j:=(4N\sin^2(\vartheta_j/2))^{-1}$ ($j=1,\dots,2N$), we have
\[
\frac{\partial}{\partial t}\Delta(s,t)=\sum_{j=1}^{2N}(-1)^{j+1}w_j\,\Delta(s,t+\vartheta_j),\qquad\sum_{j=1}^{2N}w_j=N.
\]
(The second identity is obtained by applying the first identity to the trigonometric polynomial $\sin Nt$ at $t=0$.) Put $\mathcal W(t):=\mathcal W_{\theta(t)}(a,b)$. Then $\mathcal W(t)=\int_{\R}|\Delta(s,t)|\,ds$ by \eqref{eq:KR}. Integrating the absolute value of $\Delta(s,t)-\Delta(s,\tilde t)=\int_{\tilde t}^t\frac{\partial}{\partial t'}\Delta(s,t')\,dt'$ with respect to $s$, we obtain from the first identity of the interpolation formula and Fubini's theorem, for $\tilde t\le t$,
\[
|\mathcal W(t)-\mathcal W(\tilde t)|\ \le\ \int_{\tilde t}^tY(t')\,dt',\qquad Y(t'):=\sum_{j=1}^{2N}w_j\,\mathcal W(t'+\vartheta_j).
\]
Since $\theta(t+\pi)=-\theta(t)$, the functions $\mathcal W$ and $Y$ have period $\pi$. We write $\|\cdot\|_q$ for the $L^q$ norm with respect to the measure $dt/\pi$ on $[0,\pi)$. Since $\varsigma$ is rotation invariant, $\Lc_q(a,b)=\|\mathcal W\|_q$. The norm $\|\cdot\|_q$ of a function of period $\pi$ does not change under translations, so the triangle inequality and $\sum_jw_j=N$ give $\|Y\|_q\le N\|\mathcal W\|_q$. Let $t_k:=k\pi/K$, put $\tilde{\mathcal W}(t):=\mathcal W(t_k)$ when $|t-t_k|<\pi/(2K)$, and extend it periodically with period $\pi$. Then $\|\tilde{\mathcal W}\|_q=\Lc_{q,K}(a,b)$. Let $J$ be $[t_k,t_k+\pi/(2K)]$ or $[t_k-\pi/(2K),t_k]$. For $t\in J$, H\"older's inequality gives $|\mathcal W(t)-\mathcal W(t_k)|\le\int_JY\le(\pi/(2K))^{1-1/q}(\int_JY^q)^{1/q}$, so $\int_J|\mathcal W-\tilde{\mathcal W}|^q\,dt\le(\pi/(2K))^q\int_JY^q\,dt$. Summing over $J$ gives $\|\mathcal W-\tilde{\mathcal W}\|_q\le\frac{\pi}{2K}\|Y\|_q\le\frac{\pi N}{2K}\|\mathcal W\|_q$. For $q=\infty$ as well, the same inequality follows from $|\mathcal W(t)-\mathcal W(t_k)|\le\int_JY$. The assertion follows from the triangle inequality.
\end{proof}

\recall{\restfdir}

\begin{proof}
Let $f\in\Qs$ be a unit vector that satisfies $\Mo_\beta(f)\le\Lambda$. For $t:=\|f-\Pi_Nf\|$ we have $t^2=\sum_{k>N}|\ip{h_k}f|^2\le\Lambda e^{-\beta(N+1)}\le\frac12$, so $f_N:=\Pi_Nf/\|\Pi_Nf\|$ is well defined and $\|f-f_N\|\le t+(1-\sqrt{1-t^2})\le2t$. The triangle inequality for $\En^{1/2}$, together with $\En(\Pi_Nf)\le\En(f)$ and $(1-t^2)^{-1/2}-1\le t^2\le t$, gives $\En(f-f_N)^{1/2}\le\En(f-\Pi_Nf)^{1/2}+t\,\En(f)^{1/2}$. The left-hand side of \eqref{eq:lip} in Proposition~\ref{prop:lipschitz} does not depend on $\phi$ and, by \eqref{eq:KR}, equals $\sup_\theta \mathcal W_\theta(\psi,\tilde\psi)$. Hence, applying \eqref{eq:lip} with $\psi=f$ and $\tilde\psi=f_N$, we get, for all $\theta$,
\[
\mathcal W_\theta(f,f_N)\ \le\ \En(f)^{1/2}\|f-f_N\|+\En(f-f_N)^{1/2}\ \le\ 3t\,\En(f)^{1/2}+\En(f-\Pi_Nf)^{1/2}.
\]
In the range $j\ge N+1\ge1/\beta$, the quantity $(2j+1)e^{-\beta j}$ is nonincreasing in $j$. Hence \eqref{eq:fock} gives $\En(f-\Pi_Nf)\le\Lambda(2N+3)e^{-\beta(N+1)}$. Moreover, the first assertion of Lemma~\ref{lem:expmom} gives $\En(f)\le1+\frac2\beta\log\Lambda$. Altogether, $\mathcal W_\theta(f,f_N)\le\epsilon_N/8$ for all $\theta$.

We apply this estimate with $f$ taken to be $\psi$ and $\phi$, and define $\psi_N$ and $\phi_N$ by the same formula as $f_N$. The triangle inequality for $\mathcal W_1$ then gives $|\mathcal W_\theta(\psi)-\mathcal W_\theta(\psi_N,\phi_N)|\le\epsilon_N/4$ for all $\theta$. In particular, the absolute value of the difference between $\Lc_q(\psi)$ and $\Lc_q(\psi_N,\phi_N)$ and the absolute value of the difference between $\Lc_{q,K}(\psi)$ and $\Lc_{q,K}(\psi_N,\phi_N)$ are both at most $\epsilon_N/4$. Since $K\ge\pi N$, applying Lemma~\ref{lem:equi} with $a=\psi_N$ and $b=\phi_N$ gives $|\Lc_{q,K}(\psi_N,\phi_N)-\Lc_q(\psi_N,\phi_N)|\le\frac12\Lc_q(\psi_N,\phi_N)\le\frac12\Lc_q(\psi)+\frac18\epsilon_N$. Combining the three estimates above, we get $|\Lc_{q,K}(\psi)-\Lc_q(\psi)|\le\frac12\Lc_q(\psi)+\frac58\epsilon_N$, and the assertion follows.
\end{proof}

\recall{\restrestricted}

\begin{proof}
Put $Y:=\mathscr J(\omega)\mathcal Z_0$. Here $Y$ is a finite-dimensional real subspace of $\Qs$, and since $0\in Y$ we have $\tilde\delta_\omega\le\|\ip\phi\psi\phi\|\le1$. We first consider the case $\tilde\delta_\omega^2>u$. In this case $\tilde\delta_\omega>0$. As shown in the proof of Lemma~\ref{prop:sharp}, $\kappa_*(s)\le\frac18$ and $\Gamma_*(s)\ge1$ hold for all $s\in(0,1]$, so \eqref{eq:tau} gives $\tau(s)\ge s$. Hence $u<\tilde\delta_\omega^2\le16\,\tau(\tilde\delta_\omega^2)$. In what follows we assume $\tilde\delta_\omega^2\le u$.

We prove a lower bound on the loss. By Lemma~\ref{lem:tau}, $\tau$ is nondecreasing on $(0,\bar s]$, so \eqref{eq:ubar} gives $16\,\tau(u)\le16\,\tau(\bar u')\le1$. The inequality $16\,\tau(u)\le1$ is equivalent to $\sqrt u\le\kappa_*(u)/(16\,\Gamma_*(u))$, and in particular $\sqrt u\le\kappa_*(u)/\Gamma_*(u)$ also holds. The second assertion of Lemma~\ref{prop:sharp}, Proposition~\ref{prop:fdir} and the hypothesis $\epsilon_N\le\frac14\kappa_*(u)\sqrt u$ give
\[
\Lc_{q,K}(\psi)\ \ge\ \tfrac12\,\kappa_*(u)\sqrt u-\epsilon_N\ \ge\ \tfrac14\,\kappa_*(u)\sqrt u\ >\ 0.
\]

Next, we construct an element of $\partial^\circ\Lc_{q,K}(\psi)$ that vanishes on $Y$. As stated right after \eqref{eq:LqK}, $\Lc_{q,K}$ is locally Lipschitz on $\Qs\setminus\{0\}$. Moreover, by the definition given right after \eqref{eq:LqK}, its value does not change when the argument is multiplied by a positive constant, so $\Ls_{q,K}(\omega)=\Lc_{q,K}(\Psi_\omega)$. We apply the inequality of Lemma~\ref{lem:clarkechain} in the appendix with $Z=\mathcal Z$, $\Omega_0=\Omega$, $X=\Qs$, $U=\Qs\setminus\{0\}$, $F(\omega)=\Psi_\omega$, and $g=\Lc_{q,K}$. Since $DF(\omega)=\mathscr J(\omega)$ (Lemma~\ref{lem:tangent}), we get, for all $\zeta\in\mathcal Z_0$,
\[
0\ \le\ D^\circ_\zeta\Ls_{q,K}(\omega)\ \le\ D^\circ_{\mathscr J(\omega)\zeta}\,\Lc_{q,K}(\Psi_\omega).
\]
Hence $D^\circ_h\Lc_{q,K}(\Psi_\omega)\ge0$ for all $h\in Y$, and by the second assertion of Lemma~\ref{lem:clarkebasic} in the appendix there is $y^*\in\partial^\circ\Lc_{q,K}(\Psi_\omega)$ that vanishes on $Y$. The second paragraph of the proof of Lemma~\ref{lem:netstat} remains valid when $\Lc_q$ is replaced with $\Lc_{q,K}$, so $x^*:=\|\Psi_\omega\|\,y^*$ belongs to $\partial^\circ\Lc_{q,K}(\psi)$ and vanishes on $Y$.

We prove an upper bound on the loss. Since $Y$ is finite-dimensional, there is an element $y_0$ of $Y$ with $\|\ip\phi\psi\phi-y_0\|=\tilde\delta_\omega$. Since $x^*(y_0)=0$, the definition of the generalized gradient gives $0\le D^\circ_{y_0}\Lc_{q,K}(\psi)$. The proof of the second assertion of Lemma~\ref{prop:onepoint} uses only this inequality, \eqref{eq:pertheta}, the formulas \eqref{eq:clarkeinf} and \eqref{eq:ddq} for the generalized directional derivative, and, for $q<\infty$, the positivity of the loss. The inequality \eqref{eq:pertheta} holds for this $y_0$ with $\delta_\omega$ replaced by $\tilde\delta_\omega$, by \eqref{eq:graddomre} and $\|v_{\theta,\mu}(\psi)\|\le\Gamma(\psi)$. The formulas \eqref{eq:clarkeinf} and \eqref{eq:ddq} remain valid when $\varsigma$ on $\Sph$ is replaced by the uniform probability measure on $\{\theta_0,\dots,\theta_{K-1}\}$. Indeed, in the proof of Lemma~\ref{lem:stat}, for $q=\infty$ we replace the index space of Lemma~\ref{lem:clarkemax} in the appendix by $\{\theta_0,\dots,\theta_{K-1}\}\times\Kc$, and for $q<\infty$ we replace $S$ of Lemma~\ref{lem:clarkeint} in the appendix by $\{0,\dots,K-1\}$. The former is a closed subset of $\Ac$ and the latter is a finite set, so the hypotheses of the two lemmas hold for the same reasons. The inequality $\Lc_{q,K}(\psi)>0$ for $q<\infty$ was shown above. Hence the same argument as in the proof of the second assertion of Lemma~\ref{prop:onepoint} gives $\Lc_{q,K}(\psi)\le2\Gamma(\psi)(\tilde\delta_\omega+u)$.

Finally, we combine the two estimates. By the second assertion of Lemma~\ref{lem:expmom}, $\Gamma(\psi)\le\Gamma_*(u)$, so $\frac14\kappa_*(u)\sqrt u\le2\Gamma_*(u)(\tilde\delta_\omega+u)$. From $\sqrt u\le\kappa_*(u)/(16\,\Gamma_*(u))$ we get $2\Gamma_*(u)\,u\le\frac18\kappa_*(u)\sqrt u$, so $u\le16\,(\tau(u)/u)\,\tilde\delta_\omega^2$. Since $u>0$, we have $\tilde\delta_\omega>0$, and from $\tilde\delta_\omega^2\le u$ and the fact that $\tau(s)/s$ is nonincreasing (Lemma~\ref{lem:tau}) we obtain $u\le16\,\tau(\tilde\delta_\omega^2)$.

We prove the assertion for the case $\mathcal Z_0=\mathcal Z$. Since $\mathscr J(\omega)\mathcal Z=\ran\mathscr J(\omega)$, we have $\tilde\delta_\omega=\delta_\omega$, and $\delta_\omega^2\le u$ holds by the first assertion of Proposition~\ref{thm:finite}.
\end{proof}

\section{Discussion}\label{sec:discussion}

\paragraph{On the assumption of the prior work.}
The proof of the landscape theorem of the prior work uses the overparameterization assumption in order to deduce, from the fact that at a critical point the directional derivative of the loss of the slice in a maximizing direction vanishes in every direction of the parameters, that this slice coincides with the slice of the target (Lemma 2.11 of the prior work). For networks with Gaussian units, this assumption does not hold at finite width. This assumption corresponds to the case in which the error $\delta_\omega$ of approximating the target by the range of the tangent map is $0$. Even when $\delta_\omega>0$, the infidelity of a stationary point in the neighborhood of the target is at most $\delta_\omega^2$ multiplied by a factor that is a power of a logarithm (Proposition~\ref{thm:finite}). What is needed in the infinite-width limit is not that the range contains all Schwartz functions, but that, for each of the units spanning a dense space, its distance to the range at the stationary points converges to $0$ (Theorem~\ref{thm:limit} and Theorem~\ref{thm:landscape}). About the network, the proofs of Theorem~\ref{thm:limit} and Proposition~\ref{thm:finite} use only that $\omega\mapsto\Psi_\omega$ is a $C^1$ map into $\Qs$, that $\C\psi_\omega\subset\ran\mathscr J(\omega)$, and the approximation of the target by the range. Hence, to obtain the same conclusions as these for the GELU multilayer perceptron with Gaussian envelope of the prior work, it suffices to estimate the approximation of the target by the range at stationary points. The proof of Theorem~\ref{thm:landscape} uses, in addition, that the range contains the complex subspace $\mathcal V(c)$, and the approximation by the range of each of the units spanning a dense subspace. These estimates are problems of approximation theory and are not treated in this paper.

\paragraph{Constants.}
The functions $\kappa_*(s)$ and $\tau(s)/s$ can be written as explicit expressions in $\En(\phi)$ and $E_*(s)$, and $\bar u$ is determined as the maximum of the $s\in(0,\bar s]$ that satisfy $\tau(s)\le1$ (see \eqref{eq:consts}, \eqref{eq:tau}, and \eqref{eq:ubar}). However, the estimates with these explicit constants are pessimistic. In the example of Section~\ref{sec:numerics}, the size $\bar u'$ of the neighborhood is $2.7\times10^{-11}$ for $q\ge2$ and $2.2\times10^{-14}$ for $q=1$, and the infidelities of the points that entered the neighborhood are smaller than the upper bound of Proposition~\ref{prop:restricted} by factors between $2.9\times10^{10}$ and $4.7\times10^{10}$. The proof of Lemma~\ref{lem:separate} uses crude estimates in three places. They are the following: the $\mathcal W_1$ distance is bounded from below by means of a potential that is a trigonometric function of a single frequency, the tail of the difference of the expectations of the Weyl operators in the two states is bounded by the second moment, and the integral over a ball of the squared modulus of this difference is bounded from above, direction by direction, by the supremum over the radius of the quotient of the modulus of this difference by the radius. For $q<2$, the last estimate also uses that the modulus of this difference is at most $2$, so the estimate is even cruder. In addition, the estimates of the quadratic term in Lemma~\ref{prop:sharp} and of the error term in Lemma~\ref{prop:onepoint} are also crude. The second assertion of Lemma~\ref{lem:approx} represents a function as an average of units and approximates it by independent samples, and is an estimate of the same type as the bound of \citet{barron} and the approximation of functions by random features \citep{rahimirecht,rahimirecht2}. Its constant $\|\varrho\|_{L^2(\nu)}\|\varphi_0\|$ can grow exponentially in the number of modes. In that case, the width needed to make the right-hand side of this estimate at most a given value also grows with the number of modes.

\paragraph{Open points.}
When $q=\infty$ and $\Mo_\beta(\phi)\le\Lambda$, the range of infidelity at most $\bar u$ contains no state $\psi$ satisfying $\Mo_\beta(\psi)\le\Lambda$ and $0\in\overline{\conv}\,\mathcal G(\psi)$ other than the target. Indeed, if such a $\psi$ satisfied $0<u(\psi)\le\bar u$, then there would be $v\in\mathcal G(\psi)$ with $\Re\ip v{\ip\phi\psi\phi}\ge0$, and the first assertion of Lemma~\ref{prop:onepoint} would give $\Lc_\infty(\psi)\le2\Gamma(\psi)\,u(\psi)$, which is incompatible with Lemma~\ref{prop:sharp}. Whether such points outside this range are global minimizers is treated neither in the prior work nor in this paper. For $q<\infty$, we can say nothing outside this range. Theorem~\ref{thm:limit}, Theorem~\ref{thm:landscape}, and Proposition~\ref{thm:finite} assume that the exponential moment of the stationary point is at most $\Lambda$. The exponential moment can be computed at the point that has been reached, and this assumption can be checked at that point, but we have not shown that the exponential moment stays bounded during training. Theorem~\ref{thm:limit} and Theorem~\ref{thm:landscape} also assume that the sum of the squared displacements of the centers from their initial values is bounded, and do not treat sequences in which the centers accumulate. Whether the iterates of training converge to a stationary point, and whether they stay in the neighborhood along the way, are also not treated.

\section*{Acknowledgments}
This work was supported by ERATO ``Super Quantum Entanglement'' (Grant No.\ JPMJER2402) from JST.

\bibliography{refs}
\bibliographystyle{plainnat}

\appendix
\section{The Moyal identity}\label{sec:moyal}

We prove, from \eqref{eq:weyldef}, the orthogonality relation for the matrix elements of the Weyl operators that is used in the proof of Lemma~\ref{lem:separate}. This orthogonality relation is that of \citet[Chapter~1, Proposition~(1.42)]{folland} with the variables changed to $(p,q)=(\varpi,\xi/(2\pi))$.

\begin{lemma}[Moyal identity]\label{lem:moyal}
For $f,g\in\Hs$, the function $Y_{g,f}$ in \eqref{eq:Ydef} is a bounded continuous function on $\R^n\times\R^n$ that belongs to $L^2(\R^n\times\R^n)$, and for $f_1,f_2,g_1,g_2\in\Hs$ we have
\begin{equation}\label{eq:moyal}
\int_{\R^n\times\R^n}Y_{g_1,f_1}(\xi,\varpi)\,\overline{Y_{g_2,f_2}(\xi,\varpi)}\,d\xi\,d\varpi=(2\pi)^n\,\ip{g_1}{g_2}\,\ip{f_2}{f_1}.
\end{equation}
\end{lemma}

\begin{proof}
Since $V(\xi,\varpi)$ is unitary, we have $|Y_{g,f}(\xi,\varpi)|\le\|g\|\,\|f\|$, and since $(\xi,\varpi)\mapsto V(\xi,\varpi)f$ is a continuous map into $\Hs$ (see the proof of Lemma~\ref{lem:slice}), the function $Y_{g,f}$ is continuous.

First let $f_1,f_2,g_1,g_2\in\Sc$. For $\varpi\in\R^n$ put $A_\varpi(x):=\overline{g_1(x)}\,f_1(x+\varpi)$ and $B_\varpi(x):=\overline{g_2(x)}\,f_2(x+\varpi)$. Then $A_\varpi$ and $B_\varpi$ belong to $\Sc$, and \eqref{eq:weyldef} gives
\[
Y_{g_1,f_1}(\xi,\varpi)=e^{i\langle\xi,\varpi\rangle/2}\int_{\R^n}e^{i\langle\xi,x\rangle}A_\varpi(x)\,dx,\qquad
Y_{g_2,f_2}(\xi,\varpi)=e^{i\langle\xi,\varpi\rangle/2}\int_{\R^n}e^{i\langle\xi,x\rangle}B_\varpi(x)\,dx.
\]
Fix $\varpi$. In the product $Y_{g_1,f_1}\overline{Y_{g_2,f_2}}$, the product of the factor $e^{i\langle\xi,\varpi\rangle/2}$ of absolute value $1$ and its complex conjugate equals $1$, so the Plancherel identity gives
\[
\int_{\R^n}Y_{g_1,f_1}(\xi,\varpi)\,\overline{Y_{g_2,f_2}(\xi,\varpi)}\,d\xi=(2\pi)^n\int_{\R^n}A_\varpi(x)\,\overline{B_\varpi(x)}\,dx
=(2\pi)^n\int_{\R^n}\overline{g_1(x)}\,g_2(x)\,f_1(x+\varpi)\,\overline{f_2(x+\varpi)}\,dx.
\]
The function $(x,\varpi)\mapsto\overline{g_1(x)}\,g_2(x)\,f_1(x+\varpi)\,\overline{f_2(x+\varpi)}$ is integrable on $\R^n\times\R^n$, so integrating both sides of the identity above with respect to $\varpi$ and using Fubini's theorem and the change of variables $y=x+\varpi$, we obtain
\[
\int_{\R^n\times\R^n}Y_{g_1,f_1}\,\overline{Y_{g_2,f_2}}\,d\xi\,d\varpi=(2\pi)^n\int_{\R^n}\overline{g_1(x)}\,g_2(x)\,dx\int_{\R^n}f_1(y)\,\overline{f_2(y)}\,dy=(2\pi)^n\,\ip{g_1}{g_2}\,\ip{f_2}{f_1}.
\]
Here the left-hand side is the iterated integral taken first in $\xi$ and then in $\varpi$. In the case $g_1=g_2$ and $f_1=f_2$ the integrand is nonnegative, so by Tonelli's theorem the iterated integral equals the double integral, and for $f,g\in\Sc$ we have
\begin{equation}\label{eq:moyalnorm}
\|Y_{g,f}\|_{L^2(\R^n\times\R^n)}=(2\pi)^{n/2}\,\|g\|\,\|f\|.
\end{equation}
Hence, for general $f_1,f_2,g_1,g_2\in\Sc$ as well, the product $Y_{g_1,f_1}\overline{Y_{g_2,f_2}}$ is integrable on $\R^n\times\R^n$ by the Cauchy--Schwarz inequality, and by Fubini's theorem the iterated integral equals the double integral.

Next let $f,g\in\Hs$, and take sequences in $\Sc$ with $f_k\to f$ and $g_k\to g$ (convergence in $\Hs$). Since $Y_{g,f}$ is conjugate-linear in $g$ and linear in $f$, \eqref{eq:moyalnorm} shows that $Y_{g_k,f_k}$ is a Cauchy sequence in $L^2(\R^n\times\R^n)$, and it converges to some $Y\in L^2(\R^n\times\R^n)$. On the other hand, $|Y_{g_k,f_k}-Y_{g,f}|\le\|g_k-g\|\,\|f_k\|+\|g\|\,\|f_k-f\|$ shows that $Y_{g_k,f_k}$ converges uniformly to $Y_{g,f}$. A sequence that converges in $L^2$ has a subsequence that converges almost everywhere, so $Y=Y_{g,f}$ holds almost everywhere, and hence $Y_{g,f}\in L^2(\R^n\times\R^n)$ and $Y_{g_k,f_k}\to Y_{g,f}$ (convergence in $L^2$). For each of $f_1,f_2,g_1,g_2\in\Hs$ take a sequence in $\Sc$ that converges to it in $\Hs$, and pass to the limit in \eqref{eq:moyal} for the elements of these sequences. By the continuity of the inner product of $L^2$ for the left-hand side and by the continuity of the inner product of $\Hs$ for the right-hand side, the two sides converge to the left-hand side and the right-hand side, respectively, of \eqref{eq:moyal} for $f_1,f_2,g_1,g_2$. Hence \eqref{eq:moyal} holds for all $f_1,f_2,g_1,g_2\in\Hs$.
\end{proof}

\section{Generalized gradients}\label{sec:clarke}

We prove from the definitions the properties of the generalized gradient of Clarke that are used in this paper. All of them are special cases, under the assumptions of this paper, of results in \citet{clarke}. The first and third assertions of Lemma~\ref{lem:clarkebasic} correspond to \citet[Proposition~2.1.1, Proposition~2.1.2]{clarke}, Lemma~\ref{lem:clarkemax} to the theorem on pointwise maxima \citep[Theorem~2.8.2]{clarke}, Lemma~\ref{lem:clarkeint} to the theorem on integral functionals \citep[Theorem~2.7.2]{clarke}, and Lemma~\ref{lem:clarkechain} to the chain rule \citep[Theorem~2.3.10]{clarke}. For the statements and proofs in general form, see the results cited above. Let $X$ be a real Banach space, let $X^*$ be its dual space, that is, the set of all continuous linear functionals on $X$, let $U$ be an open subset of $X$, and let $g:U\to\R$ be a locally Lipschitz function. For $x\in U$ and $\zeta\in X$ put
\begin{equation}\label{eq:clarkedef}
\begin{gathered}
D^\circ_\zeta g(x):=\limsup_{x'\to x,\ t\downarrow0}\frac{g(x'+t\zeta)-g(x')}{t},\qquad
\partial^\circ g(x):=\big\{x^*\in X^*:\ x^*(\zeta)\le D^\circ_\zeta g(x)\ \ (\zeta\in X)\big\},\\
D^+_\zeta g(x):=\lim_{t\downarrow0}\frac{g(x+t\zeta)-g(x)}{t}.
\end{gathered}
\end{equation}
Here $D^+_\zeta g(x)$ is defined when the limit on the right-hand side exists, and it is called the one-sided directional derivative. If $D^+_\zeta g(x)$ exists and equals $D^\circ_\zeta g(x)$ for all $\zeta\in X$, we say that $g$ is regular at $x$ in the sense of Clarke \citep{clarke}. The definition \eqref{eq:clarkeQ} is the case $X=\Qs$, $U=\Qs\setminus\{0\}$, and $g=\Lc_q$, and \eqref{eq:clarkeZ} is the case $X=\mathcal Z$, $U=\Omega$, and $g=\Ls_q$. Since we may take $x'=x$ in the limit superior in the definition of $D^\circ_\zeta g(x)$, if $D^+_\zeta g(x)$ exists then $D^+_\zeta g(x)\le D^\circ_\zeta g(x)$.

\begin{lemma}[Basic properties]\label{lem:clarkebasic}
Suppose that $g$ has Lipschitz constant $L$ on a neighborhood of $x$. First, $\zeta\mapsto D^\circ_\zeta g(x)$ takes finite values and satisfies $|D^\circ_\zeta g(x)|\le L\|\zeta\|$, and we have $D^\circ_{\lambda\zeta}g(x)=\lambda D^\circ_\zeta g(x)$ for $\lambda\ge0$ and $D^\circ_{\zeta_1+\zeta_2}g(x)\le D^\circ_{\zeta_1}g(x)+D^\circ_{\zeta_2}g(x)$. Second, if $Y$ is a subspace of $X$ and $D^\circ_\zeta g(x)\ge0$ for all $\zeta\in Y$, then there is $x^*\in\partial^\circ g(x)$ that vanishes on $Y$. Third, for every $\zeta_0\in X$ there is $x^*\in\partial^\circ g(x)$ with $x^*(\zeta_0)=D^\circ_{\zeta_0}g(x)$. In particular $\partial^\circ g(x)$ is nonempty and $D^\circ_\zeta g(x)=\max_{x^*\in\partial^\circ g(x)}x^*(\zeta)$.
\end{lemma}

\begin{proof}
Put $p(\zeta):=D^\circ_\zeta g(x)$. If $x'$ is sufficiently close to $x$ and $t$ is sufficiently small, then $|g(x'+t\zeta)-g(x')|\le Lt\|\zeta\|$, so $|p(\zeta)|\le L\|\zeta\|$. For $\lambda>0$ we have $t^{-1}(g(x'+t\lambda\zeta)-g(x'))=\lambda\,(\lambda t)^{-1}(g(x'+\lambda t\zeta)-g(x'))$, so $p(\lambda\zeta)=\lambda p(\zeta)$, and $p(0)=0$. Moreover, we have
\[
\frac{g(x'+t\zeta_1+t\zeta_2)-g(x')}{t}=\frac{g\big((x'+t\zeta_1)+t\zeta_2\big)-g(x'+t\zeta_1)}{t}+\frac{g(x'+t\zeta_1)-g(x')}{t},
\]
and $x'+t\zeta_1\to x$ as $x'\to x$ and $t\downarrow0$, so taking the limit superior gives $p(\zeta_1+\zeta_2)\le p(\zeta_2)+p(\zeta_1)$. This proves the first assertion.

Since $p$ is sublinear, the Hahn--Banach extension theorem shows that if a linear functional $y^*$ on a subspace $Y_0$ of $X$ satisfies $y^*\le p$ on $Y_0$, then there is a linear functional $x^*$ on $X$ that equals $y^*$ on $Y_0$ and satisfies $x^*(\zeta)\le p(\zeta)$ for all $\zeta\in X$. In this situation, applying $x^*(\zeta)\le p(\zeta)\le L\|\zeta\|$ to $\zeta$ and to $-\zeta$ gives $|x^*(\zeta)|\le L\|\zeta\|$, so $x^*$ is bounded, hence continuous, and $x^*\in\partial^\circ g(x)$. The second assertion follows by taking $Y_0:=Y$ and $y^*:=0$, because $y^*\le p$ on $Y$ by hypothesis. To prove the third assertion, take $Y_0:=\R\zeta_0$ and $y^*(\lambda\zeta_0):=\lambda p(\zeta_0)$. The inequality $y^*\le p$ holds on $Y_0$ because $y^*(\lambda\zeta_0)=p(\lambda\zeta_0)$ if $\lambda\ge0$, and because $0=p(0)\le p(\zeta_0)+p(-\zeta_0)$ gives $y^*(\lambda\zeta_0)=-|\lambda|p(\zeta_0)\le|\lambda|p(-\zeta_0)=p(\lambda\zeta_0)$ if $\lambda<0$. The extension $x^*$ belongs to $\partial^\circ g(x)$ and satisfies $x^*(\zeta_0)=y^*(\zeta_0)=p(\zeta_0)$. The last identity of the lemma follows from the fact that every $x^*\in\partial^\circ g(x)$ satisfies $x^*(\zeta)\le D^\circ_\zeta g(x)$ by the definition of $\partial^\circ g(x)$, and from the third assertion.
\end{proof}

\begin{lemma}[Maxima]\label{lem:clarkemax}
Let $T$ be a compact metrizable space, and let $g_t:U\to\R$ ($t\in T$) satisfy the following three hypotheses. First, $(t,y)\mapsto g_t(y)$ is continuous on $T\times U$. Second, for all $t\in T$, $y\in U$, and $\zeta\in X$ the limit
\[
Dg_t(y)[\zeta]:=\lim_{s\to0}\frac{g_t(y+s\zeta)-g_t(y)}{s}
\]
exists, and for each $\zeta$ the map $(t,y)\mapsto Dg_t(y)[\zeta]$ is continuous on $T\times U$. Third, every point of $U$ has a neighborhood on which all the $g_t$ have a Lipschitz constant that does not depend on $t$. For $y\in U$ put
\[
g(y):=\max_{t\in T}g_t(y),\qquad T(y):=\{t\in T:g_t(y)=g(y)\}.
\]
Then $g$ is locally Lipschitz, and for all $x\in U$ and $\zeta\in X$ the one-sided directional derivative $D^+_\zeta g(x)$ exists and
\[
D^\circ_\zeta g(x)=D^+_\zeta g(x)=\max_{t\in T(x)}Dg_t(x)[\zeta].
\]
\end{lemma}

\begin{proof}
By the first hypothesis the maximum is attained and $T(y)$ is a nonempty compact set. By the third hypothesis, every point of $U$ has a neighborhood on which $g$ has a Lipschitz constant common to all the $g_t$. Take $x\in U$ and $\zeta\in X$, and put $D_{\max}:=\max_{t\in T(x)}Dg_t(x)[\zeta]$ (the maximum is attained by the second hypothesis).

Let $t\in T(x)$. If $s>0$ is sufficiently small, then $g(x+s\zeta)-g(x)\ge g_t(x+s\zeta)-g_t(x)$, so $\liminf_{s\downarrow0}s^{-1}(g(x+s\zeta)-g(x))\ge Dg_t(x)[\zeta]$. Since $t\in T(x)$ is arbitrary, this lower limit is at least $D_{\max}$.

Next we show $D^\circ_\zeta g(x)\le D_{\max}$. Take $x_k\to x$ and $s_k\downarrow0$ such that $s_k^{-1}(g(x_k+s_k\zeta)-g(x_k))\to D^\circ_\zeta g(x)$, and take $t_k\in T(x_k+s_k\zeta)$. Since $g(x_k)\ge g_{t_k}(x_k)$, we have
\[
g(x_k+s_k\zeta)-g(x_k)\ \le\ g_{t_k}(x_k+s_k\zeta)-g_{t_k}(x_k).
\]
If $k$ is sufficiently large, the segment $\{x_k+s'\zeta:0\le s'\le s_k\}$ is contained in $U$, and by the second hypothesis $s'\mapsto g_{t_k}(x_k+s'\zeta)$ is differentiable with derivative $Dg_{t_k}(x_k+s'\zeta)[\zeta]$. Hence, by the mean value theorem, there is $s_k'\in(0,s_k)$ for which the right-hand side of the inequality above equals $s_k\,Dg_{t_k}(x_k+s_k'\zeta)[\zeta]$. Since $T$ is compact and metrizable, passing to a subsequence we may assume that $t_k\to\hat t$ for some $\hat t\in T$. By the first hypothesis and the continuity of $g$, in the identity $g_{t_k}(x_k+s_k\zeta)=g(x_k+s_k\zeta)$ the left-hand side converges to $g_{\hat t}(x)$ and the right-hand side to $g(x)$, so $\hat t\in T(x)$. By the second hypothesis, $Dg_{t_k}(x_k+s_k'\zeta)[\zeta]\to Dg_{\hat t}(x)[\zeta]\le D_{\max}$, so $D^\circ_\zeta g(x)\le D_{\max}$.

From these two estimates and the fact that the limit superior obtained by taking $x'=x$ in the definition of $D^\circ_\zeta g(x)$ is at most $D^\circ_\zeta g(x)$, we have
\[
D_{\max}\ \le\ \liminf_{s\downarrow0}\frac{g(x+s\zeta)-g(x)}{s}\ \le\ \limsup_{s\downarrow0}\frac{g(x+s\zeta)-g(x)}{s}\ \le\ D^\circ_\zeta g(x)\ \le\ D_{\max},
\]
and the assertion follows.
\end{proof}

\begin{remark}\label{rem:regular}
The identity $D^+_\zeta g(x)=D^\circ_\zeta g(x)$ is not a general property of locally Lipschitz functions. For example, the function $g(x)=-|x|$ on $\R$ satisfies $D^+_\zeta g(0)=-1$ and $D^\circ_\zeta g(0)=1$ for $x=0$ and $\zeta=1$. In this paper the identity holds because $\mathcal W_1$ can be written, by the Kantorovich duality \eqref{eq:KR}, as the maximum of a family of $C^1$ functions.
\end{remark}

\begin{lemma}[Aggregation by the $L^q$ norm]\label{lem:clarkeint}
Let $(S,\varsigma)$ be a probability space, and let $w_\theta:U\to[0,\infty)$ ($\theta\in S$) and $x\in U$ satisfy the following three hypotheses. First, for each $y\in U$ the function $\theta\mapsto w_\theta(y)$ is measurable. Second, there are a neighborhood $U_0$ of $x$ and constants $L,\bar w$ such that, for all $\theta$, the function $w_\theta$ has Lipschitz constant $L$ on $U_0$ and satisfies $w_\theta\le\bar w$ there. Third, for all $\theta\in S$ and $\zeta\in X$ the one-sided directional derivative $D^+_\zeta w_\theta(x)$ exists and equals $D^\circ_\zeta w_\theta(x)$. Let $1\le q<\infty$. For $y\in U_0$ put
\[
G(y):=\Big(\int_Sw_\theta(y)^q\,d\varsigma(\theta)\Big)^{1/q},
\]
and assume $G(x)>0$. For $\theta\in S$ put
\[
\rho_\theta:=\Big(\frac{w_\theta(x)}{G(x)}\Big)^{q-1}\quad(q>1),\qquad\rho_\theta:=1\quad(q=1).
\]
Then $G$ has Lipschitz constant $L$ on $U_0$. Moreover, for every $\zeta\in X$, the function $\theta\mapsto D^\circ_\zeta w_\theta(x)$ is bounded and measurable, the one-sided directional derivative $D^+_\zeta G(x)$ exists, and
\[
D^\circ_\zeta G(x)=D^+_\zeta G(x)=\int_S\rho_\theta\,D^\circ_\zeta w_\theta(x)\,d\varsigma(\theta).
\]
\end{lemma}

\begin{proof}
For $y,y'\in U_0$, Minkowski's inequality gives $|G(y)-G(y')|\le\|\theta\mapsto w_\theta(y)-w_\theta(y')\|_{L^q(\varsigma)}\le L\|y-y'\|$. Fix $\zeta\in X$ and put $H:=G^q$. For $y\in U_0$ and $s>0$ with $y+s\zeta\in U_0$ put
\[
Q_\theta(y,s):=\frac{w_\theta(y+s\zeta)^q-w_\theta(y)^q}{s},\qquad b_\theta(y,s):=\frac{w_\theta(y+s\zeta)-w_\theta(y)}{s}.
\]
Then $s^{-1}(H(y+s\zeta)-H(y))=\int_SQ_\theta(y,s)\,d\varsigma$. By the mean value theorem for the function $r\mapsto r^q$ on $[0,\bar w]$, there is $r_\theta(y,s)$ between $w_\theta(y)$ and $w_\theta(y+s\zeta)$ such that
\[
Q_\theta(y,s)=q\,r_\theta(y,s)^{q-1}\,b_\theta(y,s).
\]
When $q=1$ we have $Q_\theta=b_\theta$, and in what follows, when $q=1$, the power of $0$ with exponent $q-1$ is also understood to be $1$. Since $|b_\theta(y,s)|\le L\|\zeta\|$, we have $|Q_\theta(y,s)|\le C:=q\,\bar w^{q-1}L\|\zeta\|$. As $y\to x$ and $s\downarrow0$, the inequality $|w_\theta(y)-w_\theta(x)|\le L\|y-x\|$ gives $r_\theta(y,s)\to w_\theta(x)$.

Let $y=x$. By the definition of the one-sided directional derivative, $b_\theta(x,s)\to D^+_\zeta w_\theta(x)$ as $s\downarrow0$, and by the third hypothesis this limit equals $D^\circ_\zeta w_\theta(x)$. Hence $\theta\mapsto D^\circ_\zeta w_\theta(x)$ is measurable as the limit of a sequence of measurable functions, its absolute value is at most $L\|\zeta\|$, and $Q_\theta(x,s)\to q\,w_\theta(x)^{q-1}D^\circ_\zeta w_\theta(x)$. By the dominated convergence theorem, $D^+_\zeta H(x)$ exists and
\[
D^+_\zeta H(x)=\int_Sq\,w_\theta(x)^{q-1}D^\circ_\zeta w_\theta(x)\,d\varsigma(\theta).
\]
Next, take $x_k\to x$ and $s_k\downarrow0$ such that $\int_SQ_\theta(x_k,s_k)\,d\varsigma\to D^\circ_\zeta H(x)$. Applying Fatou's lemma to $C-Q_\theta(x_k,s_k)\ge0$ gives
\[
D^\circ_\zeta H(x)=\lim_{k\to\infty}\int_SQ_\theta(x_k,s_k)\,d\varsigma\ \le\ \int_S\limsup_{k\to\infty}Q_\theta(x_k,s_k)\,d\varsigma.
\]
For each $\theta$, the factor $q\,r_\theta(x_k,s_k)^{q-1}$ converges to $q\,w_\theta(x)^{q-1}\ge0$, and $b_\theta(x_k,s_k)$ is bounded and satisfies $\limsup_kb_\theta(x_k,s_k)\le D^\circ_\zeta w_\theta(x)$, so $\limsup_kQ_\theta(x_k,s_k)\le q\,w_\theta(x)^{q-1}D^\circ_\zeta w_\theta(x)$. Hence $D^\circ_\zeta H(x)\le D^+_\zeta H(x)$, and since the reverse inequality always holds, the two are equal.

Finally, $H(x)=G(x)^q>0$ and $H$ is continuous at $x$, so $H>0$ on a neighborhood of $x$. By the mean value theorem for $r\mapsto r^{1/q}$, there is $r$ between $H(y)$ and $H(y+s\zeta)$ such that
\[
\frac{G(y+s\zeta)-G(y)}{s}=\frac1q\,r^{\frac1q-1}\,\frac{H(y+s\zeta)-H(y)}{s},
\]
and $\frac1qr^{\frac1q-1}$ converges to $\frac1qH(x)^{\frac1q-1}>0$ as $y\to x$ and $s\downarrow0$. Hence $D^+_\zeta G(x)$ exists and $D^\circ_\zeta G(x)=\frac1qH(x)^{\frac1q-1}D^\circ_\zeta H(x)=\frac1qH(x)^{\frac1q-1}D^+_\zeta H(x)=D^+_\zeta G(x)$. Since $\frac1qH(x)^{\frac1q-1}\,q\,w_\theta(x)^{q-1}=\rho_\theta$, the assertion follows.
\end{proof}

\begin{definition}[Fr\'echet derivative and $C^1$ maps]\label{def:frechet}
Let $Z$ and $X$ be real Banach spaces, let $\Omega_0$ be an open subset of $Z$, and let $F:\Omega_0\to X$ be a map. We say that $F$ is Fr\'echet differentiable at $\omega\in\Omega_0$ if there is a continuous linear map $DF(\omega)$ from $Z$ to $X$ such that
\[
\lim_{\zeta\to0}\frac{\|F(\omega+\zeta)-F(\omega)-DF(\omega)\zeta\|}{\|\zeta\|}=0.
\]
The map $DF(\omega)$ is called the Fr\'echet derivative of $F$ at $\omega$, and its value at $\zeta\in Z$ is written $DF(\omega)\zeta\in X$. If $F$ is Fr\'echet differentiable at every point of $\Omega_0$ and $\omega\mapsto DF(\omega)$ is continuous with respect to the operator norm, then $F$ is called a $C^1$ map.
\end{definition}

\begin{lemma}[Composition with a $C^1$ map]\label{lem:clarkechain}
Let $X$, $U$, and $g$ be as at the beginning of this section. Let $Z$ be a real Banach space, let $\Omega_0$ be an open subset of $Z$, and let $F:\Omega_0\to U$ be a $C^1$ map. Then $g\circ F$ is locally Lipschitz on $\Omega_0$, and for all $\omega\in\Omega_0$ and $\zeta\in Z$ the generalized directional derivative of $g\circ F$ at $\omega$ in the direction $\zeta$ is at most the generalized directional derivative of $g$ at $F(\omega)$ in the direction $DF(\omega)\zeta$. That is,
\[
D^\circ_\zeta(g\circ F)(\omega)\ \le\ D^\circ_{DF(\omega)\zeta}\,g\big(F(\omega)\big).
\]
Moreover, if $D^+_{DF(\omega)\zeta}\,g(F(\omega))$ exists and equals $D^\circ_{DF(\omega)\zeta}\,g(F(\omega))$, then $D^+_\zeta(g\circ F)(\omega)$ exists and equality holds in the inequality above. In particular, if $0\in\partial^\circ(g\circ F)(\omega)$, then there is $x^*\in\partial^\circ g(F(\omega))$ that vanishes on the range $\{DF(\omega)\zeta:\zeta\in Z\}$ of $DF(\omega)$.
\end{lemma}

\begin{proof}
Let $\omega\in\Omega_0$, and let $B\subset\Omega_0$ be a ball centered at $\omega$. We first show that, for $\omega_1,\omega_2\in B$,
\begin{equation}\label{eq:mvi}
\big\|F(\omega_2)-F(\omega_1)-DF(\omega)(\omega_2-\omega_1)\big\|\ \le\ \sup_{\omega'\in B}\|DF(\omega')-DF(\omega)\|\,\|\omega_2-\omega_1\|.
\end{equation}
Write $v$ for the vector inside the norm on the left-hand side of \eqref{eq:mvi}. By the Hahn--Banach theorem, there is a continuous linear functional $x^*\in X^*$ with $\|x^*\|=1$ and $x^*(v)=\|v\|$. Since $B$ is convex, we have $\omega_1+t(\omega_2-\omega_1)\in B$ for $0\le t\le1$, and the real-valued function
\[
\chi(t):=x^*\Big(F\big(\omega_1+t(\omega_2-\omega_1)\big)-t\,DF(\omega)(\omega_2-\omega_1)\Big)
\]
is differentiable, because $x^*$ is continuous and linear, with $\frac{d}{dt}\chi(t)=x^*\big((DF(\omega_1+t(\omega_2-\omega_1))-DF(\omega))(\omega_2-\omega_1)\big)$. By the mean value theorem, there is $t_0\in(0,1)$ with $\|v\|=\chi(1)-\chi(0)=\frac{d}{dt}\chi(t_0)$, and since $\|x^*\|=1$, the derivative $\frac{d}{dt}\chi(t_0)$ is at most the right-hand side of \eqref{eq:mvi}. Since $DF$ is continuous at $\omega$, if the radius of $B$ is taken sufficiently small then $\sup_{\omega'\in B}\|DF(\omega')-DF(\omega)\|$ is finite, and \eqref{eq:mvi} and the triangle inequality give, for $\omega_1,\omega_2\in B$,
\begin{align*}
\|F(\omega_2)-F(\omega_1)\|&\le\big\|F(\omega_2)-F(\omega_1)-DF(\omega)(\omega_2-\omega_1)\big\|+\|DF(\omega)(\omega_2-\omega_1)\|\\
&\le\Big(\sup_{\omega'\in B}\|DF(\omega')-DF(\omega)\|+\|DF(\omega)\|\Big)\|\omega_2-\omega_1\|.
\end{align*}
That is, $F$ is Lipschitz on $B$. Hence $g\circ F$ is locally Lipschitz. Let $L$ be a Lipschitz constant of $g$ on a neighborhood of $F(\omega)$, and take $\zeta\in Z$. If $\zeta=0$, both sides of the asserted inequality are $0$, so we assume $\zeta\ne0$. As $\omega'\to\omega$ and $t\downarrow0$, the points $\omega'$ and $\omega'+t\zeta$ belong to the ball centered at $\omega$ with radius $2(\|\omega'-\omega\|+t\|\zeta\|)$, and this radius is positive and converges to $0$. Hence this ball is contained in $\Omega_0$ if $\omega'$ is sufficiently close to $\omega$ and $t$ is sufficiently small. Taking $B$ to be this ball and applying \eqref{eq:mvi} to $\omega_1=\omega'$ and $\omega_2=\omega'+t\zeta$, we see, since $DF$ is continuous at $\omega$, that the element $e\in X$ defined by $F(\omega'+t\zeta)=F(\omega')+t\,DF(\omega)\zeta+t\,e$ satisfies $\|e\|\to0$. Hence
\[
\frac{g(F(\omega'+t\zeta))-g(F(\omega'))}{t}\ \le\ \frac{g\big(F(\omega')+t\,DF(\omega)\zeta\big)-g(F(\omega'))}{t}+L\|e\|,
\]
and since $F(\omega')\to F(\omega)$, taking the limit superior gives the asserted inequality. If $0\in\partial^\circ(g\circ F)(\omega)$, then $0\le D^\circ_\zeta(g\circ F)(\omega)\le D^\circ_{DF(\omega)\zeta}\,g(F(\omega))$ for all $\zeta\in Z$, so applying the second assertion of Lemma~\ref{lem:clarkebasic} with $Y$ taken to be the range of $DF(\omega)$ gives the last assertion.

We prove the assertion on equality. Let $\zeta\in Z$, and for $t>0$ define $e_t\in X$ by $F(\omega+t\zeta)=F(\omega)+t\,DF(\omega)\zeta+t\,e_t$. Then $\|e_t\|\to0$ as $t\downarrow0$. The absolute value of the difference between $t^{-1}(g(F(\omega+t\zeta))-g(F(\omega)))$ and $t^{-1}(g(F(\omega)+t\,DF(\omega)\zeta)-g(F(\omega)))$ is at most $L\|e_t\|$, so by hypothesis $D^+_\zeta(g\circ F)(\omega)$ exists and equals $D^\circ_{DF(\omega)\zeta}\,g(F(\omega))$. A one-sided directional derivative is at most the generalized directional derivative, so $D^\circ_{DF(\omega)\zeta}\,g(F(\omega))\le D^\circ_\zeta(g\circ F)(\omega)$, and together with the inequality already proved this gives equality.
\end{proof}

\section{List of notation}\label{sec:notation}

We collect the notation used throughout this paper together with the places where it is defined. Notation used only within a single proof is not listed. For $E_*$, $\Gamma_*$, $\kappa_*$, $\tau$, and $\Gamma$, which are defined as functions, we write the argument when referring to a value. For the infidelity, we write $u:=u(\psi)$ at the beginning of a statement and omit the argument. For the symbols $\mathcal W_\theta(a,b)$, $\Lc_q(a,b)$, and $u(a,b)$, which concern two states, the second argument is omitted when it is the target $\phi$. The symbol $\varepsilon$ is the positive number in \eqref{eq:net} and is a different letter from the $\epsilon$ in $\epsilon_N$ of \eqref{eq:epsN}.

\newcommand{\notationrows}[2]{\par\smallskip\noindent{\footnotesize\textbf{#1}\par\nopagebreak\noindent
\begin{tabular}{@{}p{39mm}p{80mm}p{36mm}@{}}
\toprule
Symbol & Meaning & Defined in\\
\midrule
#2
\bottomrule
\end{tabular}\par}}

\notationrows{Spaces and sets}{
$\Hs$, $\ip fg$ & $L^2(\R^n)$ and its inner product & \eqref{eq:hilbert}\\
$\Qs$, $\En$, $\|\cdot\|_{\Qs}$ & space of elements of finite energy, energy, norm & \eqref{eq:energy}\\
$\Sc$, $\N_0$ & space of Schwartz functions, $\{0,1,2,\dots\}$ & Section~\ref{sec:quadslice}, Section~\ref{sec:results}\\
$\Sph$ & unit sphere of $\R^n\times\R^n$ (the set of all directions $\theta=(\xi,\varpi)$) & Definition~\ref{def:slice}\\
$\Kc$, $\Ac$ & set of all Kantorovich potentials $\mu$, $\Sph\times\Kc$ & \eqref{eq:potentials}\\
$\Kc_\theta(\psi)$, $\Ac(\psi)$ & set of all potentials that attain the supremum, active set & \eqref{eq:active}\\
$\mathcal G(\psi)$, $\overline{\conv}\,\mathcal G(\psi)$ & set of active gradients, its closed convex hull & \eqref{eq:vdef}\\
$\mathcal Z$, $\mathcal Z^*$, $\Omega$ & parameter space, set of all real-linear functionals on it, set of all points with $\Psi_\omega\ne0$ & Definition~\ref{def:rbf}, Definition~\ref{def:stat}\\
$\mathcal V(c)$, $\Pi_{\mathcal V(c)}$ & span of the units, orthogonal projection onto it & \eqref{eq:Vc}\\
$\mathcal Z_0$ & real subspace of $\mathcal Z$ (the set of all directions in which stationarity is assumed) & Proposition~\ref{prop:restricted}\\
$\Hs_N$, $\Pi_N$ & for $n=1$, span of the Hermite functions of photon number at most $N$, orthogonal projection onto it & \eqref{eq:HN}\\
$\mathcal D$, $\mathcal D_N$ & set of all finite linear combinations of the Hermite functions $h_k$, set of all linear combinations of those $h_k$ with $|k|=N$ & \eqref{eq:ladder}\\
}

\notationrows{Operators}{
$x_j$, $\partial_j$, $p_j$, $\hat N$ & multiplication, partial derivative, $-i\partial_j$, total photon number & Section~\ref{sec:quadslice}\\
$X_{\xi,\varpi}$, $X_\theta$, $E_\theta$ & quadrature, its spectral measure & Definition~\ref{def:slice}\\
$V(\xi,\varpi)$ & Weyl operator $e^{iX_{\xi,\varpi}}$ & Definition~\ref{def:slice}, \eqref{eq:weyldef}\\
$M_{\theta,\mu}$ & $\mu(X_\theta)$ & \eqref{eq:potentials}\\
$\mathscr J(\omega)$, $\ran\mathscr J(\omega)$ & tangent map, its range & \eqref{eq:tangent}\\
$\hat a_j$, $\hat a_j^*$ & annihilation operator, creation operator & \eqref{eq:ladder}\\
}

\notationrows{Vectors and parameters}{
$\phi$, $\psi$ & target, state (unit vectors in $\Qs$) & Section~\ref{sec:quadslice}, Section~\ref{sec:loss}\\
$\varphi_y$, $\Psi_\omega$, $\psi_\omega$ & unit, network state before normalization, network state & \eqref{eq:net}\\
$\omega=(a,c)$, $a_j$, $c_j$, $m$ & point of the network, amplitude, center, width & Definition~\ref{def:rbf}\\
$\zeta$ & element of $\mathcal Z$ (direction in the parameter space) & Definition~\ref{def:rbf}\\
$h_k$, $|k|$ & orthonormal basis consisting of products of Hermite functions, $k_1+\dots+k_n$ & Section~\ref{sec:results}\\
$g_\psi$, $\Gamma(\psi)$ & error direction, coefficient determined by the energies of the target and of the error direction & \eqref{eq:udef}\\
$v_{\theta,\mu}(\psi)$ & $\frac12$ times the gradient of $\ell_{\theta,\mu}$ on the unit sphere & \eqref{eq:vdef}\\
$c_1,c_2,\dots$, $\omega_k$, $m_k$, $c^{(k)}$, $R$ & sequence of initial centers, bounded sequence of stationary points, its widths and centers, upper bound on the displacement & Definition~\ref{def:seq}, \eqref{eq:seqbound}\\
$c'$, $\psi_*$ & centers of a point of the network when written alongside the initial centers, limit point & Definition~\ref{def:seq}, Theorem~\ref{thm:limit}\\
}

\notationrows{Measures}{
$\Pm^\theta_f$, $F^\theta_f$ & slice, its distribution function & \eqref{eq:slicedef}\\
$\varsigma$ & rotation-invariant probability measure on $\Sph$ & Definition~\ref{def:loss}\\
$\nu$, $\varrho$ & distribution of the initial centers, coefficient that expresses the target as a superposition of units & \eqref{eq:superpos}\\
$\Pr$, $\mathbb E$ & probability with respect to the initial centers, expectation & proof of Lemma~\ref{lem:approx}\\
}

\notationrows{Functionals and losses}{
$\Mo_\beta$ & exponential moment of the photon number & \eqref{eq:Mbeta}\\
$\mathcal W_\theta(a,b)$, $u(a,b)$ & 1-Wasserstein distance $\mathcal W_1$ between the slices, infidelity & \eqref{eq:Wtheta}\\
$\ell_{\theta,\mu}$ & integral of the potential $\mu$ against the difference of the slices & \eqref{eq:ell}\\
$\rho_\theta$ & weight in the formula for the generalized directional derivative of the loss with $q<\infty$ & \eqref{eq:rho}\\
$Y_{g,f}$ & $(\xi,\varpi)\mapsto\ip g{V(\xi,\varpi)f}$ & \eqref{eq:Ydef}\\
$\Lc_q(a,b)$, $\Lc_q(\psi)$, $\Ls_q(\omega)$ & loss, loss as a function of the point of the network & \eqref{eq:Lq}, \eqref{eq:Lsq}\\
$\theta_k$, $\Lc_{q,K}$, $\Ls_{q,K}$ & equally spaced angles, loss over finitely many angles & \eqref{eq:LqK}\\
$\dist$, $\delta_\omega$, $\tilde\delta_\omega$ & distance in $\Hs$, error of approximating the target by the range of the tangent map, the same error for the image of a subspace & Section~\ref{sec:nets}, \eqref{eq:deltaomega}, \eqref{eq:deltatilde}\\
$D^\circ$, $\partial^\circ$ & generalized directional derivative, generalized gradient & \eqref{eq:clarkeQ}, \eqref{eq:clarkeZ}, \eqref{eq:clarkedef}\\
}

\notationrows{Constants and thresholds}{
$n$, $q$, $\sigma$, $\varepsilon$ & number of modes, exponent of the loss, positive numbers in \eqref{eq:net} & Section~\ref{sec:quadslice}, Definition~\ref{def:loss}, Definition~\ref{def:rbf}\\
$\beta$, $\Lambda$ & index of the exponential moment $\Mo_\beta$, upper bound on the exponential moment & \eqref{eq:Mbeta}, Section~\ref{sec:results}\\
$r$, $\gamma$, $\kappa$ & constants determined by $n$ and $q$ alone & \eqref{eq:exponents}\\
$E_*(s)$, $\Gamma_*(s)$, $\kappa_*(s)$ & upper bound on the energy of the error direction, coefficients determined by it & \eqref{eq:consts}\\
$\tau(s)$, $\bar s$, $\bar u$, $\bar u'$ & function giving the upper bound on the infidelity, right endpoint of the interval on which Lemma~\ref{lem:tau} shows that $\tau$ is nondecreasing, thresholds for the infidelity (sizes of the neighborhood of the target) & \eqref{eq:tau}, \eqref{eq:ubar}\\
$K$, $N$, $\epsilon_N$ & number of angles, photon-number cutoff, error due to the cutoff & \eqref{eq:LqK}, \eqref{eq:epsN}\\
$\eta$, $\delta_m$ & probability level, probabilistic upper bound on $\dist(\phi,\mathcal V(c'))$ & \eqref{eq:deltam}\\
$\alpha$ & global phase & Lemma~\ref{prop:sharp}\\
}

\end{document}